\documentclass[useAMS,usenatbib]{mn2e}

\usepackage{times}
\usepackage{rotating}
\usepackage{amsmath}
\usepackage{amssymb}
\usepackage[dvipsnames]{xcolor}
\usepackage{graphicx}
\usepackage{graphics}
\usepackage{upgreek}
\usepackage{booktabs}
\usepackage{mathtools, cuted}
\usepackage{hyperref}
\hypersetup{
     colorlinks   = true,
     citecolor    = blue
}

\def\apj{{ ApJ}}
\def\apjl{{ApJL}}
\def\apjs{{ ApJS}}

\def\aap{{ A\&A}}

\def\mnras{{ MNRAS}}
\def\araa{{ ARA\&A}}

\def\nat {{ Nature}}

\def\physrep{{ Physics Reports}}

\def\prd{{ Phys. Rev. D}}

\def\mr{\mathrm}
\def\pi{\uppi}
\def\mc{\mathcal}
\def\d{\mathrm{d}}
\def\para{\parallel}
\def\rg{r_{\rm g}}
\def\rt{r_{\rm T}}
\def\ri{r_{\rm I}}
\def\rp{r_{\rm p}}
\def\ra{r_{\rm a}}
\def\rmb{r_{\rm mb}}
\def\rtr{r_{\rm tr}}
\def\rth{r_{\rm th}}
\def\rin{r_{\rm in}}
\def\rdec{r_{\rm dec}}
\def\bv{\bar{v}}
\def\bthe{\bar{\theta}}
\def\bphi{\bar{\phi}}
\def\mui{\mu_{\rm I}}
\def\b{\boldsymbol}
\def\t{\widetilde}
\def\eps{\mc{E}}
\def\tan{\mr{tan}}
\def\msun{\mr{M_{\odot}}}
\def\mp{m_{\rm p}}
\def\me{m_{\rm e}}
\def\epse{\epsilon_{\rm e}}
\def\epsB{\epsilon_{\rm B}}

\makeatletter
\newcommand{\lambdabar}{{\mathchoice
  {\smash@bar\textfont\displaystyle{0.25}{1.2}\lambda}
  {\smash@bar\textfont\textstyle{0.25}{1.2}\lambda}
  {\smash@bar\scriptfont\scriptstyle{0.25}{1.2}\lambda}
  {\smash@bar\scriptscriptfont\scriptscriptstyle{0.25}{1.2}\lambda}
}}
\newcommand{\smash@bar}[4]{%
  \smash{\rlap{\raisebox{-#3\fontdimen5#10}{$\m@th#2\mkern#4mu\mathchar'26$}}}%
}
\makeatother

\newcommand{\myemail}{wenbinlu@caltech.edu}

\title[TDE Stream Self-Intersection]{Self-intersection of the
  Fallback Stream in Tidal Disruption Events}
\author[W. Lu \& C. Bonnerot]
  {Wenbin Lu$^{1}$\thanks{\myemail} and Cl{\'e}ment
    Bonnerot$^{1}$\thanks{bonnerot@tapir.caltech.edu}\\ 
$^1$TAPIR, Mail Code 350-17, California Institute of Technology,
Pasadena, CA 91125, USA\\
}

\begin{document}
\label{firstpage}
\maketitle

\begin{abstract}
We propose a semi-analytical model for the self-intersection of the
fallback stream in tidal disruption events (TDEs). When the 
initial periapsis is less than about 15 gravitational radii, a large fraction
of the shocked gas is unbound in the form of a collision-induced
outflow (CIO). This is because large apsidal
precession causes the stream to self-intersect near the local escape speed at
radius much below the apocenter. The rest of the fallback gas is left
in more tightly bound orbits and quickly joins the accretion flow. We propose
that the CIO is responsible for reprocessing the hard emission from the 
accretion flow into the optical band. This picture naturally explains the large
photospheric radius (or low blackbody temperature) and typical
line widths for optical TDEs. 
We predict the CIO-reprocessed spectrum in the infrared to be
$L_{\nu}\propto \nu^{\sim0.5}$, shallower than a blackbody.
The partial sky coverage of the CIO also provides a unification of
the diverse X-ray behaviors of optical TDEs. According to this
picture, optical surveys filter out a large fraction of TDEs with
low-mass blackholes due to lack of a reprocessing layer, and the
volumetric rate of optical TDEs is nearly flat wrt. the blackhole mass in the
range $M\lesssim 10^7\msun$. This filtering also causes the optical TDE
rate to be lower than the total rate by a factor of $\sim$10 or
more. When the CIO is  decelerated by the ambient medium, radio
emission at the level of that in ASASSN-14li is produced, but the
timescales and peak luminosities can be highly diverse.
Finally, our method paves the way for global
simulations of the disk formation process by injecting gas at the
intersection point according to the prescribed velocity and density
profiles.




\end{abstract}


\section{Introduction}
Tidal disruption events (TDEs) hold promise for probing the otherwise
dormant supermassive blackholes (BHs) at the centers of most
galaxies \citep{1988Natur.333..523R}. The story starts with simple initial
conditions: a star, of certain mass and radius, approaches the BH on a
parabolic orbit 
of certain specific angular momentum. The star can be treated as a point 
mass until it reaches the tidal radius where the tidal forces exceed the star's
self-gravity. The hydrodynamical disruption phase, despite its
complexity, is understood to at least order-unity level, thanks to
decades of analytical and numerical studies
\citep[e.g.,][]{1982ApJ...262..120L, 1983A&A...121...97C, 1988Natur.333..523R,
  1989ApJ...346L..13E, 1993ApJ...410L..83L, 2000ApJ...545..772A, 
  2009MNRAS.392..332L, 2013MNRAS.435.1809S,
  2013ApJ...767...25G, 2017MNRAS.469.4483T, 2019arXiv190208202G,
  2019arXiv190303898S, 2019arXiv190309147G}. The result is that the 
post-disruption stellar debris acquires a spread of specific orbital
energy, which is roughly given by the gradient of the BH's gravitational
potential across the star at the tidal radius. This means that roughly
half of the stellar debris is unbound and the other half is left in
highly eccentric bound orbits. 

\begin{figure}
  \centering
\includegraphics[width = 0.35\textwidth,
  height=0.22\textheight]{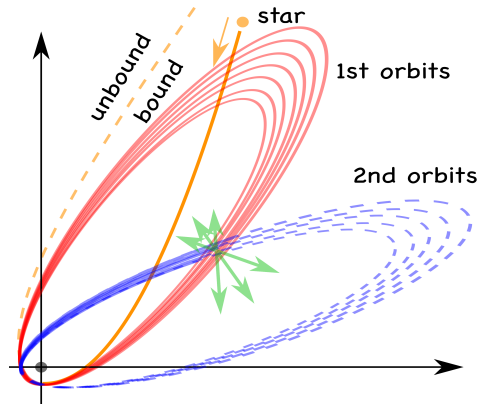}
\caption{The star was initially in a parabolic orbit (orange
  curve). After the tidal 
  disruption, the bound materials are in highly eccentric elliptical
  orbits of different semimajor axes (red curves) but have nearly the same apsidal
  precession angle per orbit. Materials in their second
  orbits (blue curves) collide with what is still in the first
  orbit. The subject of this paper is to study the dynamics of the
  shocked gas after the collision. 
}\label{fig:orbit}
\end{figure}

After the disruption phase, the star is tidally stretched into a very long
thin stream and the evolution of the stream structure in the
transverse and longitudinal 
directions are decoupled \citep{1994ApJ...422..508K}. Thus, the system
enters the free-fall phase where each stream segment follows its own
geodesic like a test particle \citep{2016MNRAS.459.3089C}. Then, after
passing the apocenters of the highly eccentric orbits, the bound
debris falls back towards the BH at a rate given by the 
distribution of specific energy \citep{1989ApJ...346L..13E,
  Phinney89}. Due to relativistic apsidal precession, the bound
debris, after passing the  pericenter, collides violently with the
still in-falling stream (see 
Fig. \ref{fig:orbit}). It has been shown that shocks at the 
self-intersection point is the main cause of orbital energy
dissipation and the subsequent formation of an  accretion disk
\citep{1988Natur.333..523R, 1994ApJ...422..508K, 2013MNRAS.434..909H,
  2014ApJ...783...23G, 2015ApJ...804...85S, 
  2016MNRAS.455.2253B}. However, the aftermath of 
the self-intersection is an extremely complex problem, which depends
on the interplay among magnetohydrodynamics, radiation, and
general relativity in 3D. No numerical simulations to date have been able to  
provide a deterministic model for TDEs with realistic star-to-BH
mass ratio and high eccentricity \citep[see][for a
review]{2018arXiv180110180S}. Many simulations consider either an
intermediate-mass BH \citep[e.g.][]{2014ApJ...783...23G, 2015ApJ...805L..19E, 
  2015ApJ...804...85S, 2016MNRAS.458.4250S} or the disruption of a
low-eccentricity (initially bound) star
\citep[e.g.][]{2016MNRAS.455.2253B, 2016MNRAS.461.3760H}. It is
unclear how to extrapolate the simulation results to realistic 
configurations and provide an answer to the following questions:
How long does it take for the bound gas to form a circular
accretion disk (if at all)? How much radiative energy is released from the
system? What fraction of the radiation is emitted in the optical, UV
or X-ray bands?

The hope lies in the rapidly growing sample of
TDE candidates discovered by recent UV-optical surveys,
such as GALEX \citep{2008ApJ...676..944G, 2009ApJ...698.1367G}, SDSS
\citep{2011ApJ...741...73V}, Pan-STARRS \citep{2012Natur.485..217G,
  2014ApJ...780...44C, 2017ApJ...843..106B}, PTF \citep{2014ApJ...793...38A,
  2017ApJ...844...46B, 2018ApJS..238...15H}, ASAS-SN
\citep{2014MNRAS.445.3263H, 2016MNRAS.455.2918H}, and ZTF
\citep{2018arXiv180902608V}, see the open TDE catalog
\href{http://tde.space}{http://tde.space}. These events have 
highly diverse properties in terms of peak optical luminosities,
lightcurve shapes, emission line profiles, and optical/X-ray flux
ratios. Still, they provide a number of important clues for
understanding the dynamics of UV-optical selected TDEs: (1)
the photospheric radius of the (thermal) 
optical emission is typically $\sim$$10^{14}$--$10^{15}\rm\,cm$; (2) the
typical widths of H$\alpha$ and/or He$\,$II emission lines in the
optical band and CIV, NV, SiIV aborption lines in the UV band
\citep[e.g.][]{2018arXiv180907446B} are of order
$\sim$$10^4\rm\,km/s$; (3) the rise/fade timescale is of order 
$\sim$months\footnote{We note a few exceptions such as iPTF16fnl
  \citep{2017ApJ...844...46B} and
  ASASSN-15lh \citep{2016Sci...351..257D, 2016NatAs...1E...2L}. We
  also note that current optical surveys are biased  
  against detecting very fast ($\lesssim$week) and very long
  ($\gtrsim$year) transients, so the rise/fade timescales of detected
  events may not representative for the entire TDE family.}; (4) the
total energy radiated in the UV-optical band is typically
$\lesssim$$10^{51}\rm\,erg$, which is much smaller than 
the energy budget of the system ($\gtrsim$$10^{53}\rm\,erg$ even for
disruption of low-mass stars).

The photospheric radius is much larger than the tidal radius (of order
$\sim$$10^{13}\rm\,cm$), and the velocity inferred from line widths is
much smaller than the Keplerian/escape velocity near the tidal
radius. These properties are inconsistent with the
wind-reprocessed emission from a circularized accretion disk near the
tidal radius \citep{2009MNRAS.400.2070S, 2015ApJ...805...83M}. The low 
radiative efficiency in the optical band is known as the  
``missing energy'' puzzle \citep{2015ApJ...806..164P,
  2016MNRAS.455..859S, 2018ApJ...865..128L}, whose solution depends on 
the source of the optical emission. Based on the 
arguments that the photospheric radius is of the same order as the
semimajor axis of the most bound orbit and that the line width
roughly agrees with the Keplerian velocity at the same radius,
\citet{2015ApJ...806..164P} proposed that the optical emission is
powered by the dissipation of orbital energy by stream
self-intersection. An alternative phonomelogical model proposed
by \citet{2016MNRAS.461..948M} is that only a small fraction $f_{\rm
  in}\ll 1$ of the fall-back gas actually accretes onto the BH and the
rest $(1 - f_{\rm in})$ is blown away by the gravitational energy
released from the accreting gas. In this model, if the energy efficiency of
accreting gas is $\eta_{\rm acc} = 0.1\eta_{\rm
  acc,-1}$, then the accretion fraction of order $f_{\rm in}\sim
10^{-2}\eta_{\rm acc, -1}^{-1}$. However, these models do not consider
the detailed dynamics of the stream self-intersection and disk
formation.

In this paper, we consider the stream kinematics in a semi-analytical
way and explore the diverse consequences of the stream
self-intersection. This approach is similar to
\citet{2015ApJ...812L..39D} who studied the location and
gas velocity at the self-intersection point in a post-Newtonian
way (only considering the lowest-order apsidal precession). However, we
evolve the system in full general relativity before and after the 
self-intersection and study the properties of the shocked
gas that are unbound, accreting, and plunging. More importantly,
instead of assuming inelastic collision as in
\citet{2015ApJ...812L..39D}, we use the realistic equation of state for 
radiation-dominated gas to model the intersection, motivated by the
local simulation of colliding streams by
\citet{2016ApJ...830..125J}. Thus, our approach provides a more
comprehensive and self-consistent picture of the dynamics and
multiwavelength emission from TDEs.

This paper is organized as follows. In \S 2, we calculate the location
of the self-intersection point and the velocities of the two streams
before the collision. In \S 3, we perform hydrodynamical simulation of
the collision process. In \S 4, we consider the fate of the shocked
gas after the self-intersection. Implications of TDE dynamics on the
multiwavelength observations will be considered in \S 5. We discuss a
number of issues in our modeling in \S 6. A summary is provided in \S 7.
Unless otherwise specified, we use geometrical units where the
gravitational constant and speed of light are $G=c=1$.

\section{Self-Intersection of the Fallback Stream}
We consider a star of mass $M_* = m_*\msun$ and radius $R_* =
r_*R_{\odot}$ interacting with a BH of mass $M = 10^6M_6\msun$. The
gravitational radius of the BH is $\rg \equiv M$. We
take the pericenter of the star's initial orbit to be $\rp =
\rt/\beta$, where $\beta$ is a free impact parameter describing the
depth of penetration and the $\rt$ is the Newtonian Roche tidal radius defined as
\citep{1975Natur.254..295H}
\begin{equation}
  \label{eq:2}
    {r_{\rm T}\over \rg} \equiv {R_*\over r_{\rm g}} \left( {M
      \over M_*} \right)^{1/3} = 46.7\, M_6^{-2/3} m_*^{-1/3} r_*.
\end{equation}
The lower limit of the impact parameter $\beta_{\rm min}$ is of
order unity, but to obtain its exact value corresponding to marginal
disruption, one must take into account relativistic tidal forces and
realistic stellar structure/rotation (these will be discussed later
in \S 5.3). After the disruption, the
stellar debris attains a spread of specific orbital 
energy for the stellar debris 
$\eps\in(-\eta_{\rm max}\eps_{\rm T}, +\eta_{\rm max}\eps_{\rm T})$,
where we have defined the Newtonian tidal energy
\begin{equation}
  \label{eq:3}
  \eps_{\rm T}\equiv {\rg R_*\over r_{\rm T}^2}
  =2.13\times10^{-4}M_6^{1/3}m_*^{2/3} r_*^{-1}, 
\end{equation}
and $\eta_{\rm max}$ is a constant of order unity containing the
uncertainties due to stellar structure/rotation and the detailed
relativistic disruption process. The Newtonian orbital period of 
the leading edge ($\eta = \eta_{\rm max}$) is $P_{\rm min}\simeq
(41\mr{\,d})\, \eta_{\rm max}^{-3/2} M_6^{1/2}m_*^{-1}r_*^{3/2}$.

The bound materials corresponding to $\eps<0$ form an elongated
thin stream which collides with itself due to apsidal precession
(Fig. \ref{fig:orbit}). Since the width of the stream is much smaller
than the pericenter radius \citep[e.g.,][]{1994ApJ...422..508K,
  2016MNRAS.459.3089C}, a given stream segment, characterized by its
specific energy $\eps=\eta \eps_{\rm T}$ and pericenter radius
$\rp=\rt/\beta$ ($\eta\leq\eta_{\rm max}$ and $\beta\gtrsim 1$ are free
parameters), moves along a certain geodesic until it collides with the
still in-falling gas. Note that we define $\eps$ and $\rp$ based on
Newtonian quantities $\eps_{\rm T}$ and $\rt$ only for convenience
reason, our treatment of the orbital kinematics is
fully general relativistic.

In this paper, we consider the simplest case of a non-spinning BH (the
effects of BH spin will be discussed in \S 6). In spherical
coordinates for the Schwarzschild spacetime, the initial position of
the stream segment is $(t = 0,\  
r = r_{\rm p},\ \theta = 0,\ \phi = 0)$ and the proper time of the
stream segment starts as $\tau=0$. We align the orbital plane with the
equatorial plane of the coordinate system, so $\dot{\theta}\equiv \d
\theta/\d\tau = 0$. The specific angular momentum of is
given by
\begin{equation}
  \label{eq:4}
  \ell = r_{\rm p}\sqrt{(1 + \eps)^2/\mu_{\rm p} - 1},\ \mu_{\rm p}
  \equiv 1 - 2r_{\rm g}/r_{\rm p}, 
\end{equation}
where $1 + \eps$ is the total energy including rest mass. Hereafter,
the time derivative of any quantity $Q$ with respect to the stream
segment's proper time $\tau$ is denoted by $\dot{Q}$. Measuring the
proper time in units of $\rg$, we write the geodesic equations
\begin{equation}
  \label{eq:5}
  \begin{split}
\dot{t} &= {1 + \eps \over 1 - 2\rg/r},\ \dot{\phi} = \ell \rg/r^2,\\
      \dot{r}^2 &= (1 + \eps)^2 - \left(1 -
          {2r_{\rm g}\over r}\right) 
\left(1 + {\ell^2\over r^2}\right),\\
\ddot{r} & = -{\rg^2 \over r^2} + {\ell^2\rg \over r^3}\left(
1 - {3\rg\over r}\right).
  \end{split}
\end{equation}
We use a Leapfrog method to integrate the above geodesic
equations with timestep $\delta \tau = \rg/30$. Since these two
colliding flows have similar specific energies $\eta_1\approx \eta_2$,
the radius for self-intersection $\ri$ is approximately given by
$\phi(\ri)=\pi$.

\begin{figure}
  \centering
\includegraphics[width = 0.48\textwidth,
  height=0.65\textheight, trim=0.0cm 0.4cm 0.0cm
  .4cm]{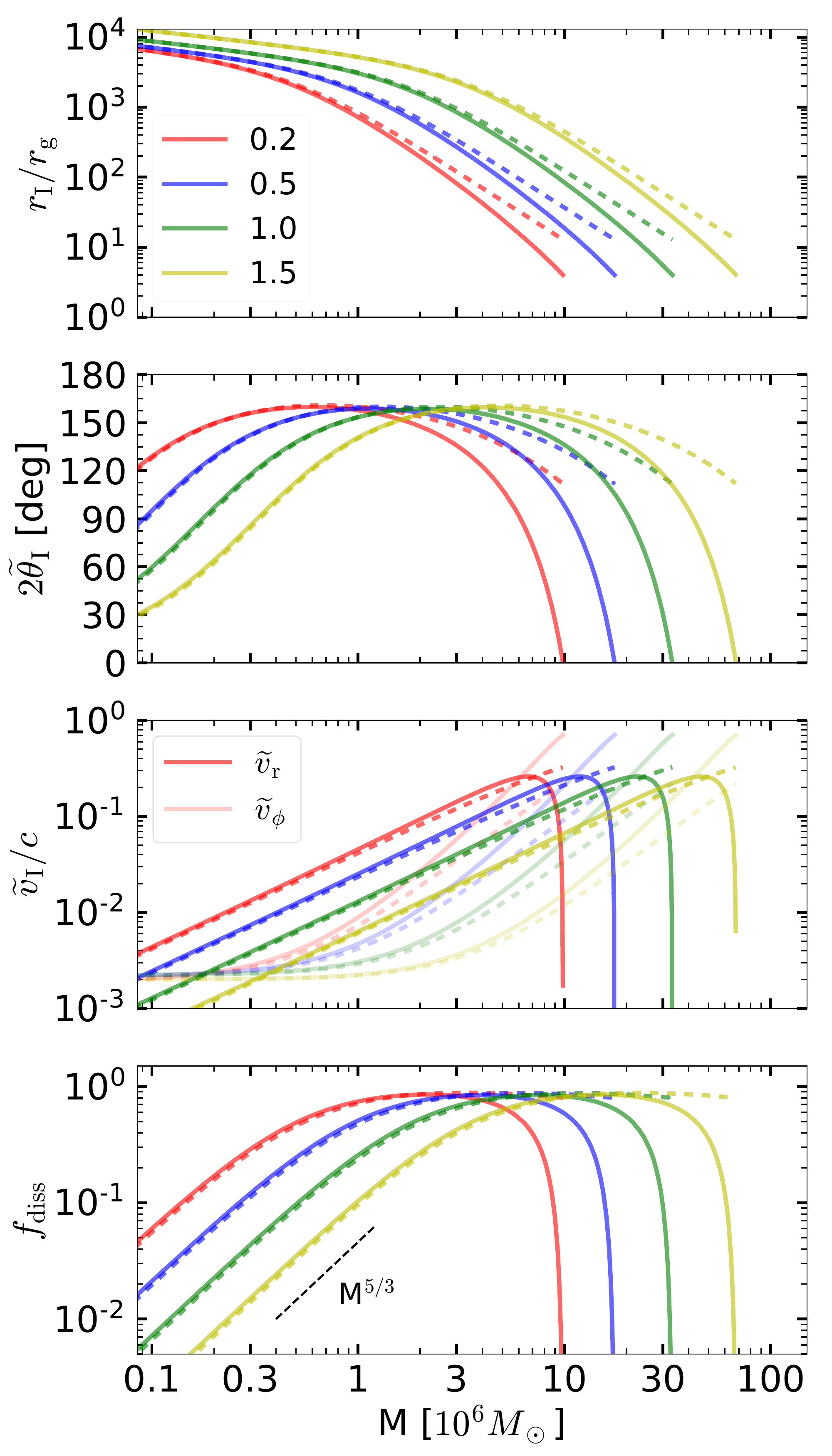}
\caption{The self-intersection radius, full angle, velocity, and
  efficiency of orbital energy dissipation as a function of the BH's
  mass. The red, blue, green, and yellow curves are for stellar masses of
  $m_*=0.2$, $0.5$, $1.0$, $1.5$ (shown in the legend of the first
  panel). For all panels, we use the same the impact parameter
  $\beta = 1.0$ ($\rp = \rt/\beta$) and orbital energy parameter $\eta
  = 1.0$ ($\eps = \eta \eps_{\rm T}$). The solid curves are from general
  relativistic (GR) calculations and the dashed curves are the
  corresponding post-Newtonian results given by
  \citet{2015ApJ...812L..39D}. The cut-off of each curve on the high
  BH mass end is due to direct capture of the star by the BH. In the
  third panel, we show the radial component $\t{v}_{\rm r}$ in darker
  curves and transverse component $\t{v}_{\phi}$ in lighter curves,
  both measured by a local stationary observer at the intersecting
  point.
}\label{fig:intersection}
\end{figure}

For a stationary observer at the intersecting point $\ri$, we
define $\mu_{\rm I} \equiv 1 - 2\rg/\ri$, so the local differential
length in the radial direction is $\d \t{r} = \mu_{\rm I}^{-1/2} \d r$ and the local
differential time is $\d \t{t} = \mu_{\rm I}^{1/2}\d t = \mu_{\rm
  I}^{-1/2}(1+\eps)\d \tau$. In the following, we consider the
stream-intersection process in the comoving
frame of a local stationary observer at radius $\ri$ (LSO frame
hereafter), in which any quantity $Q$ is denoted with a tilde
$\t{Q}$. Then, the radial and transverse velocities of the colliding
streams in the LSO frame are  
\begin{equation}
  \label{eq:7}
  \begin{split}
      \t{v}_{\rm r} &= {\d \t{r}\over \d \t{t}} = \mu_{\rm I}^{-1} {\d
  r\over \d t} = {\dot{r}(\ri)\over 1 +\eps},\\
\t{v}_{\rm \phi} &= \ri {\d \t{\phi}\over \d \t{t}}  = {\ri \mu_{\rm
  I}^{1/2}\over 1 + \eps} \dot{\phi}(\ri).
  \end{split}
\end{equation}
The intersecting half angle $\t{\theta}_{\rm I}$ in the LSO frame is
given by
\begin{equation}
  \label{eq:8}
  \tan\,\t{\theta}_{\rm I} = {\t{v}_{\rm r}\over \t{v}_{\rm \phi}} =
  {\dot{r}(\ri) \over \mu_{\rm I}^{1/2} \ri \dot{\phi}(\ri)}.
\end{equation}
In Fig. \ref{fig:intersection}, we compare the self-intersection radius, angle, and
velocities from our general relativistic calculations with
the corresponding lowest-order post-Newtonian results by
\citet{2015ApJ...812L..39D}. We take $\beta=1.0$ and $\eta=1.0$ to be
our fiducial parameters. We consider four different stellar
masses of $m_*=0.2$, $0.5$, $1.0$, $1.5$ and the corresponding zero-age
main-sequence stellar radii $r_*=0.23$, $0.46$, $0.89$, $1.63$ are taken from
\citet{1996MNRAS.281..257T} assuming solar metallicity, with errors of
a few percent. As expected, we find that, for more massive BHs, the 
self-intersection occurs closer to the event horizon and the
intersecting velocity is larger (the interaction is more
violent).

If one \textit{assumes} that two colliding flows have equal
cross-sections and that the collision 
is \textit{completely inelastic}, then the radial 4-velocity 
component $\dot{r}(\ri)$ gets dissipated and that the transverse
4-velocity component $\ri\dot{\phi}(\ri)$ survives. 
In this case, we can quantify the efficiency of 
orbital energy dissipation by defining
\begin{equation}
  \label{eq:9}
 f_{\rm diss} = \left( {\mui^{1/2} \over \sqrt{1 - \t{v}_{\rm r}^2 -
       \t{v}_{\rm \phi}^2}} - {\mui^{1/2} \over \sqrt{1 -
       \t{v}_{\rm \phi}^2}}\right) (1 - \mui^{1/2})^{-1},
\end{equation}
which describes the change in orbital energy
divided by the gravitational binding energy at radius $\ri$.
This (maximum possible) dissipation efficiency is shown in the fourth
panel of Fig. \ref{fig:intersection}. In the low BH mass limit $M_6\ll
1$, the dissipation of orbital energy by shocks is extremely weak and
we asymptotically have 
$f_{\rm diss}\propto M^{5/3}$ (marked as a black dashed line in the
fourth panel), which agrees with the result of
\citet{2017MNRAS.464.2816B}. In those cases, if the circularization is
still dominated by stream 
self-intersection, then the orbit stays highly eccentric 
for roughly $f_{\rm diss}^{-1}$ rounds and hence the circularization
timescale is roughly $f_{\rm diss}^{-1} P_{\rm min}\propto
M^{-7/6}$ (since $P_{\rm min}\propto M^{1/2}$). Other mechanisms,
e.g. the magneto-rotational instability, may cause angular momentum
exchange and drive circularization on a shorter timescale
\citep{2018ApJ...856...12C}. As we discuss later in \S 5.1, TDEs by
low-mass BHs typically generates long-lasting eccentric accretion
disks which produce long-duration transients. On the other hand, for
high-mass BHs $M_6\gtrsim 1$,  stream intersection causes strong
dissipation of orbital energy and hence the orbit may quickly circularize.

In the next section, we show that completely inelastic collision,
as assumed by e.g. \citet{2015ApJ...812L..39D},
is a poor description of the stream dynamics, because the shocked gas
is highly optically thick and hence evolves in a nearly adiabatic
manner \citep{2016ApJ...830..125J}.
\section{Hydrodynamical Simulations of the Self-Intersecting Shocks}
We numerically simulate the stream-stream collision
in a special inertia frame described as follows. 
In the LSO frame, the
4-velocity of the outward-moving stream is $(\t{u}_{\alpha}) =
(\t{u}_t, \t{u}_r, \t{u}_\theta, \t{u}_\phi) = \t{\Gamma} (1, \t{v}_r,
0, \t{v}_\phi)$, where the Lorentz factor is $\t{\Gamma}\equiv (1 -
\t{v}^2)^{-1/2}$ and $\t{v}^2 = \t{v}_{\rm r}^2 +
\t{v}_\phi^2$. Our simulation box is centered at the self-intersecting
point and is moving at velocity $\t{v}_{\phi}$ with respect to the local
stationary observer in the $\hat{\phi}$ direction. Thus, in the
comoving frame of the simulation box (hereafter the SB frame), the two
streams collide head-on with each of them moving at 4-velocity
$\bar{u}_{\rm c} = \t{\Gamma}\t{v}_r$ and Lorentz factor
$\bar{\Gamma}_{\rm c} = (1 + \t{\Gamma}^2\t{v}_{\rm r}^2)^{1/2}$,
which means
\begin{equation}
  \label{eq:10}
  \bar{v}_{\rm c} = {\t{\Gamma}\t{v}_r\over \sqrt{1 + \t{\Gamma}^2\t{v}_r^2}}.
\end{equation}
Hereafter, any quantity $Q$ in the SB frame is denoted with an
overhead bar $\bar{Q}$. In all possible cases, the
incoming velocities of the streams in the SB frame are
sub-relativistic $\bar{v}_{\rm c}< 0.3c$ (see
Fig. \ref{fig:intersection}). Due to extreme stretch and adiabatic
cooling, the initial 
streams are dynamically cold with sound speed much less than the bulk velocity
$\bar{v}_{\rm c}$. Another property of the initial streams is that the
transverse size is much less than the orbital size $\sim$$\ri$
\citep{1994ApJ...422..508K, 2016MNRAS.459.3089C,
  2017MNRAS.464.2816B}. These properties of the problem enable us to
use a single non-relativistic 
hydrodynamic simulation in a flat spacetime to capture the structure
of the shocked gas, which is self-similar within a region of size $\ll\ri$.

In our simulation, we use an adiabatic ideal gas equation of state $P\propto
\rho^{4/3}$. This is motivated by: (1) the high-density shocked gas is
radiation pressure dominated, and (2) 
the shocked gas is highly optically thick before most of the heat is
converted back to bulk motion via $PdV$ work
\citep{2016ApJ...830..125J}. The radiative efficiency of the shocked
gas is estimated in Appendix B. Since we are concerned with the fate
of the majority of the gas with fallback time 
$\lesssim 10 P_{\rm min}$, the adiabatic assumption is a good one. For
simplicity, we assume that the two colliding streams have the same
cross-section and that there is no offset in the transverse
direction. We will discuss the validity of these assumptions and
consequences of relaxing them in \S6.

\begin{figure}
  \centering
\includegraphics[width = 0.48\textwidth,
  height=0.45\textheight]{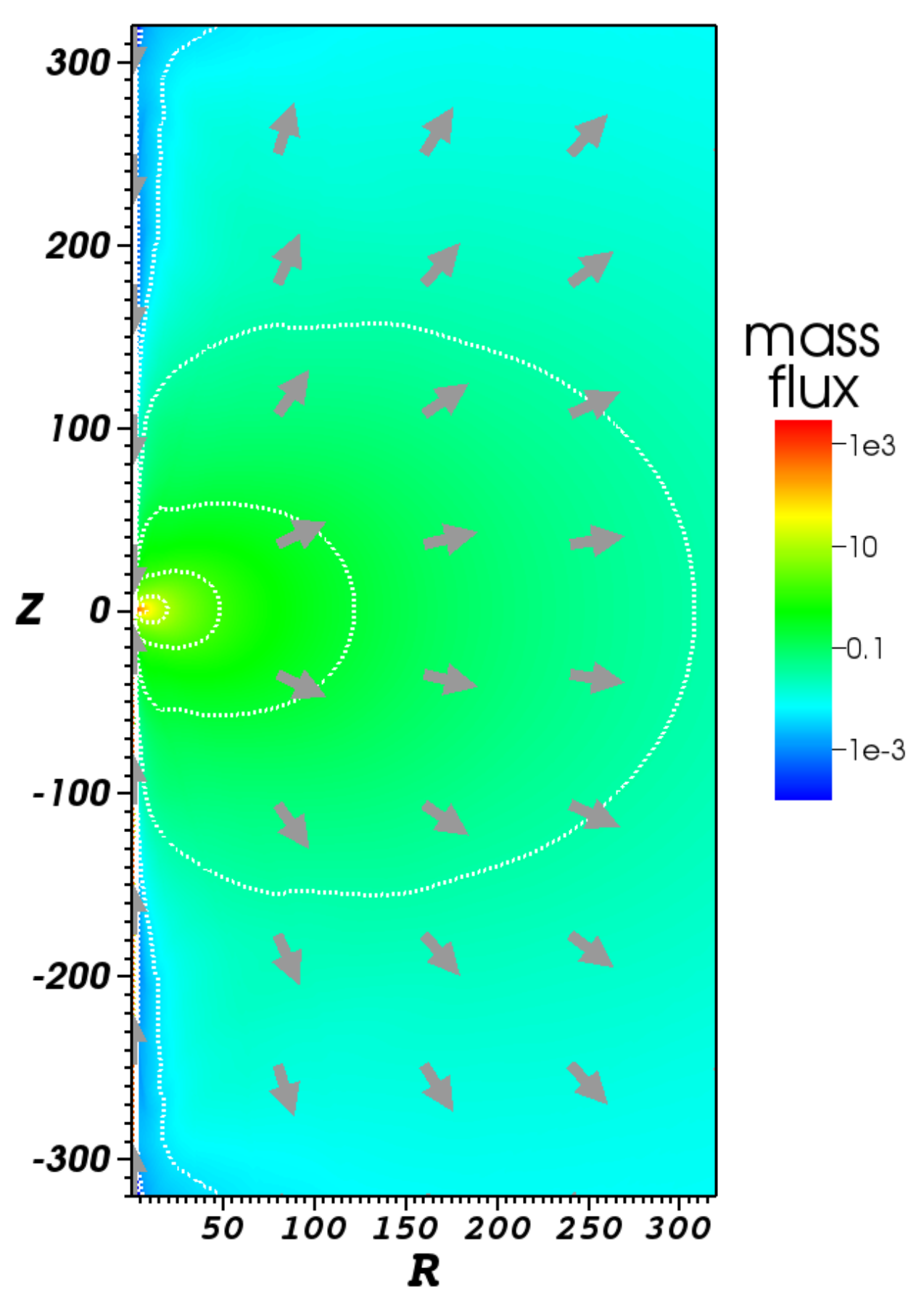}
\caption{Quasi-steady profiles of the velocity vector field (arrows) and mass flux
  $\bar{\rho}\bar{v}$ (color image and contours) time t = 0.8, which
  is 7.5 times the shock-crossing time of the entire simulation domain
  (or 1200 times the shock-crossing time of the initial streams). At
  large distances $\gg 10$ from the shocks, velocity vectors are
  nearly in the radial direction in most directions. This figure is
  generated with the open source visualization tool VisIt by \citet{HPV:VisIt}.
}\label{fig:massflux}
\end{figure}

We perform the simulation with the non-relativistic
hydrodynamics module of PLUTO \citep{2007ApJS..170..228M}, solving the
mass and momentum conservation equations in 2D cylindrical coordinates ($R,\,
z$). The $R$-axis corresponds to the $\hat{\phi}$ direction in
the BH rest frame, and the $z$-axis is parallel to the $\hat{r}$
direction in the BH rest frame. The size of our simulation box is $0\leq R 
\leq320$ and $-320\leq z\leq 320$. Two identical steady streams are 
injected in the form of top-hat jets moving in opposite directions
at $z=-320$ and $z=320$ in the radius range 
$0\leq R\leq 1$. The other boundary conditions are as follows: $R=0$
axis-symmetric, $R=320$ outflow, $z=-320$ and $z=320$ outflow 
(except for the inner cylinder $R\leq 1$ where the streams are injected).
The resolution\footnote{We also ran the same simulation with lower
  resolution $\delta R =\delta z= 0.25$, and found the results to be
  similar. } is $\delta R =\delta z= 0.125$ ($N_{\rm R} = N_{\rm 
  z}/2 = 2560$), which means the initial stream is resolved by 8 cells
in the transverse direction. The initial streams have mass density 1,
pressure $(4/3)^{-1}$, and velocities $\pm3000$ (all in 
machine units, since the problem 
is scale-free in the non-relativistic limit). Since the adiabatic sound speed of
the stream is unity, the Mach number is $3000\gg 1$ and hence the
streams are effectively cold. The pressure of the ambient medium
matches that of the streams. The mass density of the ambient medium is
extremely small $10^{-8}$, so the shocked gas expands as if in vacuum.
We run the simulation with time step $\delta t\simeq 6\times10^{-6}$
for a sufficiently long time $t = 0.8$ (or 7.5 times the domain
crossing time) so that the structure of the shocked
gas within a sphere of radius 320 has relaxed to a nearly stable
configuration.

\begin{figure}
  \centering
\includegraphics[width = 0.48\textwidth,
  height=0.25\textheight]{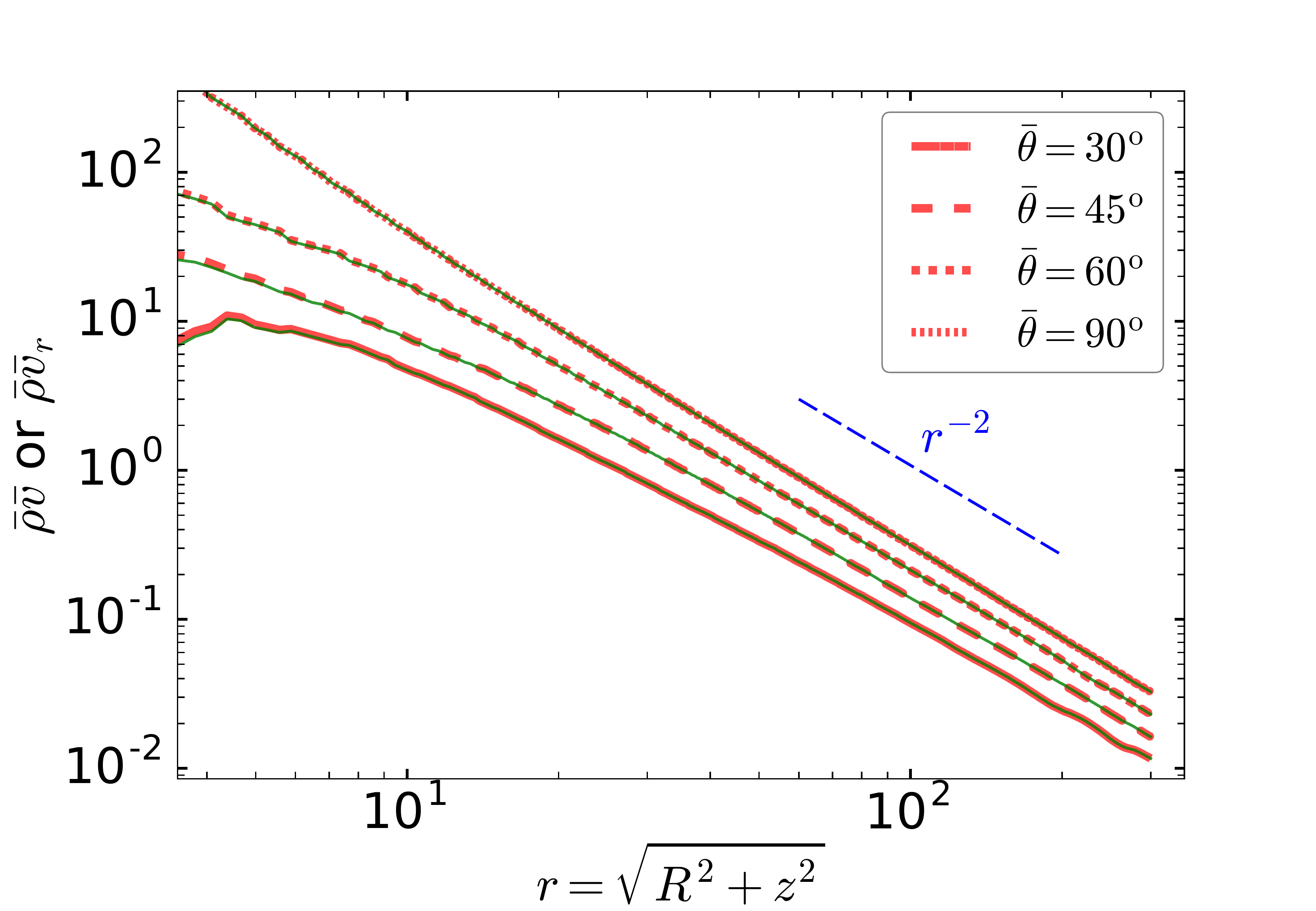}
\caption{Thick red curves show the radial profiles of the mass flux
  $\bar{\rho}\bar{v}$ at different polar angles $\bar{\theta}\equiv
  \mr{arctan}(R/z) = 30^{\rm o},\ 
  45^{\rm o},\ 60^{\rm o}$, and $90^{\rm o}$ at t = 0.8. We also
  over-plot the radial component of the mass flux
  $\bar{\rho}\bar{v}_r$ in thin green curves, which overlap with the
  red curves. This means that at sufficiently large 
  radii $r > 3$, the velocity vectors of the expanding shocked gas are
  nearly in the radial direction. We can also see that the mass flux
  is higher near the equatorial plane ($\bar{\theta}\sim 90^{\rm o}$)
  than that near the poles ($\bar{\theta}\lesssim 30^{\rm o}$). The
  mass flux profiles asymptotically approach the inverse-square law since
  $\bar{v}\approx\,$const and $\bar{\rho}\propto r^{-2}$.
}\label{fig:massflux-radial}
\end{figure}

\begin{figure}
  \centering
\includegraphics[width = 0.48\textwidth,
  height=0.25\textheight]{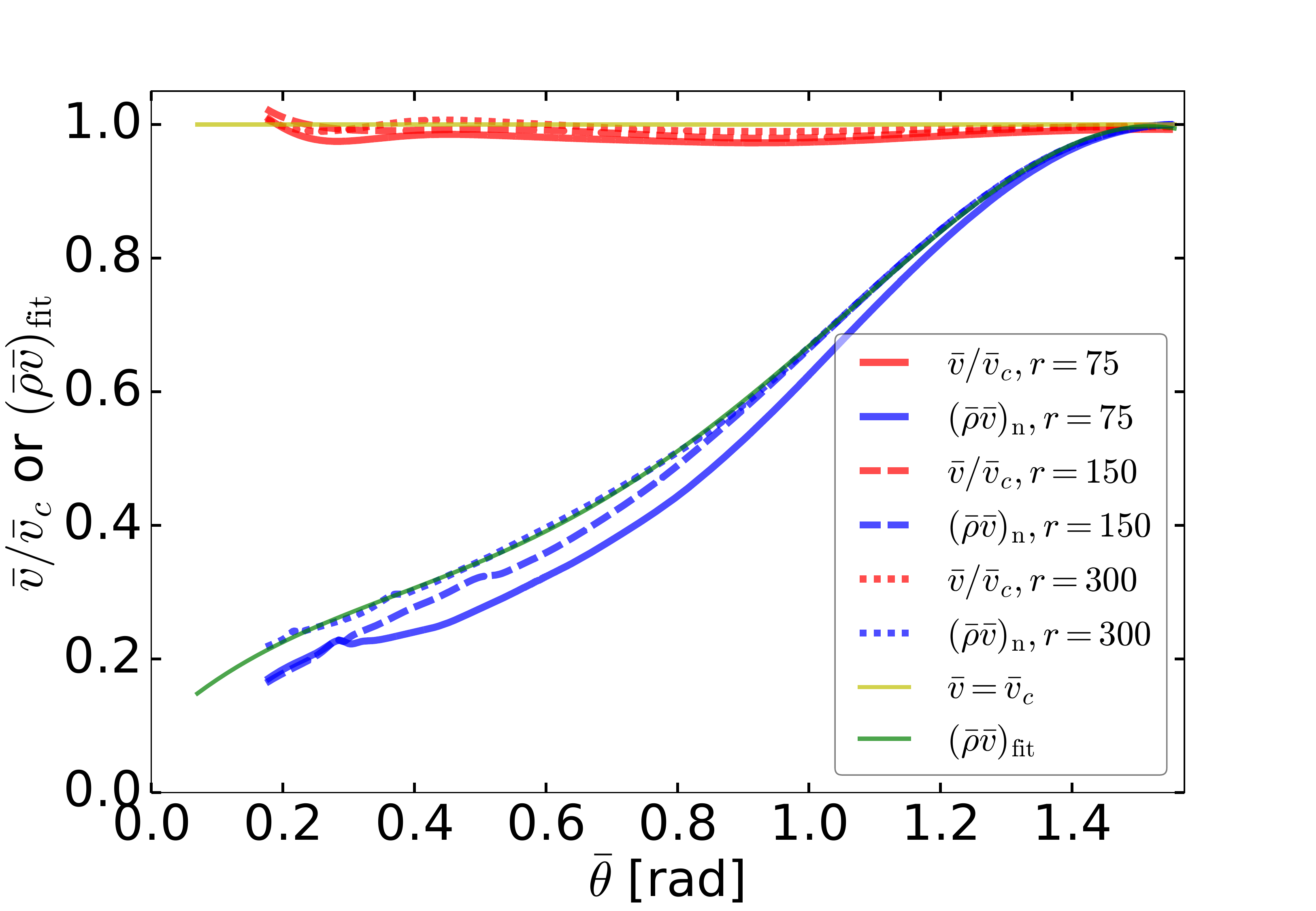}
\caption{The angular profiles of the velocity and mass
  flux at three spheres of radii $r = \sqrt{R^2 + z^2} =75$, $150$ and
  $300$. The polar angle $\bar{\theta}$ is defined as $\bar{\theta} =
  \mr{arctan}(R/z)$. The velocity profiles are normalized by the speed
  of the incoming stream $\bar{v}_c = 3000$. At all three radii, the 
  velocity profiles are nearly flat with $\bar{v}\approx \bar{v}_c$ at
  most angles, so we simplify them as an isotropic
  function $\bar{v}(\bar{\theta})=\bar{v}_c$ (yellow curve). The mass
  flux profiles shown here are simply peak-normalized to
  illustrate the equatorial concentration near
  $60^{\rm o}\lesssim \bar{\theta}\leq 90^{\rm o}$. We fit the mass
  flux profile at $r=300$ with a fourth-order polynomial
  function (green curve), which is then re-normalized in
  eq. (\ref{eq:14}) such that $2\pi\int_0^{\pi}\bar{\rho}_{\rm n}
  \sin\bthe \d\bthe = 1$.
}\label{fig:v-rho-profile}
\end{figure}

The large-scale structure of the system at $t=0.8$ is shown in
Fig. \ref{fig:massflux}. We see that the two streams collide at $z=0$
and the shocked gas expands in a roughly spherical way to radii much
larger than the stream width (which equals to unity). In
Fig. \ref{fig:massflux-radial}, we show the 
radial profiles of the mass flux at different polar angles
$\bar{\theta} = 30^{\rm o},\ 45^{\rm o},\ 60^{\rm o}$, and $90^{\rm
  o}$. The angular profiles of the velocity and mass flux at three
different radii $r=75$, $150$ and $300$ are shown in
Fig. \ref{fig:v-rho-profile}. We can see that, at large distances from
the shocks $r\gtrsim 100$, the velocity profile is very flat but the
mass flux is heavily concentrated near the equatorial plane $60^{\rm
  o}\lesssim \bar{\theta}\leq 90^{\rm o}$. In the following, we
simplify the velocity angular profile as isotropic
\begin{equation}
  \label{eq:12}
  \bar{v} (\bar{\theta})= \bar{v}_c,
\end{equation}
and hence the density angular profile is the same as the mass flux
profile. We use a fourth-order polynomial fit to the normalized density
angular profile given by
\begin{equation}
  \label{eq:14}
  \begin{split}
      \bar{\rho}_{\rm n} &= \sum_{i=0}^{4} q_i x^i,\
q_0 = 1.051\times10^{-2},\ q_1 = 0.1103,\\
q_2 &=-0.2017,\ q_3 = 0.2434,\ q_4 = -0.08436,
  \end{split}
\end{equation}
where $x\equiv \mr{min}(\bthe,\, \pi - \bthe)$ and
$2\pi\int_0^{\pi}\bar{\rho}_{\rm n} \sin\bthe \d\bthe = 1$.

In the following, we Lorentz transform the velocity and mass flux
angular profiles of the shocked gas from the SB frame back to the LSO
frame. For a fluid element moving with 
speed $\bv_c$ in an arbitrary ($\bthe, \bphi$) direction ($\bthe$ being the
polar angle and $\bphi$ being the azimuthal angle), we write its
four-velocity in Cartesian components $(\bar{u}_{\alpha}) =
\bar{\gamma}(1, \bv_x, \bv_y,
\bv_z)$, where $\bar{\gamma} = (1 - \bv^2_c)^{-1/2}$, $\bv_x =
\bv_c\sin\bthe \cos \bphi$, $\bv_y = \bv_c\sin\bthe \sin \bphi$, and
$\bv_z = \bv_c\cos\bthe$. The simulation box is moving at velocity
$\t{v}_\phi$ and the corresponding Lorentz factor is $\t{\Gamma}_\phi\equiv (1 -
\t{v}_\phi^2)^{-1/2}$, so the 4-velocity in the LSO frame is
\begin{equation}
  \label{eq:6}
  \begin{split}
      \t{u}_t &= \t{\Gamma}_\phi \bar{u}_t +\t{\Gamma}_\phi
      \t{v}_\phi  \bar{u}_x,\ \  \t{u}_y =\bar{u}_y,\\
 \t{u}_x &= \t{\Gamma}_\phi \t{v}_\phi \bar{u}_t +
 \t{\Gamma}_\phi \bar{u}_x, \ \  \t{u}_z = \bar{u}_z.
  \end{split}
\end{equation}
Then, the specific angular momentum and specific energy of this fluid
element in the Schwarzschild spacetime are
\begin{equation}
  \label{eq:16}
  \ell(\bthe, \bphi) = \ri \sqrt{\t{u}_x^2 + \t{u}_y^2},\ 1 +
  \eps(\bthe, \bphi) = \mui^{1/2} \t{u}_t,
\end{equation}
In the next section, we discuss
what fraction of the shocked gas is unbound, plunging, or accreting.


\section{Fate of the shocked gas after the self-intersection}
When the shocked gas expands to a distance much greater than the
stream width, the internal pressure becomes low enough that the motion
of individual fluid elements is approximately ballistic. If the geodesic
reaches infinity or inside the event horizon, we call the fluid
element ``unbound'' or ``plunging'', respectively. Those fluid
elements with bound but non-plunging geodesics are denoted as
``accreting.'' The geodesic of a fluid element moving in the 
($\bthe, \bphi$) direction at $\ri$ has specific angular momentum
$\ell(\bthe, \bphi)$ and specific energy $1+ \eps(\bthe, \bphi)$,
which are given by eq. (\ref{eq:16}). We note that the marginally
bound parabolic orbit for the 
Schwarzschild spacetime has specific angular momentum $\ell_{\rm mb} =
4\rg$ and pericenter radius $\rmb = 4\rg$ \citep{1972ApJ...178..347B},
so the stream self-intersection radius must always be greater than
$4\rg$ (see Fig. \ref{fig:intersection}).

\begin{figure}
  \centering
\includegraphics[width = 0.48\textwidth,
  height=0.25\textheight]{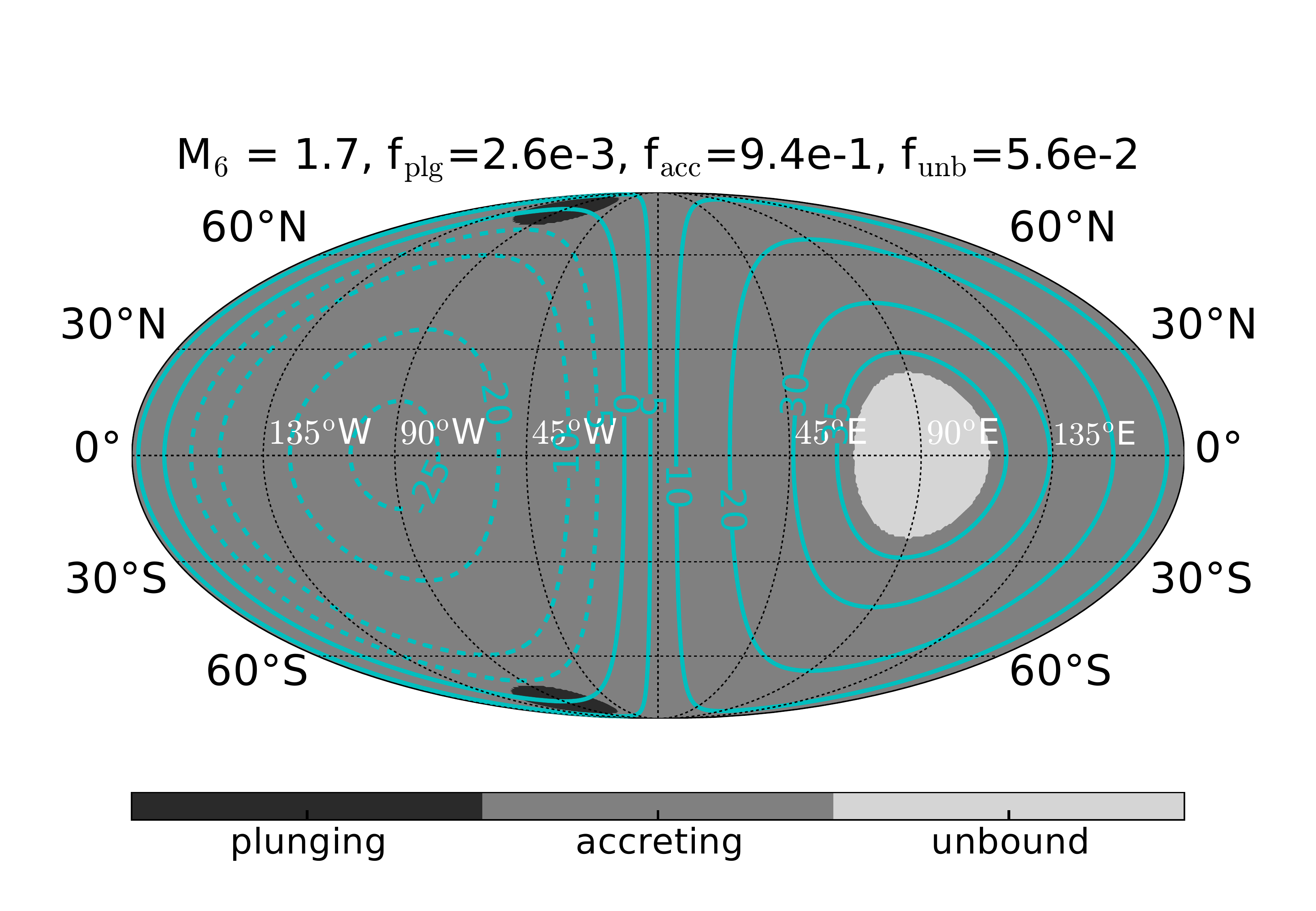}
\includegraphics[width = 0.48\textwidth,
  height=0.25\textheight]{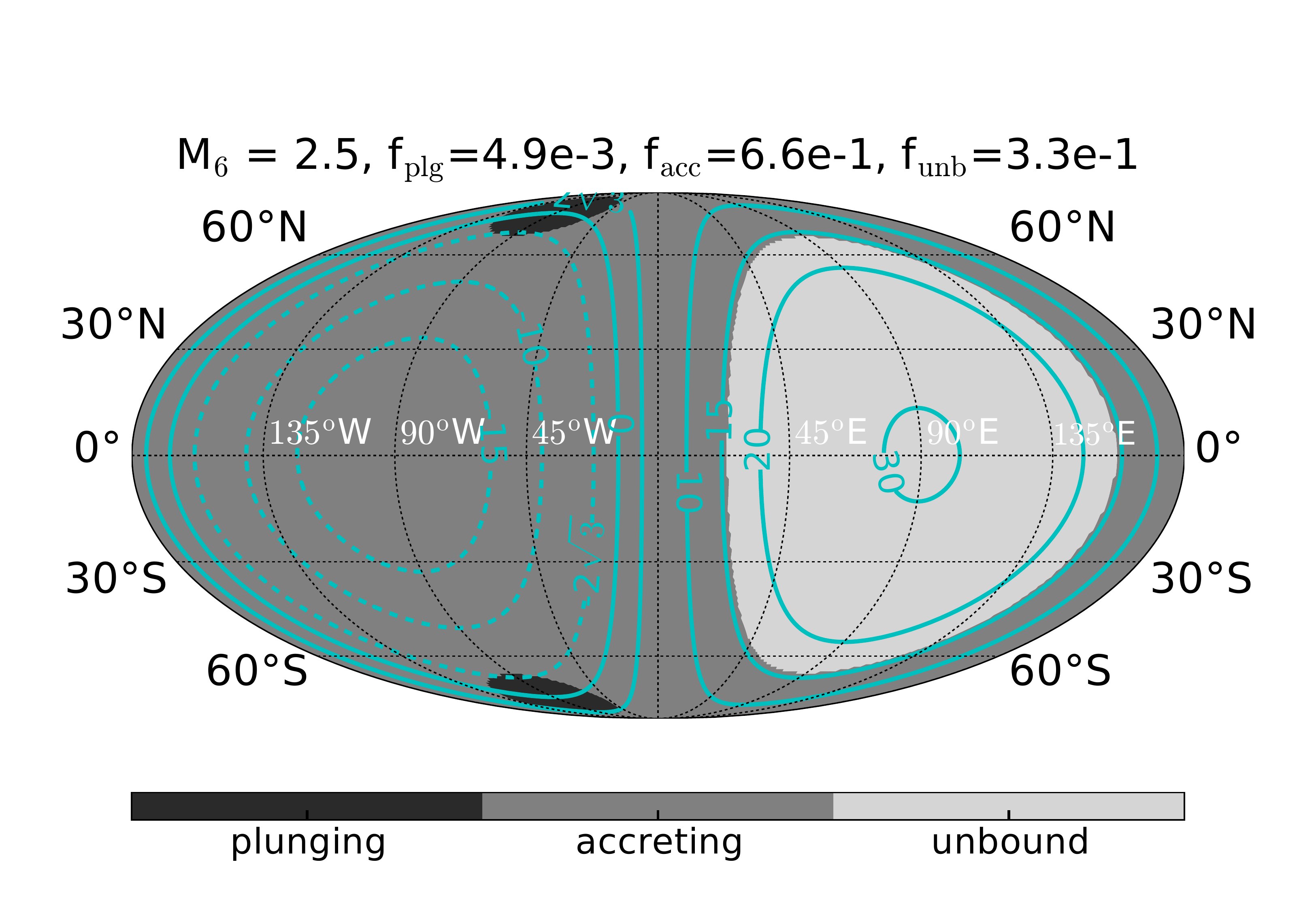}
\caption{Fate of the shocked gas expanding in different 
  directions ($\bthe$, $\bphi$) in the simulation box frame ---
  plunging (black), accreting (dark grey), and unbound (light
  gray). Here the polar angles $\bthe=0$ and $\pi$ correspond to
  latitudes $90^{\rm o}$ N and $90^{\rm o}$ S,
  respectively. The azimuthal angles $\bphi=0$ and $\pi$ correspond to
  longitudes $90^{\rm o}$ E and $90^{\rm o}$ W. The mass
  fraction for the three different fates ($f_{\rm 
    plg}$, $f_{\rm acc}$, $f_{\rm unb}$) are shown in the map title. The upper and
  lower panels are for BH masses $M_6=1.7$ and $2.5$, respectively. 
  We fix the star's mass $m_*=0.5$, impact parameter $\beta=1.0$
  ($\rp = \rt/\beta$) and orbital energy parameter $\eta = 
  1.0$ ($\eps = \eta\eps_{\rm T}$). The \textcolor{cyan}{cyan} contours are for the
  specific angular momentum in units of $\rg$ \textit{projected} in
  the direction of the star's initial orbital angular momentum. The
  specific angular momenta of the pre-disruption star are $\ell_*
  = \rp\sqrt{1/\mu_{\rm p}-1} = 9.0\rg$ and $10.2\rg$ for BH masses of
  $M_6=1.7$ and $2.5$, respectively. We see that the collision causes significant
   angular momentum redistribution.
}\label{fig:fatemap1}
\end{figure}



\begin{figure}
  \centering
\includegraphics[width = 0.48\textwidth,
  height=0.25\textheight]{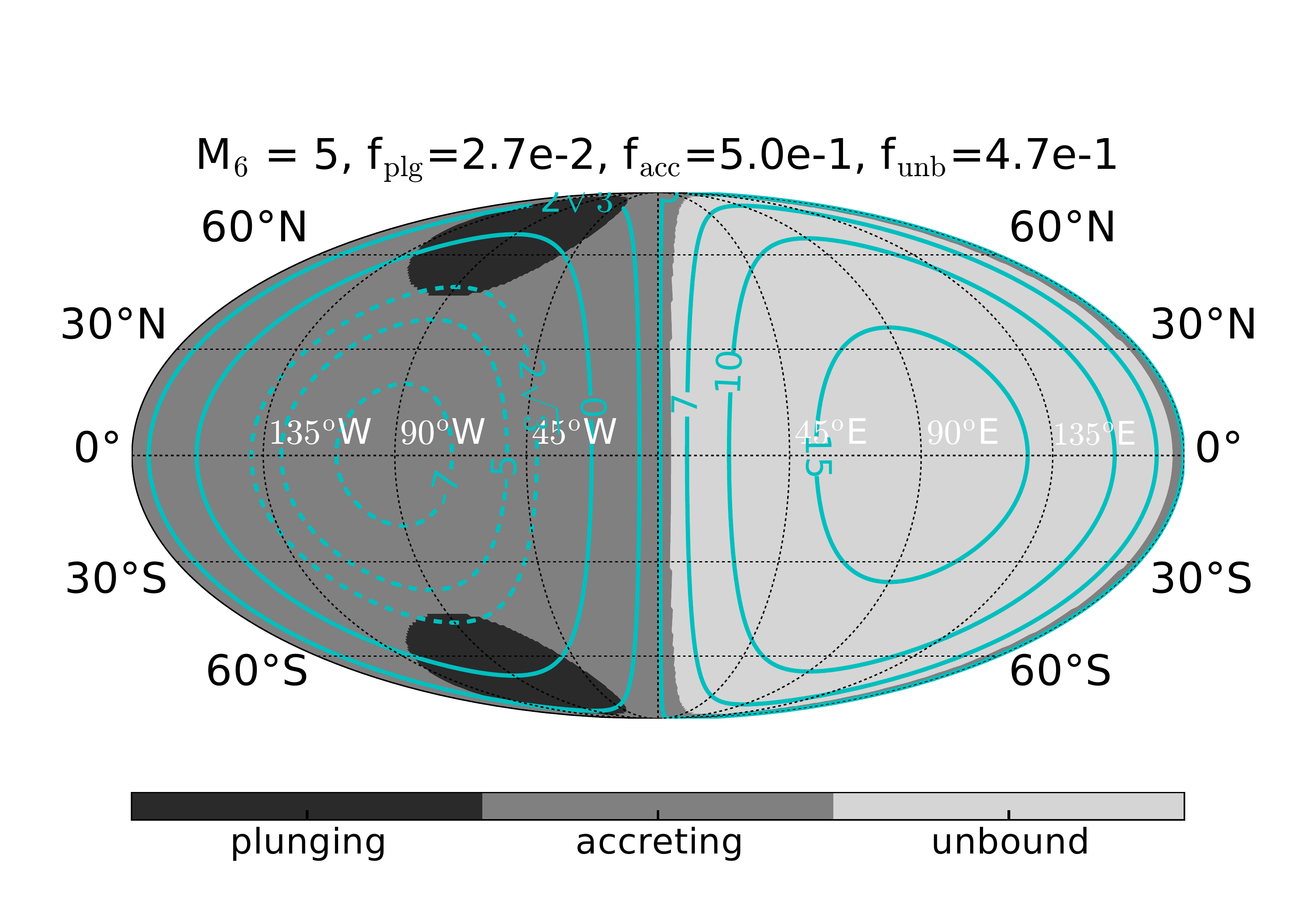}
\includegraphics[width = 0.48\textwidth,
  height=0.25\textheight]{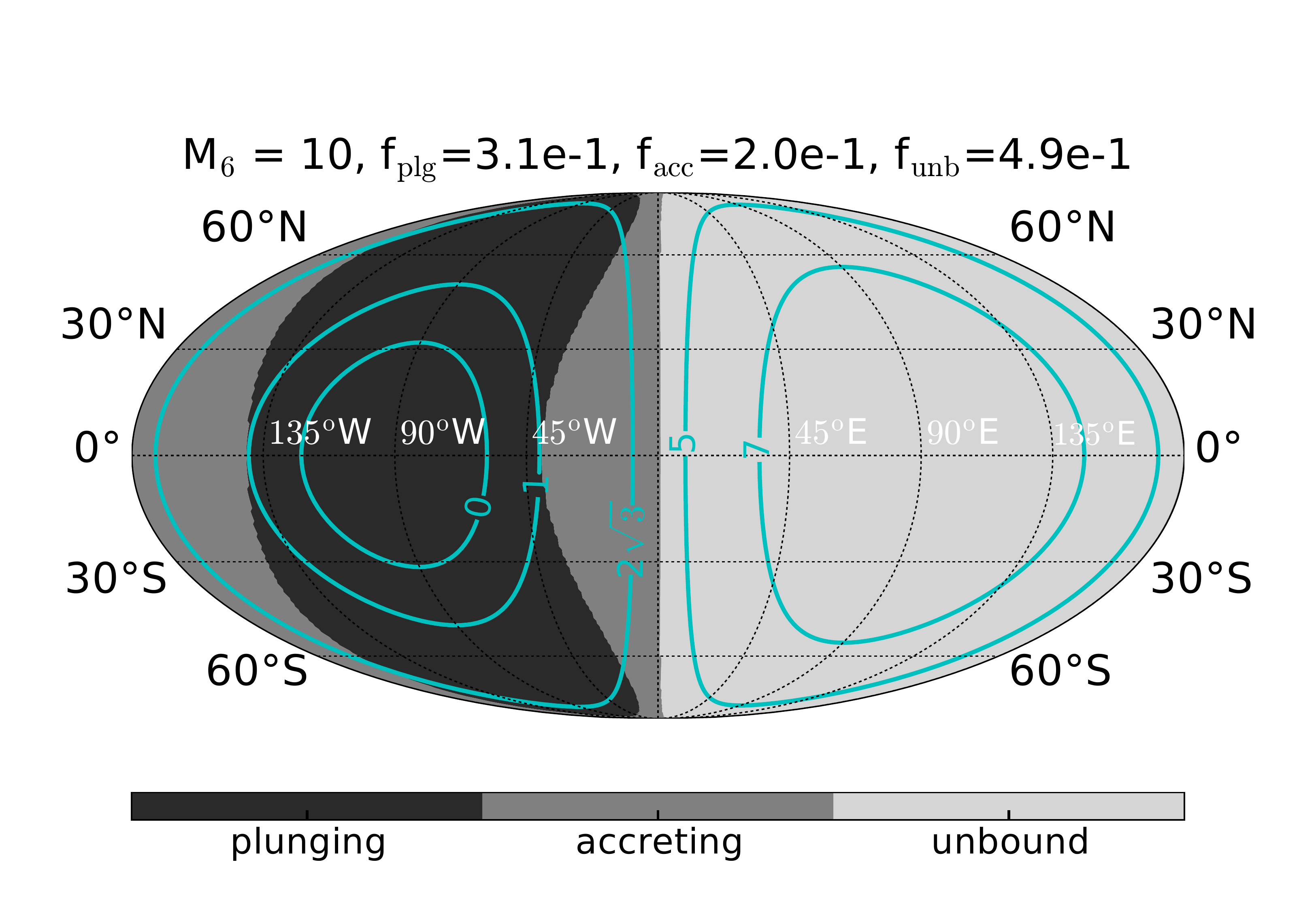}
\caption{The same as Fig. \ref{fig:fatemap1}, but for different BH masses
  $M_6 = 5$ (upper panel) and $10$ (lower panel). The
  specific angular momenta of the pre-disruption star are $\ell_*
  = 12.7\rg$ and $16.0\rg$ for the two cases of BH masses $M_6=5$ and
  $10$, respectively.
}\label{fig:fatemap2}
\end{figure}

In Fig. \ref{fig:fatemap1}
and \ref{fig:fatemap2}, we show the \textit{Mollweide projection} map
of fate in terms of the polar angle $\bthe$ and azimuthal 
angle $\bphi$ in the simulation box frame. Here the polar angle
$\bthe=0$ and $\pi$ correspond to latitudes $90^{\rm o}$ N and
$90^{\rm o}$ S, respectively. The azimuthal angle $\bphi=0$ and
$\pi$ correspond to longitudes $90^{\rm o}$ E and
$90^{\rm o}$ W. The unbound (``unb''), accreting (``acc''), and
plunging (``plg'') regions are shown in light grey, dark grey, and black,
respectively. For the four cases with different BH masses $M_6=1.7,\ 2.5,\ 5,$
and $10$, we fix the star's mass $m_*=0.5$, impact parameter
$\beta=1.0$ ($\rp = \rt/\beta$) and orbital energy parameter $\eta =
1.0$ ($\eps = \eta\eps_{\rm T}$). The \textcolor{cyan}{cyan} contours
show the distribution of specific angular momentum (in units of $\rg$)
\textit{projected}  in the direction of the star's initial orbital
angular momentum. Note that, when determining the fate of a certain
fluid element, we take into account its total specific energy and
total specific angular momentum. Subsequently, the out-of-plane
component of the angular moment will be further damped by shocks
within the ``accreting'' gas. If cooling is efficient, then more gas
is expected to plunge into the horizon. We also note that not
all fluid elements marked as ``plunging'' will necessarily fall
directly into the horizon. For instance, those moving in the
($\bthe\sim 0$, $\bphi\sim \pi/2$) direction (near the north pole
of Fig. \ref{fig:fatemap1} and \ref{fig:fatemap2}) will most likely
run into the ``accreting'' gas. The detailed dynamical evolution of
the bound gas is studied in a separate work
\citep{2019arXiv190605865B}. 

\begin{figure}
  \centering
\includegraphics[width = 0.48\textwidth,
  height=0.35\textheight]{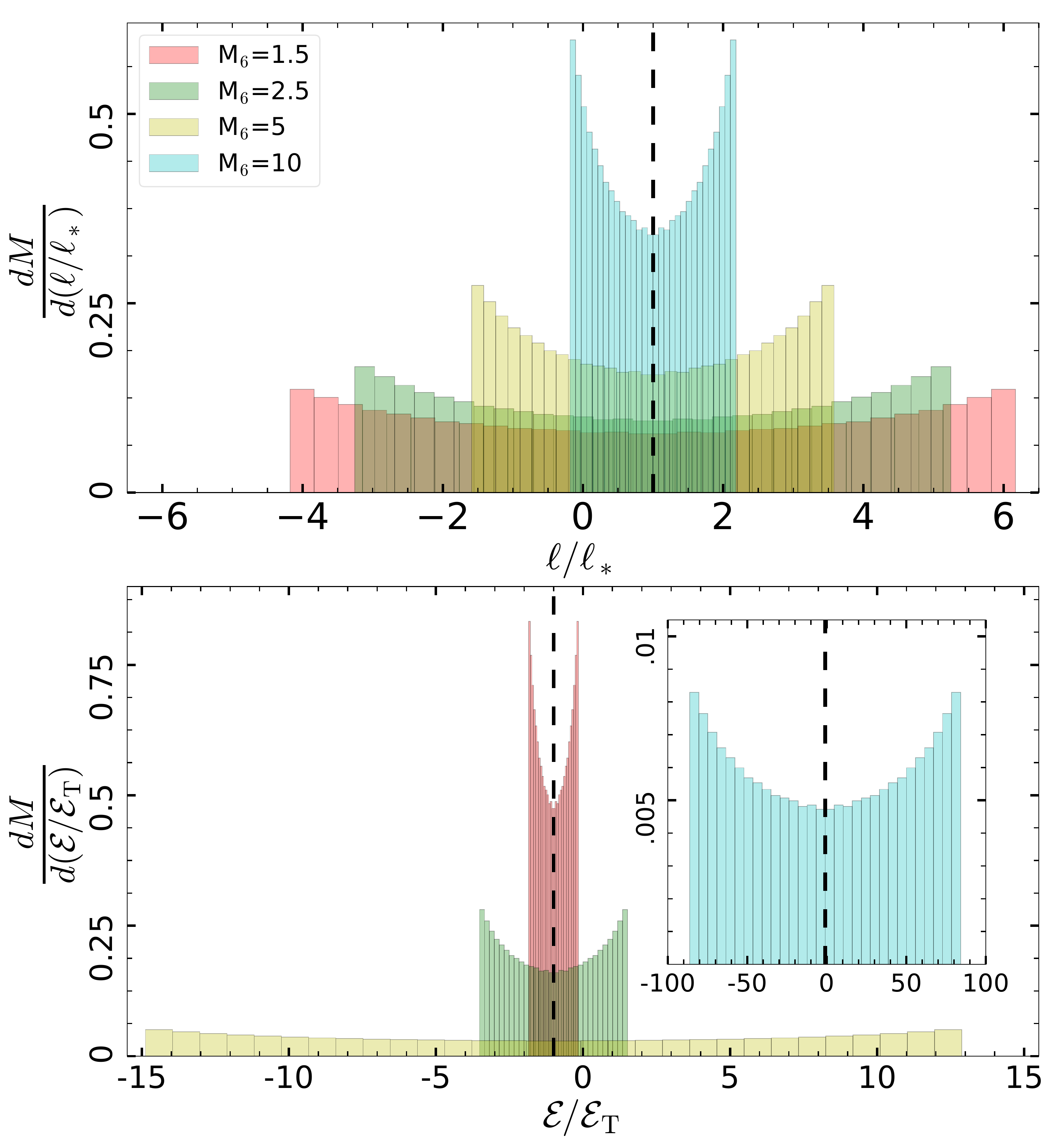}
\caption{Distributions of specific angular momentum projected along
  the star's initial angular momentum (upper panel) and specific
  orbital energy (lower panel) after the stream
  self-intersection, for four cases with different BH masses
  $M_6=1.5$, 2.5, 5.0, and 10 (while keeping $\beta = 1.0$, 
  $\eta=1.0$, and $m_*=0.5$ fixed). Each bin of the histograms is
  weighted by the gas mass (fraction) within its indicated specific
  energy or angular momentum range. The overall normalization is such
  that the integration over all bins is unity. Before the collision, the stream
  has specific angular momentum $\ell_0\approx \ell_* = \rp
  \sqrt{1/\mu_{\rm p}-1}$ and orbital energy $-\eps_{\rm T}$. Then,
  the stream collision leads to large spreads of specific angular
  momentum and orbital energy centered around the initial
  values. The fluid elements with $\ell<0$ have counter-rotating
  orbits and those with $\eps>0$ are unbound. In the $M_6=10$ case
  (subplot in the lower panel), the fastest moving unbound gas has
  speed $v_{\rm max}\simeq 0.3$.
}\label{fig:eps-ell-histogram}
\end{figure}

In Fig. \ref{fig:eps-ell-histogram}, we show the mass-weighted
distributions of specific angular momentum projected along the star's
initial angular momentum and specific orbital energy, for four cases
with $M_6=1.5$, 2.5, 5, and 10 (while keeping $\beta = 1.0$, 
  $\eta=1.0$, and $m_*=0.5$ fixed). Before the
collision, the stream has specific angular momentum $\ell_0\approx
\ell_* = \rp \sqrt{1/\mu_{\rm p}-1}$ and orbital energy $-\eps_{\rm
  T}$. The collision causes a spread of specific angular momentum by
$\Delta \ell/\ell_*\sim$ a few, and the corresponding spread in the
Keplerian circularization radius is about a factor of 10. In
some cases (e.g. $M_6=1.5$ and 2.5), a large fraction of shocked gas is in
counter-rotating orbits ($\ell<0$) and will subsequently collide with the
forward-rotating gas ($\ell>0$) at a wide range of radii. The spread
in specific orbital energy after the collision is very sensitive to the BH mass, with
$\Delta \eps/\eps_{\rm T}\sim 1$ for $M_6=1.5$ but 
$\Delta \eps/\eps_{\rm T}\sim 100$ for $M_6=10$. For the $M_6=1.5$
case, there is no unbound gas. For BH mass $M_6\gtrsim 2$, a large
fraction of the shocked gas is unbound ($\eps>0$). For a highly
eccentric Keplerian orbit, the eccentricity is given by $e =
\sqrt{1-2|\eps|\ell^2/\rg^2} \approx 1 -
|\eps|\ell^2/\rg^2$. We can see that strong shocks due to
self-intersection increase the product $|\eps|\ell^2$ by one order of
magnitude or more, and hence the accreting fraction of gas should
quickly circularize.

\begin{figure}
  \centering
\includegraphics[width = 0.48\textwidth,
  height=0.25\textheight]{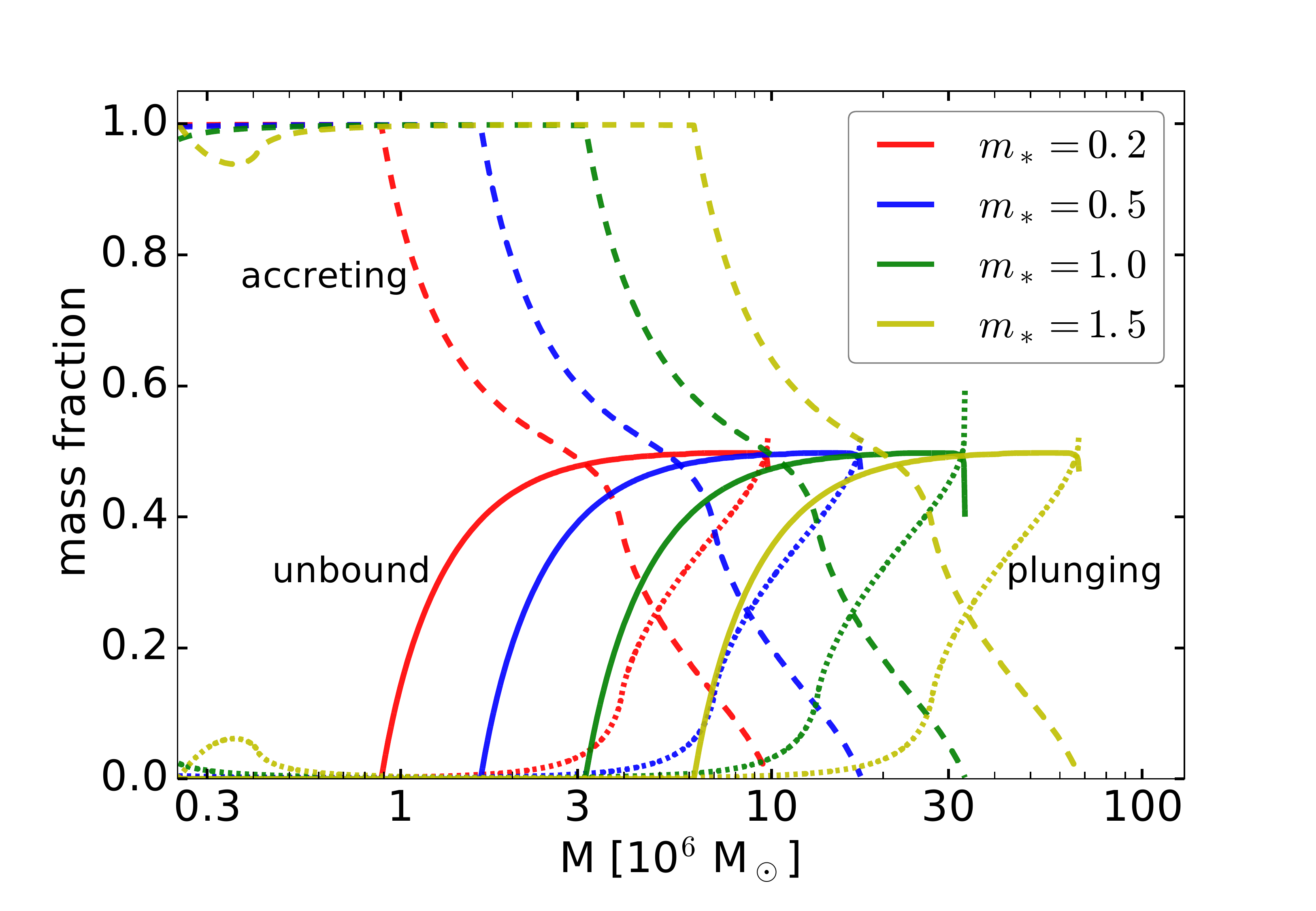}
\caption{The mass fractions of the unbound (solid), plunging (dashed),
  and accreting (dotted) gas for different stellar masses $m_*=0.2$,
  $0.5$, $1.0$, and $1.5$. In all cases, we fix the impact parameter
  $\beta=1.0$ ($\rp = \rt/\beta$) and orbital energy parameter $\eta =
  1.0$ ($\eps = \eta\eps_{\rm T}$). The small bump (or dip) in the
  plunging (or accreting) fraction for $m_*=1.5$  and
  $M_6\sim 0.3$ is because the velocity before the collision has
  comparable $\hat{r}$ and $\hat{\phi}$ components:
  $v_r\sim v_{\phi}$. See Fig. \ref{fig:fatemap3} for the map
  of fate for this case.
}\label{fig:fate-fraction}
\end{figure}

In Fig. \ref{fig:fate-fraction}, we show the mass fractions of the
unbound, accreting, and plunging gas as a function of BH mass, for
four stellar masses $m_*=0.2$, $0.5$, $1.0$, and $1.5$
(while keeping $\beta=1.0$ and $\eta=1.0$ fixed). We find that, above a
critical BH mass (to be quantified shortly), the unbound
fraction quickly rises from 
$0\%$ to a maximum of $50\%$. At the same time, the accreting
fraction drops from $100\%$ to $50\%$. For higher
mass BHs, the plunging fraction\footnote{Note that the plunging fraction
is non-zero even for very low BH masses, this is because the collision
broadens the angular momentum distribution such that part of the
shocked gas has almost zero angular momentum (see the upper panel of
Fig. \ref{fig:eps-ell-histogram}). The small bump (or dip)
in the plunging (accreting) fraction for the $m_*=1.5$ case near BH masses
$M_6\sim 0.3$ is because the velocity before the collision has
comparable $\hat{r}$ and $\hat{\phi}$ components:
$v_r\sim v_{\phi}$ (see the third panel of
Fig. \ref{fig:intersection}). We show the map of fate for $m_*=1.5$
and $M_6=0.3$ in Fig. \ref{fig:fatemap3} in the Appendix.} quickly
rises at the expense of the dropping accreting 
fraction, while the unbound fraction stays roughly unchanged at
$\sim$$50\%$. There is a maximum mass $M_{\rm max}$ above which no
accretion is possible, because the entire star plunges into the event
horizon.

\begin{figure}
  \centering
\includegraphics[width = 0.48\textwidth,
  height=0.5\textheight]{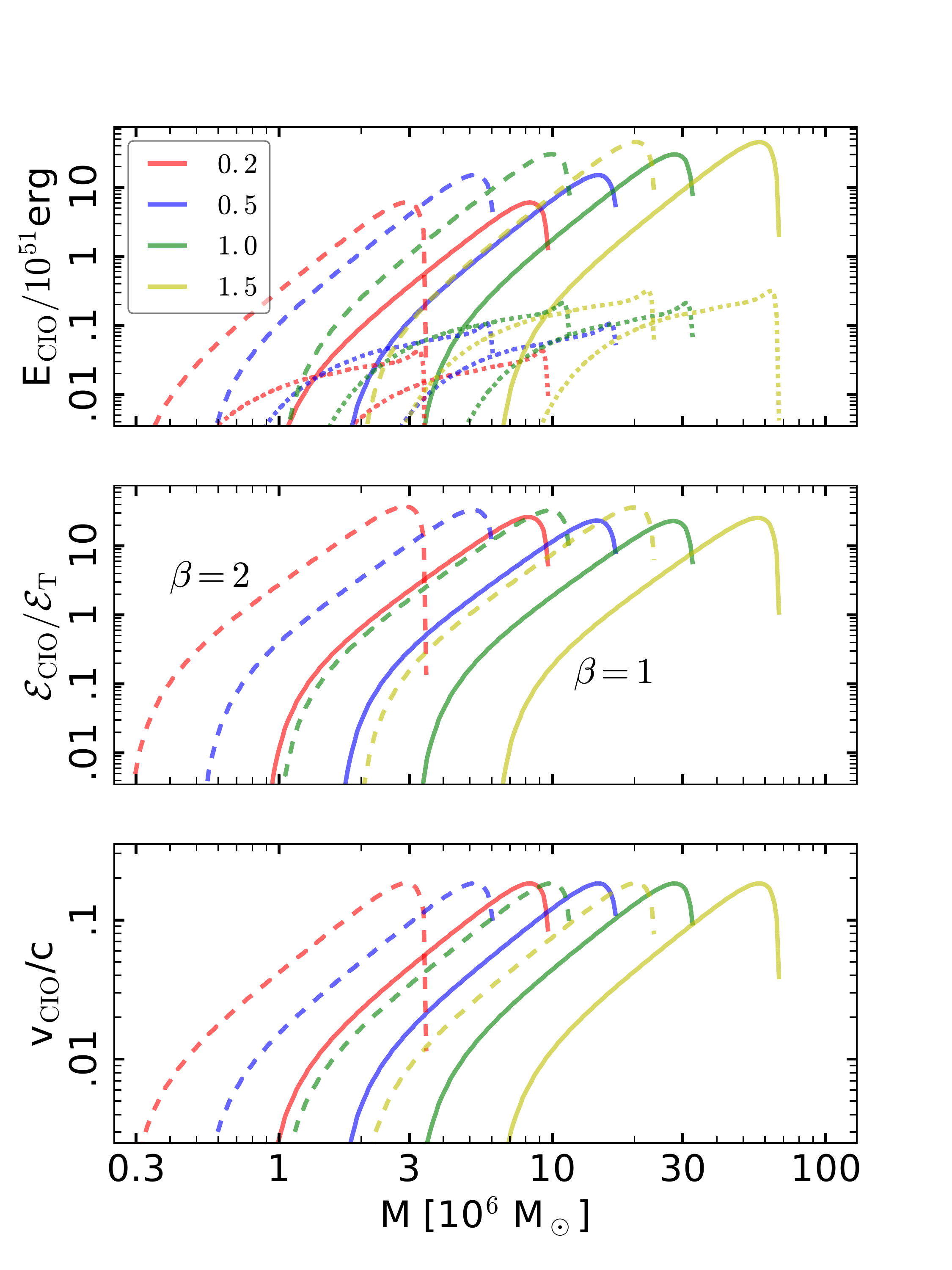}
\caption{The asymptotic total kinetic energy (upper panel), mean specific
  energy (middle panel), and mean speed (lower panel) of the CIO for a number
  of cases with different stellar masses ($m_*=0.2$, 0.5, 1.0, and 1.5
  as marked in the legend) and impact parameters $\beta=1$ (solid
  curved) and $2$ (dashed curves). In the upper panel, the dotted
  lines show the kinetic energy dissipated by collisions
  within the CIO (see Fig. \ref{fig:CIOtraj1} and
  \ref{fig:CIOtraj2}), which is only a small fraction of the total
  kinetic energy. We find that, when the unbound 
  fraction is high $f_{\rm unb}\gtrsim 20\%$, the total kinetic
  energy spans a wide range from $\sim$$10^{50}\rm\,erg$ up to
  $\sim$$10^{52}\rm\,erg$. When the (mass-weighted) mean specific
  energy is greater than $\eps_{\rm T}$, the CIO is more powerful than
  the unbound tidal debris. The (mass-weighted) mean speed varies from 
  $\sim$$0.01c$ to $\sim$$0.1c$.
}\label{fig:Ecio}
\end{figure}

In this paper, we call the unbound fraction of the shocked gas the
``collision-induced outflow'' (CIO), which has important observational
consequences (see the next section). The launching
of CIO can be understood in the following Newtonian picture. If the
intersection occurs at $\rp\ll \ri\lesssim \ra/3$ 
($\ra$ being the apocenter radius), then the two streams typically
collide at a large angle $2\theta_{\rm I}\gtrsim 90^{\rm o}$ near the local escape
speed $|v_{\rm r}\hat{r} + v_{\rm \phi}\hat{\phi}|=\sqrt{v_{\rm
    r}^2 + v_{\rm \phi}^2} \simeq v_{\rm 
  esc}\simeq \sqrt{2\rg/\ri}$. The radial velocity component is
dissipated by shocks, and then the shocked gas adiabatically expands
at speed $\sim$$v_{\rm r}$ in a roughly spherical manner in the SB frame
moving at velocity $v_{\rm \phi}\hat{\phi}$. Going back to the LSO
frame, we find the fastest moving shocked gas 
in the ($\bthe=\pi/2$, $\bphi=0$) direction with velocity
$(v_{\rm  r} + v_{\rm \phi})\hat{\phi}$ and 
speed $v_{\rm  r} + v_{\rm \phi}> \sqrt{v_{\rm r}^2 + v_{\rm
    \phi}^2}\simeq v_{\rm esc}$. We see that CIO is a generic feature
of gas streams colliding near the local escape speed of the intersection
point\footnote{This feature was captured in the simulations
  of deeply penetrating TDEs by
\citet{2015ApJ...805L..19E, 2016MNRAS.458.4250S,
  2016ApJ...830..125J}. In many other works, the gas streams do not
collide near the local escape speed of the collision point, either
because the aspidal precession is so weak (due to small $M/M_*$ ratio)
that the collision occurs near the apocenter
\citep[e.g. ][]{2015ApJ...804...85S} or because the star is  
  initially in a bound orbit with too low eccentricity
  \citep[e.g.][]{2016MNRAS.455.2253B, 2016MNRAS.461.3760H}.
}. In Fig. \ref{fig:Ecio}, we show the asymptotic kinetic energy and
mass-weighted mean speed of the CIO for a number of cases, assuming
the total amount of unbound mass to be $f_{\rm unb} m_*\msun$.

We define the critical BH mass $M_{\rm cr}$ above which the mass fraction 
of unbound gas is more than $20\%$, i.e. $f_{\rm unb} (M_{\rm cr})=
20\%$. After exploring an extensive grid of
parameters (see Fig. \ref{fig:Mcrit} in the Appendix), we find
\begin{equation}
  \label{eq:19}
  M_{\rm cr} \simeq (4.6\times10^{6}\,\msun)\, \eta^{0.4}
  \beta^{-1.7} m_*^{-1/2} r_*^{3/2}. 
\end{equation}
This can be translated to a critical pericenter radius
\begin{equation}
  r_{\rm p,cr}/\rg\simeq 17\, \eta^{-4/15}\beta^{2/15},
\end{equation}
below which $f_{\rm unb}\gtrsim 20\%$. If we choose the critical unbound
fraction to be $30\%$ (instead of $20\%$), the scalings in the above
equations stay the same but the normalization changes to $M_{\rm
  cr}\simeq 5.3\times10^{6}\,\msun$ 
(and $r_{\rm p,cr}/\rg\simeq 15$). The precise value
of the critical unbound fraction is unimportant, because $f_{\rm unb}$
is very sensitive to the BH mass near $M\sim M_{\rm cr}$.

The maximum Schwarzschild BH mass for tidal disruption to
occur outside the event horizon can be estimated by requiring 
$\rp = \rt/\beta>\rmb = 4\rg$ (since typically $|\eps|\ll 1$, see eq. \ref{eq:3}),
\begin{equation}
  \label{eq:13}
  M_{\rm max} \simeq (4.0\times10^7\,\msun)\, \beta^{-3/2}
  m_*^{-1/2} r_*^{3/2}.
\end{equation}
Note that the maximum BH mass
depends on the minimum impact 
parameter $\beta_{\rm min}$ at which the relativistic tidal forces
exceeds the star's self-gravity. In the limit $R_*\ll
4\rg$, the local gravitational-field gradients can described by the
relativistic tidal tensor in Fermi normal coordinates. For the
Schwarzschild spacetime, the criterion for marginal tidal disruption
can be written as
\begin{equation}
  \label{eq:15}
  \rp \approx \rmb\approx 5^{1/3}\xi_*\rt,
\end{equation}
where the $5^{1/3}$ factor comes from relativistic tidal
stretching\footnote{It can be shown that, for the case where the
  star's initial angular momentum is parallel to the spin of a Kerr
  BH,  the $5^{1/3}$ factor stays the same for arbitrary spin. This is
  because the eigenvalues of the tidal tensor
  depends on the ratio $(\ell_{\rm mb} - a)/\rmb \equiv 1$
  \citep{2012PhRvD..85b4037K}, where $\ell_{\rm mb}$ is the angular
  momentum of the marginally bound parabolic orbit and $-1<a<1$ is the
  dimensionless BH spin ($a<0$ corresponds to retrograde orbits).
} in the radial direction \citep{2012PhRvD..85b4037K} and the
parameter $\xi_*\sim 1$ accounts for the internal structure of the star. We
will discuss the choice of $\xi_*$ for different stellar masses in \S 
5.3 on TDE demographics. The marginal
disruption case occurs when $\ell \approx 
4\rg$ and $\rp\approx 4\rg$, which gives the minimum impact
parameter $\beta_{\rm min} \approx 5^{-1/3}\xi_*^{-1}\simeq
0.6\,\xi_*^{-1}$ and the maximum 
mass for non-spinning BHs associated with TDEs $M_{\rm max, 
  Sch} \approx (8.9\times10^{7}\mr{\msun})\, \xi_*^{3/2} m_*^{-1/2}
r_*^{3/2}$. We can see that relativistic tidal forces are slightly
better at disrupting stars than in the Newtonian approximation. In
realistic situations, the precise $\beta_{\rm min}$ (and hence $M_{\rm max}$)
will depend on the stellar structure, BH's spin, star's spin, and the
misalignment between the star's  
orbital and the BH's spin angular momenta, etc. Fortunately, the
precise value of $\beta_{\rm min}$ may not be important from the
observational point of view, because those TDEs with BH
mass close to $M_{\rm max}$ should be quite dim due to their low
accreting fraction (most gas is either unbound or plunging, see
Fig. \ref{fig:fate-fraction}).

\section{Observations}
In previous sections, we have described a semi-analytical model for
the TDE dynamics, including the fluid properties at the stream
self-intersection point and the fate of the shocked gas moving in
different directions. In this section, we first discuss the
circularization of the fallback stream and the formation of accretion
disk in \S 5.1 and then we consider the observational implications of
the unbound gas (when $f_{\rm unb}\neq0$) in \S 5.2. TDE demographics
will be discussed in \S 5.3.

\subsection{Circularization of the fallback mass}
For main-sequence stars disrupted by low-mass BHs ($M_6\ll 1$), the
stream self-crossing occurs near the apocenter and the shocks only
dissipate a small fraction of the orbital energy
\citep[see the fourth panel of Fig. \ref{fig:intersection},
and][]{2016MNRAS.455.2253B, 2018arXiv180608093C}. After exploring an
extensive grid of parameters (see Fig. \ref{fig:fdiss} in the
Appendix), we find the dissipation efficiency in eq. (\ref{eq:9}) can
be written in the following analytical form for $M_6\lesssim 1$
\begin{equation}
  \label{eq:20}
  f_{\rm diss} = 1 - {1\over 1 + x},\ \mr{where}\ x = 0.27\,
  \eta^{-1}\beta^3 M_6^{5/3} m_*^{1/3}r_*^{-2}.
\end{equation}
Note that in the limit $x\ll 1$ ($f_{\rm diss}\approx x$), if the dissipation of
orbital energy is only due to stream intersection, 
then the circularization timescale can be roughly estimated by $f_{\rm
diss}^{-1}P_{\rm min}\simeq (152\mr{\,d})\, \eta^{-1/2}\beta^3 M_6^{-7/6}
m_*^{-4/3} r_*^{7/2}$. For an average star ($m_*\lesssim1$) disrupted by
low-mass BHs $M_6\ll 1$, this timescale is much longer than the
durations of typical TDEs discovered in recent UV-optical surveys. We
also see that tidal disruptions of red giants ($r_*\gg 1$) will likely
have very long circularization timescales as well, unless they are in deeply
penetrating orbits ($\beta \gg 1$).

If MHD turbulence develops rapidly, shear due to magnetic stresses
may cause dissipation of orbital energy at a rate (per unit mass)
$\dot{\eps}_{\rm vis} \sim \alpha \Omega_\ell v_{\rm A}^2$
\citep{2017MNRAS.467.1426S}, where 
$\alpha\sim 0.1$ is the viscous parameter \citep{1973A&A....24..337S},
$\Omega_\ell(r) = \ell_*/r^2$ is the local orbital angular frequency,
and $v_{\rm A} =
\sqrt{B^2/4\pi \rho^2}$ is the Alfv{\'e}n speed. Due to conservation of
flux along the stream, the magnetic field strength evolves with radius
as $B\propto H(r)^{-2}$, where $H(r)$ is the stream thickness at
radius $r$. Right after the disruption, the marginally bound
part of the stream moves as $r\propto t^{2/3}$ and the stream length
stretches\footnote{For nearly radial orbits, the Newtonian equation of
  motion is $r(t) = r_0 + \int^t_0 \d t\sqrt{2(\rg/r + \mc{E})}$,
  where $\mc{E}$ is the binding energy and $r_0$ is the initial
  position. In the limit $\mc{E}\approx 0$ 
  (marginally bound) and $r\gg r_0$, we have $r\propto
  t^{2/3}$. Consider two fluid elements with 
  the same initial position $r_0$ but slightly different binding energy $\Delta
  \mc{E}$. After expanding for time $t$, they are separated by a
  distance $\Delta r = \Delta \mc{E} \int_0^t \d t [2(\rg/r +
  \mc{E})]^{-1/2}$. In the limit $\mc{E} \approx 0$ and $r\gg r_0$, we
  have $\Delta r\propto t r^{1/2} \propto r^2$. } as $r^2$, so we obtain  
the  stream density evolution $\rho\propto 
r^{-2}H(r)^{-2}$ and hence the total viscous heating $\Delta \eps_{\rm vis}
\sim \dot{\eps}_{\rm vis} t \propto
r^{3/2} H(r)^{-2}$. The evolution of stream thickness $H(r)$ may be highly
complex, depending on tidal forces, self-gravity, magnetic fields, and recombination
of hydrogen \citep{1994ApJ...422..508K, 2014ApJ...783...23G,
  2016MNRAS.459.3089C}. In the limiting case of
equilibrium between self-gravity and gas pressure, we have $H\propto
\sqrt{P/\rho^2} \propto \rho^{-1/6}$ and hence $H(r)\propto r^{1/2}$
\citep{2016MNRAS.459.3089C}, where
we have taken a polytropic index of 5/3 which is appropriate before
magnetic pressure overwhelms gas pressure or the
recombination\footnote{For $\beta\sim 1$, recombination occurs at
  radius $r_{\rm rec}\sim \sqrt{T_0/10^4\mr{\,K}}\, \rt \sim 30\,\rt$,
  where $T_0\sim GM_*/kR_* \sim 10^7\rm\,K$ ($k$ being the Boltzmann
  constant) is the gas temperature right after tidal 
  disruption.} of hydrogen. In this regime, $\Delta \eps_{\rm vis} \propto
r^{1/2}$ and most dissipation occurs near the largest radii at which the
scaling $H(r)\propto r^{1/2}$ holds. In the other limit where tidal
forces dominate over self-gravity, we have $H(r)\propto r$ and $\Delta
\eps_{\rm vis}\propto r^{-1/2}$, which means that most dissipation
occurs near the smallest radii at which the scaling $H(r)\propto r$ holds.

For $\beta\sim\eta \sim 1$, we provide a conservative estimate of the
viscous dissipation by 
assuming the scaling $\Delta \eps_{\rm vis} \propto r^{1/2}$ up to the
apocenter radius of the most tightly bound orbit $r_{\rm a}$ and obtain
\begin{equation}
  \label{eq:18}
  \Delta \eps_{\rm vis, max}\sim \alpha \,  v_{\rm  A}^2({\rt})
  \sqrt{r_{\rm a}\over \rt}.
\end{equation}
Since $\alpha \sqrt{\ra/\rt}\sim \alpha (M/M_*)^{1/6}\sim 1$, we
obtain $\Delta \eps_{\rm vis, max}\sim v_{\rm  A}^2({\rt})$, where
$v_{\rm A}$ is the Alfv{\'e}n speed near radius $\rt$. The
magnetic field may be amplified in the tidal disruption process due to
forced differential rotation \citep{2017MNRAS.469.4879B} and the total
magnetic energy may 
be written as $f_{\rm B}GM_*^2/R_*$, where $GM_*^2/R_*$ is
the work done by tidal forces and $f_{\rm B}\ll 1$ is the conversion
efficiency. Then, we obtain $\Delta \eps_{\rm vis, 
  max} \lesssim f_{\rm B}GM_*/R_*\sim 2\times10^{-6}f_{\rm B}\ll
\eps_{\rm T}$, where $\eps_{\rm T}$ 
is the typical orbital energy of the stream (eq. \ref{eq:3}). Therefore, the
dissipation of orbital energy due to viscous shear is highly
inefficient over the orbital timescale.

We also note that dissipation by the nozzle shock operating
near the pericenter may also be inefficient, because the ratio between the
velocity components perpendicular and inside the 
star's orbital plane is of order $H/r\ll 1$. However, this picture may
be changed by strong apsidal precession (which causes oblique
compression) if the pericenter is close to the horizon $\rp\lesssim
10\rg$. 

Therefore, we conclude that TDEs by low-mass BHs ($M_6\ll1$) have
circularization timescale $t_{\rm cir}\gg P_{\rm min}$, which is much
longer than the typical duration of the current sample of TDEs
discovered by UV-optical surveys. This has important implications on
TDE demographics, which will be discussed in \S 5.3.

In the following, we focus on TDEs by relatively high-mass BHs 
$M_6\gtrsim 1$ where the dissipation of orbital kinetic energy is
dominated by stream self-intersection because $f_{\rm diss}\sim 1$. As
shown in Fig. \ref{fig:eps-ell-histogram}, the distributions of 
specific angular momentum and orbital energy are broadened by
the collision. The eccentricity $e\simeq 1 - |\eps|\ell^2/\rg^2$ drops
to the level of $1-e\sim 0.1$ due to the increase of the product
$|\eps|\ell^2$ by typically one order of magnitude. Subsequently, the
bound gas (and a small fraction of the unbound gas, see \S 5.2) will
collide violently at a wide range of 
radii between $\sim$$\rp$ and $\sim$$\ri$ over a timescale
$\Omega_{\rm K}^{-1}(\ri) = c^{-1}\sqrt{\ri^3/\rg} < P_{\rm
  min}$, where $\Omega_{\rm K}(\ri)$ is the Keplerian angular
frequency for a circular orbit at $\ri$. Thus, orbital circularization
due to exchange of energy and angular momentum occurs rapidly after
the initial stream self-intersection. The detailed dynamics is
highly complicated due to the interplay among gas, radiation 
(providing cooling), and magnetic fields (providing viscosity). This
is explored in a separate work \citep{2019arXiv190605865B}.

The most interesting situation is when a significant fraction of the
shocked gas becomes unbound in the form of CIO, which occurs for BH
masses in the range $M_{\rm cr} <M< M_{\rm max}$. The CIO carries away
(positive) energy of $\eps\gtrsim \eps_{\rm T}$ (see
Fig. \ref{fig:eps-ell-histogram} and \ref{fig:Ecio}), and the rest of the
shocked gas is left in more tightly bound orbits. The (positive)
angular momentum carried away by the CIO is a factor of a few greater
than that before the collision, so the remaining bound gas typically
have negative angular momentum and hence rotates in the opposite
direction of the initial star. Due to subsequent shocks, the
counter-rotating gas will quickly settle into circular orbits not at
radius $2\rp$ but with a radial spread of at least one order of
magnitude (even without viscosity).


\subsection{Collision-induced Outflow (CIO)}
For BH masses in the range $M_{\rm cr}< M<M_{\rm
  max}$, we find a large fraction of gas is launched in the form of a
wide-angle CIO. In the following, we first study the morphology of
the CIO (\S 5.2.1), and then discuss the observational
implications of the CIO, including reprocessing the extreme-UV (EUV)
or soft X-ray
disk emission into the optical band (\S 5.2.2) and 
radio emission from the shock driven into the ambient medium
(\S 5.2.3).

\begin{figure}
  \centering
\includegraphics[width = 0.48\textwidth,
  height=0.22\textheight]{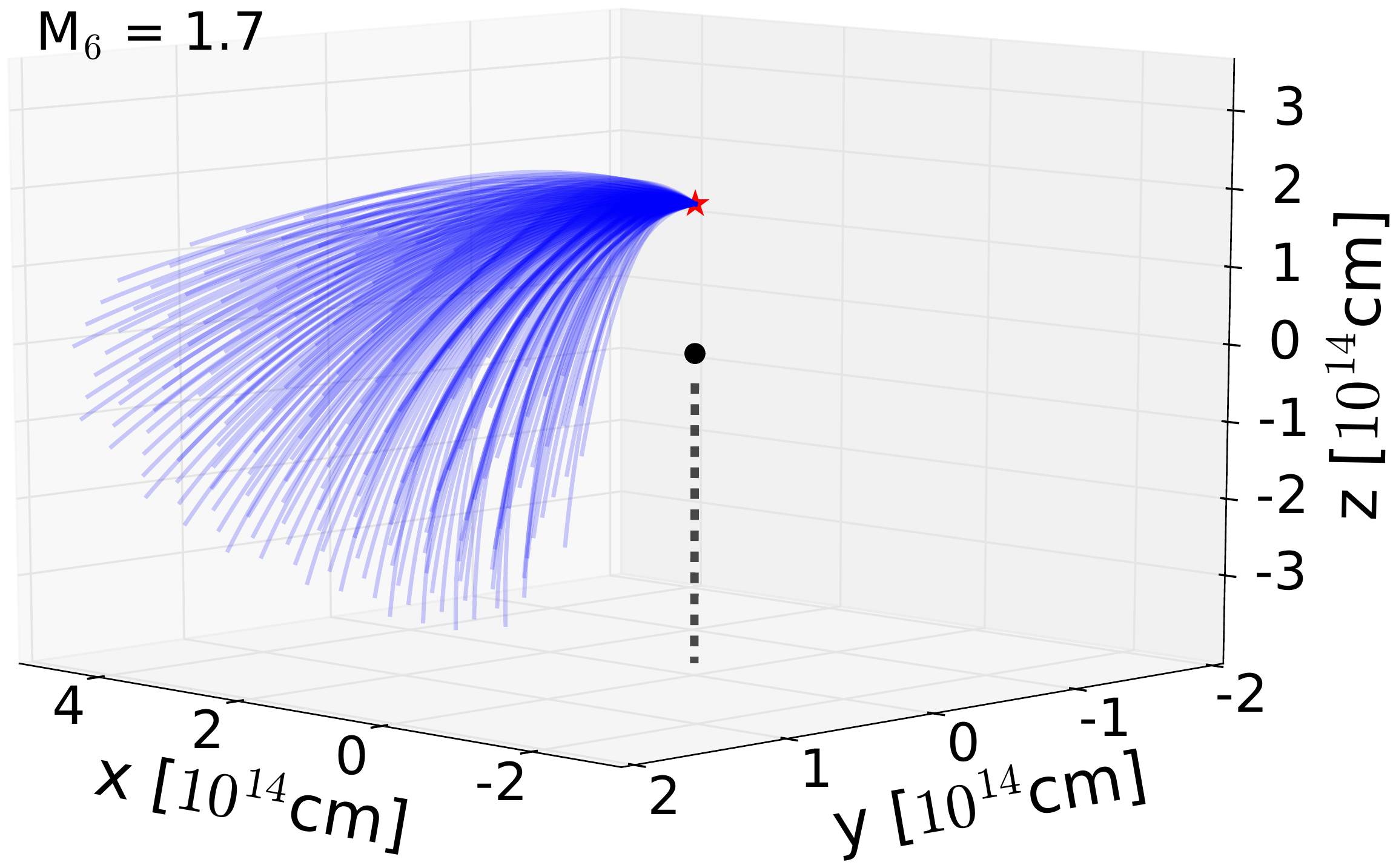}
\includegraphics[width = 0.48\textwidth,
  height=0.22\textheight]{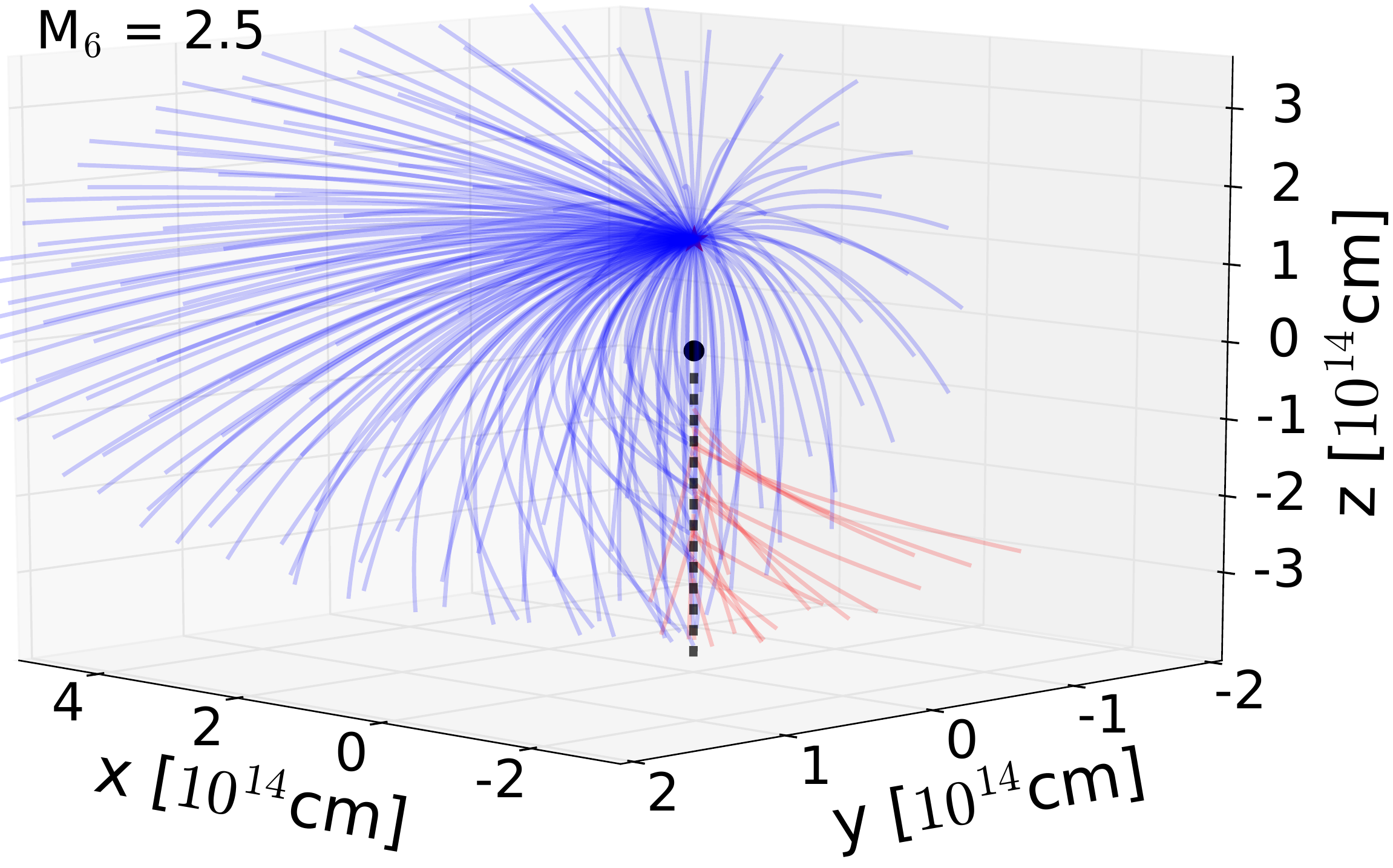}
\caption{The free-fall trajectories of $\sim$200 fluid elements nearly
  uniformly distributed within the unbound cone, for two cases with
  $M_6=1.7$ (upper panel), $M_6=2.5$ (lower panel). The other
  parameters $\beta=1.0$, $\eta = 1.0$, and $m_*=0.5$ are fixed. The
  integration time (after the stream intersection) for the $M_6=1.7$ case is
  $1.9\,$d and for the $M_6=2.5$ case it is $2.2\,$d. For the
  $M_6 = 2.5$ case, a small fraction of the fluid elements collide
  with others within the integration time and their free-fall
  trajectories after the collision (shown in red) are
  inaccurate. The blue curves show the trajectories without or before
  collisions. The stream self-intersection point is marked by a red
  star (at $x=y=0$, $z=\ri$) and the BH is marked by a black circle
  (at $x=y=z=0$).
}\label{fig:CIOtraj1}
\end{figure}

\subsubsection{Morphology of the CIO}
\begin{figure}
  \centering
\includegraphics[width = 0.48\textwidth,
  height=0.22\textheight]{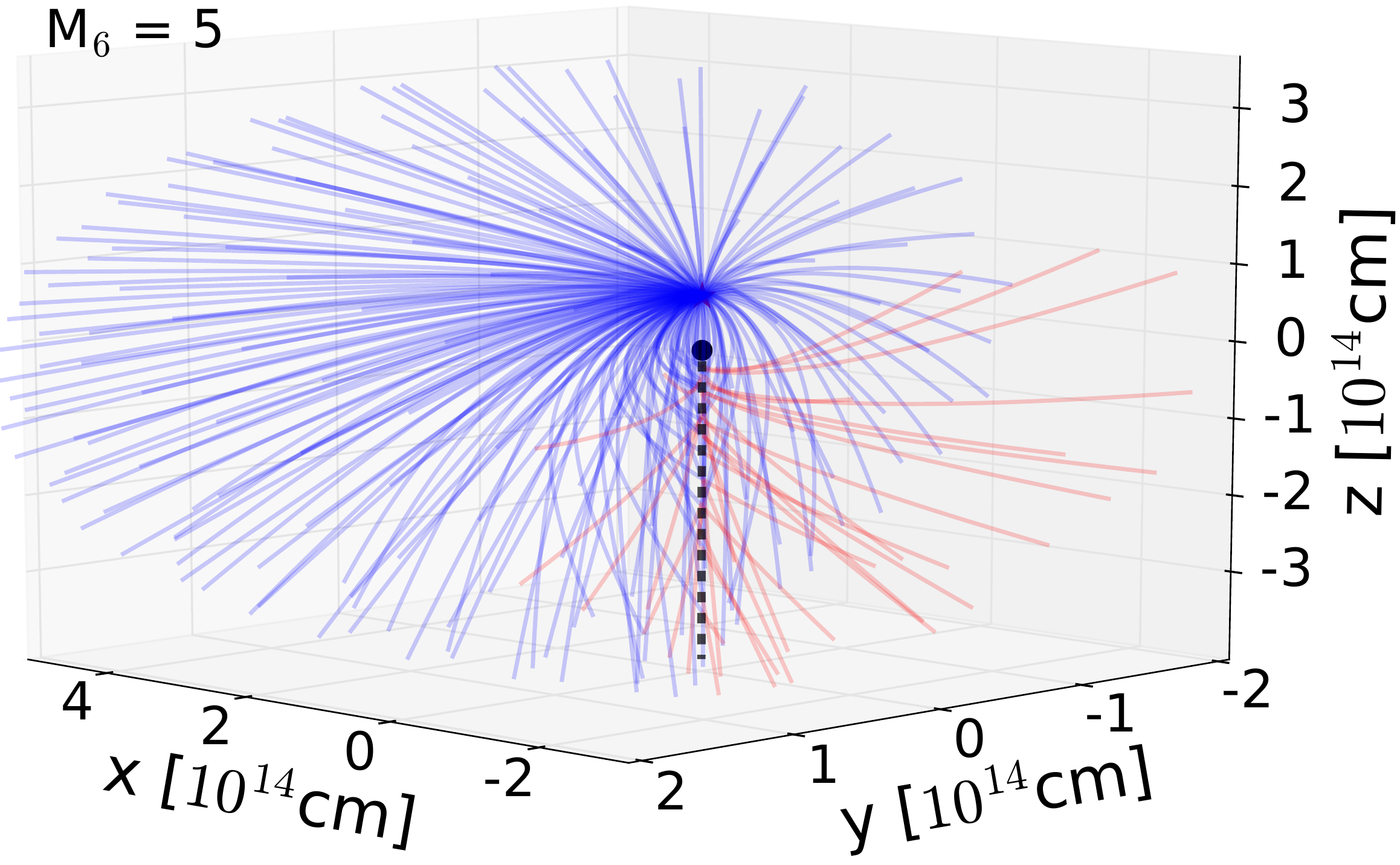}
\includegraphics[width = 0.48\textwidth,
  height=0.22\textheight]{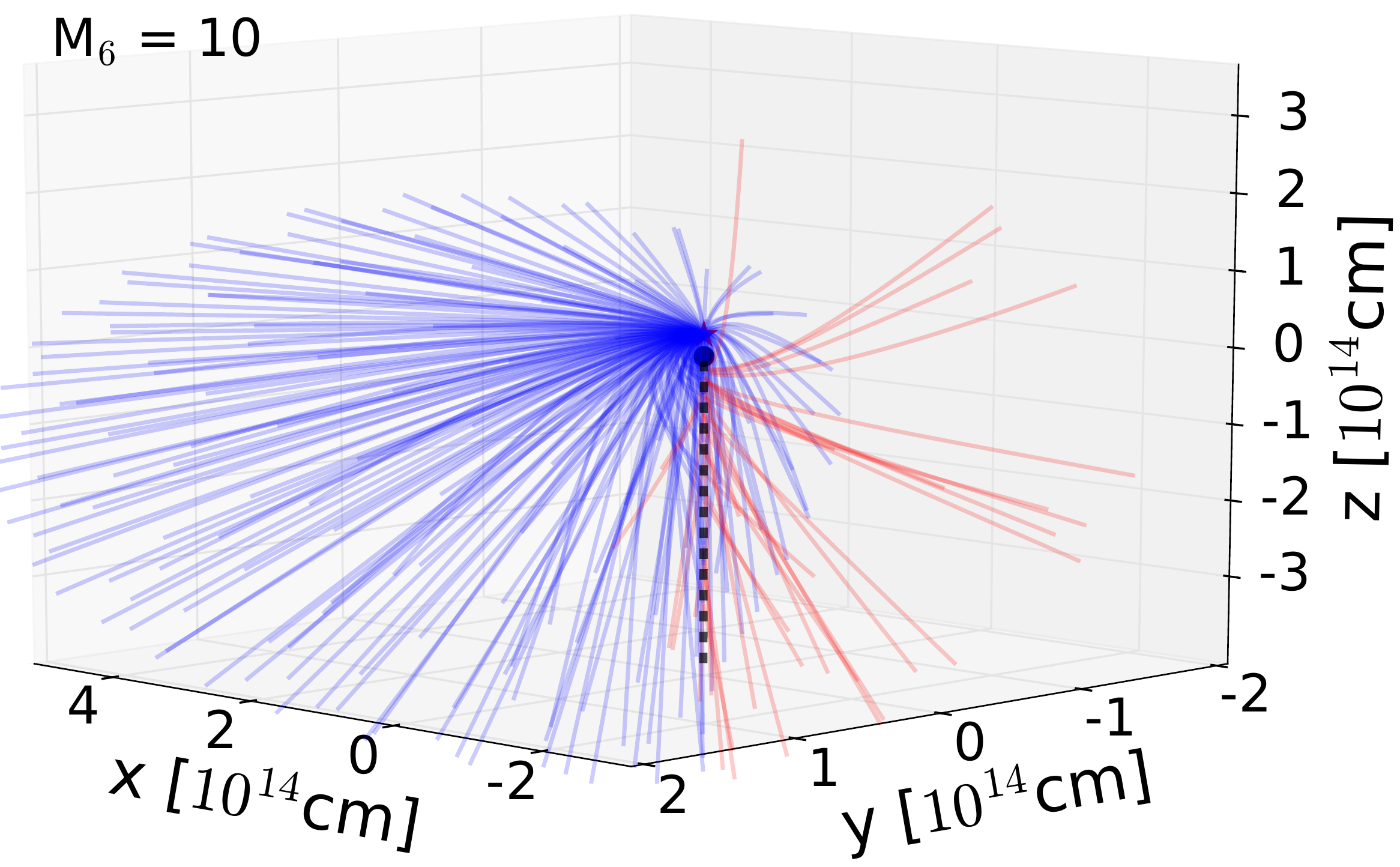}
\caption{The same as Fig. \ref{fig:CIOtraj1} but for
  $M_6=5$ (upper panel), $M_6=10$ (lower panel). The
  integration time (after the stream intersection) for the $M_6=5$ case is
  $2.2\,$d and for the $M_6=10$ case it is $2.5\,$d. Blue curves show
  the free-fall trajectories without or before collisions and red
  curves show the (inaccurate) trajectories after the collisions.
}\label{fig:CIOtraj2}
\end{figure}

We discretize the unbound cone uniformly into $\sim$200 beams
and then integrate the geodesics of each beam over longer
timescales. We ignore the internal pressure of the CIO based on strong
adiabatic cooling during the expansion. The CIO morphologies are shown
in Figs. \ref{fig:CIOtraj1} and \ref{fig:CIOtraj2}. We see that, within
a distance of $\sim$$10^{14}\,$cm, the unbound gas expands into
complex morphology and covers a large fraction of the sky viewed from the
BH. A small portion of the unbound gas will collide with each 
other (all along the negative z axis due to BH's gravitational
focusing) and further dissipate their kinetic energy via shocks. For
the low BH mass (but $M>M_{\rm cr}$ or $f_{\rm unb}\gtrsim
20\%$) cases, the self-intersection point is far from the event horizon and
the gas in the unbound cone is ejected mildly above the local escape
speed, so their trajectories are strongly affected by the BH's
gravity (see Fig. \ref{fig:CIOtraj1}). For the high BH
mass cases, the violent shocks at the intersection point launches
unbound gas well above the local escape speed, so their trajectories
are almost a straight line (see Fig. \ref{fig:CIOtraj2}). 

In the Newtonian picture (appropriate at distances $\gg \rg$), the peak
mass fallback rate of the stream can be estimated
\begin{equation}
  \label{eq:21}
  \dot{M}_{\rm fb,max} \simeq (3\mr{\,\msun\,yr^{-1}})\, \eta_{\rm max}^{3/2}
  M_6^{-1/2} m_*^2 r_*^{-3/2},
\end{equation}
where we have assumed a flat mass distribution over specific energy
between $-\eta_{\rm max}\eps_{\rm T}$ and  $\eta_{\rm max}\eps_{\rm
  T}$ after tidal disruption \citep[e.g.][]{1989ApJ...346L..13E,
  2013ApJ...767...25G}. This peak fallback rate lasts for a duration
roughly given by the period of the most bound orbit $P_{\rm min}\simeq
(41\mr{\,d})\, \eta_{\rm max}^{-3/2} M_6^{1/2}m_*^{-1}r_*^{3/2}$.
During this time, the time-averaged mass feeding rate to the self-intersecting point
is $\dot{M}_{\rm max}\simeq \dot{M}_{\rm fb,max}$ (from both colliding
streams). We note that this feeding rate is not constant but modulated
by twice the free-fall timescale $2t_{\rm ff}(\ri) =
2\sqrt{\ri^3/\rg}/c = (3.7\mr{\,d}) (\ri/10^3\rg)^{3/2}M_6$ in the
Newtonian picture, because each segment 
of length $\sim$$2\ri$ will collide with the next segment of the same
length. From Figs. \ref{fig:intersection} and \ref{fig:fate-fraction},
we see that for BH masses $M\gtrsim M_{\rm cr}$, the intersection
radius is much below the apocenter radius 
$\ri\ll \ra$, so the modulation timescale is much less than the
orbital period $2t_{\rm ff}(\ri) \ll P_{\rm min}$. This discrete mass
injection may modulate the optical lightcurve during the early rise
segment but not near the peak, because
the CIO has highly complex structure with a broad velocity
distribution such that the optical flux near the peak is contributed
by multiple shells (via photon diffusion, see \S 5.2.2). On the other
hand, if the inner accretion disk is not blocked by
the large CIO column for some viewing angles, then the X-ray
lightcurve may be strongly affected by the variable mass feeding rate
to the accretion disk, provided that the viscous timescale is
comparable or shorter than $2t_{\rm ff}(\ri)$. We also note that
hydrodynamic interaction between the fallback stream and the accretion
flow may modify the stream's trajectory and cause the modulation to be
non-periodic. 

In the next subsection, we show that the CIO generates the
optical emission from TDEs. We take an order-of-magnitude approach by
assuming that  the mass outflowing rate from the self-intersecting
point to be steady and the unbound gas expands in a roughly spherical
manner at radius $\gg r_{\rm I}$. We ignore the hydrodynamical effects
of the wind/radiation from the inner accretion
disk. Modeling the full radiative hydrodynamics is left for future works.

\subsubsection{Optical emission from TDEs}
Optically bright TDEs came as a surprise because the
radiation from the inner disk has characteristic temperature
\begin{equation}
  \label{eq:25}
  \left(\lambda_{\rm Edd}L_{\rm Edd}\over 4\pi \rt^2 \sigma_{\rm
      SB}\right)^{1\over 4} \simeq 2.6\times10^{5}\mr{\,K} {\lambda_{\rm
      Edd}^{1/4} M_6^{1/12} m_*^{1/6} \over r_*^{1/2}},
\end{equation}
where $\sigma_{\rm SB}$ is the Stefan-Boltzmann constant and
$\lambda_{\rm Edd}$ is the Eddington factor given by the disk
bolometric luminosity over the Eddington luminosity $L_{\rm Edd}\simeq
1.5\times 10^{44}\,M_6\rm\,erg\,s^{-1}$ (for solar metallicity). TDEs
selected as UV-optical transients have photospheric radii
$\sim$$10^{14}${--}$10^{15}\,\mr{cm}\gg \rt$ and color temperatures
$\sim$a few$\times10^4\,\mr{K}$ much less than that given by
eq. (\ref{eq:25}). In the following, we show that the CIO
naturally provides the long-sought ``reprocessing layer'' which
absorbs the higher frequency radiation from the 
inner disk and re-emits at lower frequencies
\citep{1997ApJ...489..573L, 2014ApJ...783...23G,
  2016MNRAS.461..948M}.

We study the temperature structure of the CIO by
assuming a steady-state spherically symmetric structure heated from
the bottom at radius $r_{\rm in}\sim\,$a few$\times\ri$. We assume the
\textit{received} heating power to be $L_{\rm in}$, 
which could be in the form of hard emission or
wind from the accretion flow\footnote{The evolution of
  the EUV and soft X-ray luminosity from the inner accretion disk
  and its wind kinetic power on timescale $\lesssim 1$yr is still
  uncertain due to our limited 
  understanding of multi-dimensional super-Eddington accretion flow,
  analytically \citep{1979MNRAS.187..237B, 
    1994ApJ...428L..13N, 2004MNRAS.349...68B} or numerically
  \citep{2014MNRAS.439..503S, 2014MNRAS.441.3177M,
    2017arXiv170902845J}. We remain agnostic about the
  heating source's nature and make the (highly simplied) assumption that
  the velocity and density profiles of the CIO are not strongly
  modified by the energy injection. This assumption breaks down when
  the energy injection significantly accelerates the CIO, which should
be studied in future works.}.

When the CIO reaches distances $\gg \ri$, for a crude estimate, we
assume the density and velocity  
distributions of the outflowing gas to be roughly uniform within a
cone of solid angle $\Omega$. Then, the density profile is given by
\begin{equation}
  \label{eq:22}
  \begin{split}
    \rho(r) &\simeq {\dot{M}_{\rm max} f_{\rm unb} \over \Omega
  r^2 v} \simeq 1.5\times10^{-12}\mr{\,g\,cm^{-3}} {K \over
  r_{14}^2}, \\
K &\equiv {2\pi\over \Omega} {f_{\rm unb}\over 0.5} {\eta_{\rm max}^{3/2}
m_*^2\over M_6^{1/2} r_*^{3/2} v_9},
  \end{split}
\end{equation}
where $f_{\rm unb}$ is the unbound fraction
(Fig. \ref{fig:fate-fraction}), $v =
10^9v_9\rm\,cm\, s^{-1}$ is the mass-weighted
mean velocity (see the third panel of Fig. \ref{fig:Ecio}), and $K$ is
the dimensionless ``wind constant'' which depends on many
parameters.

The photon-trapping radius $\rtr$, where photon diffusion time equals
to the dynamical expansion time, is given by the scattering optical
depth $\tau_{\rm s}\simeq \kappa_{\rm s}\rho(\rtr) \rtr \simeq c/v$, i.e.
\begin{equation}
  \label{eq:23}
      \rtr \simeq 1.7\times10^{14}Kv_9\mr{\,cm},
\end{equation}
where $\kappa_{\rm s} = 0.34\rm\,cm^2\, g^{-1}$ is the Thomson
scattering opacity for solar metallicity. Here we have assumed that
the Rosseland-mean opacity roughly equals to the scattering
opacity. The scattering photospheric radius $r_{\rm scat} = 5.1\times
10^{15}\rm\,cm$ is typically not important in determining the optical
appearance of a TDE.

In the radius range $\rin<r <r_{\rm tr}$, photons are 
advected by the expanding wind and the radiation energy density
evolves as $U(r)\propto \rho^{4/3}\propto r^{-8/3}$
\citep{2009MNRAS.400.2070S}. Above the radius $\rtr$, photons
rapidly diffuse away from the local fluid. Since the diffusive 
flux is given by $F_{\rm diff}\simeq Uc/\tau_{\rm s}\propto r^{-2}$,
we have $U(r)\propto r^{-3}$ for $r>r_{\rm tr}$. The
normalization for the above scalings for radiation energy density is
given by $L_{\rm in} = 4\pi r_{\rm in}^2 U(\rin) v$, which means
\begin{equation}
  U(r) = 8.0\times10^5 \mr{\,erg\,cm^{-3}} {L_{\rm
    in,44} r_{\rm in,14}^{2/3} \over r_{14}^{8/3} v_9} \mr{min}\left[1,
\left(\rtr\over r\right)^{1/3} \right].
\end{equation}
We assume that the radiation is well
thermalized near $\rin$, so the radiation spectrum is nearly a
blackbody up to $\rtr$ and the radiation temperature profile is
\begin{equation}
  T(r) \simeq 1.0\times10^5\,\mr{K}\, {L_{\rm
    in,44}^{1/4} r_{\rm in,14}^{1/6}\over r_{14}^{2/3} v_9^{1/4}}\ \
(\mr{for}\ r<\rtr).
\end{equation}
At larger radii $r>\rtr$, the temperature profile depends on whether
the majority of photons get thermalized due to a combination of
scattering and absorption. In the following, we describe a
semi-analytical way of capturing the effect of frequency-dependent
thermalization.

At each radius, we define a blackbody temperature $T_{\rm BB}\equiv
(U/a)^{1/4}$, which is the temperature the radiation field would have if
LTE is achieved. Since the emissivity and absorption opacity are
strongly frequency-dependent (due to bound-free edges and lines), it
is difficult to achieve an equilibrium between emission and absorption
\textit{at all frequencies}. Instead, we define a rough LTE criterion
\citep[see][]{2010ApJ...725..904N} which is applicable at $r>\rtr$,
\begin{equation}
  \eta(r)\equiv {U(r)c/4\pi \over \int \d \nu \,
    \mr{min}\left[B_{\nu}(T_{\rm BB}), j_\nu(T_{\rm BB}) \, ct_{\rm diff}
    \right]}, 
\end{equation}
where $B_{\nu}(T_{\rm BB})$ is the Planck function at temperature
$T_{\rm BB}$, $j_{\nu}(T_{\rm BB})$ is obtained from
Cloudy\footnote{Version 17.01 of the code last described 
    \citet{2017RMxAA..53..385F}.} by
assuming the gas is under a thermal radiation bath of temperature
$T_{\rm BB}$, and the diffusion time is given by $t_{\rm diff} =
\tau_{\rm s} r/c$. Then we dopt a critical value $\eta_{\rm crit} =
5/4$ such that the radiation is considered to be in LTE at radii where
$\eta(r)<\eta_{\rm  crit}$ and non-LTE otherwise. This critical value
means that equilibrium between emission and absorption is achieved at
about 80\% of the frequencies near the peak of the overall spectrum. Thus, the
frequency-averaged thermalization radius $\rth$ is given by
\begin{equation}
\label{eq:rth}
  \eta(\rth) = \eta_{\rm  crit}.
\end{equation}

\begin{figure*}
  \centering
\includegraphics[width = 0.48\textwidth,
  height=0.44\textheight, trim=0.0cm 1.cm 0.0cm
  1.cm]{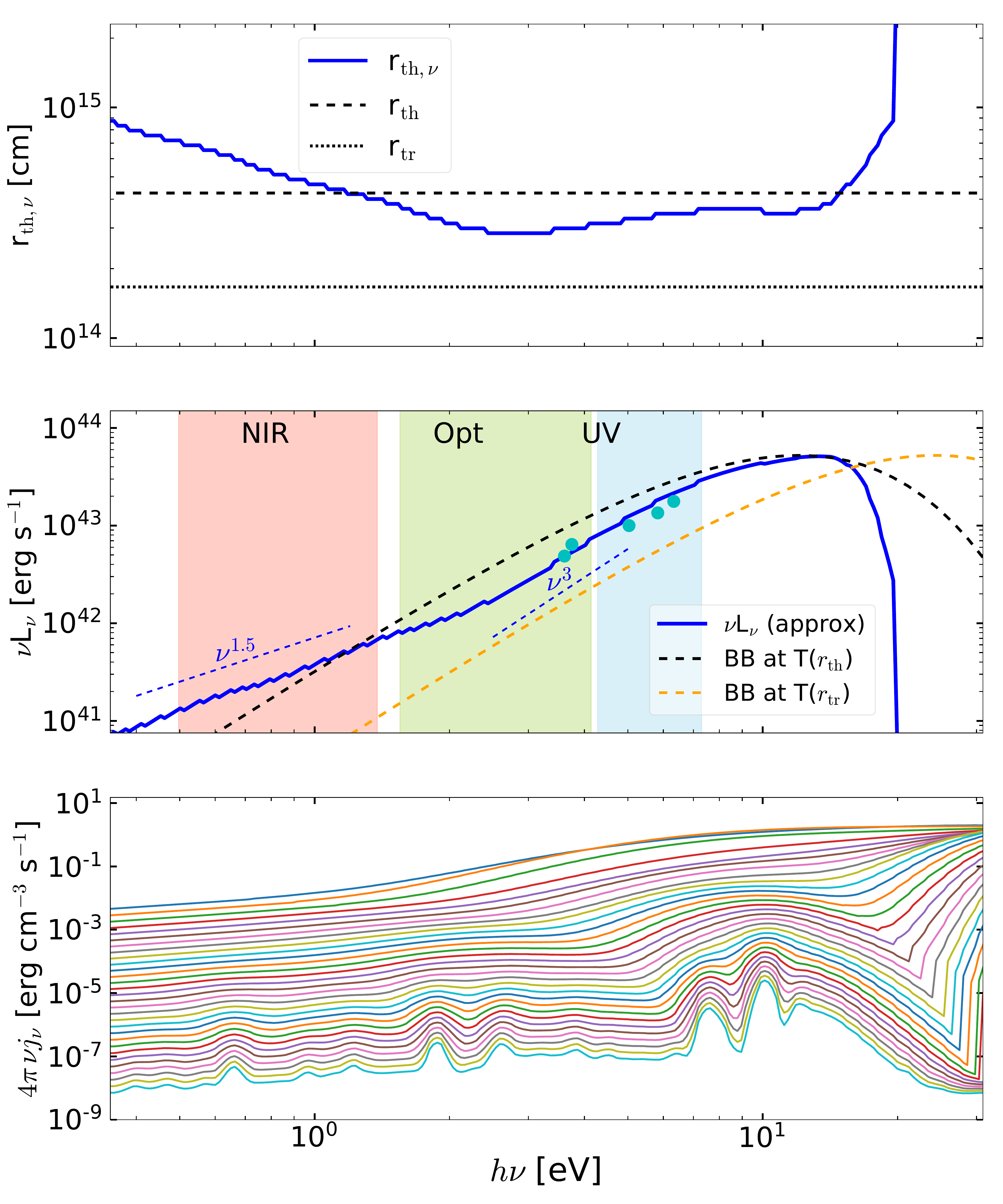} 
\includegraphics[width = 0.48\textwidth,
  height=0.44\textheight, trim=0.0cm 1.cm 0.0cm
  1.cm]{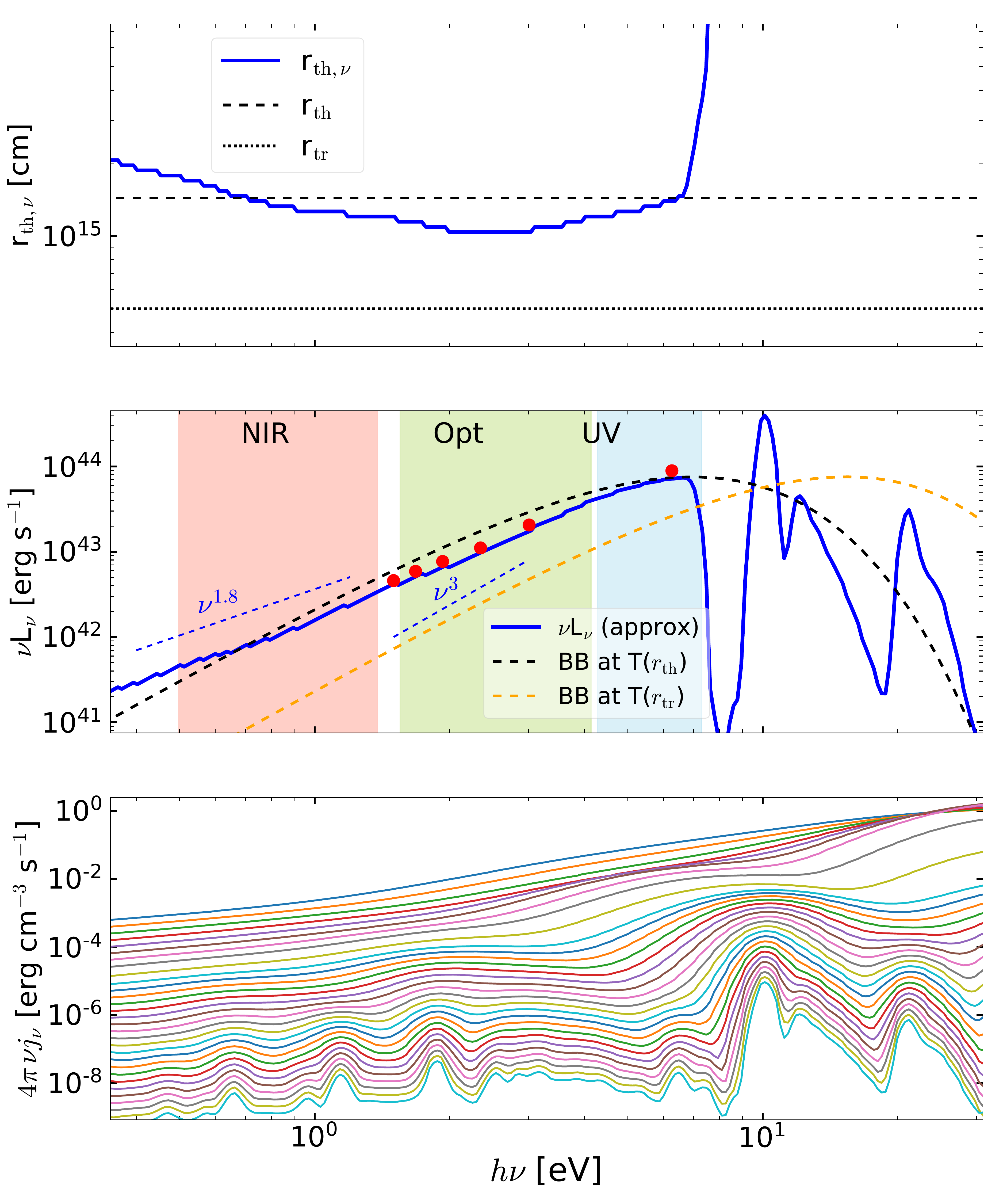} 
\caption{Two possible TDE cases with parameters $(K, v_9, r_{\rm in,14}, L_{\rm
    in,44}) = (1, 1, 1, 1)$ (left panels) and $(3, 1, 1, 3)$ (right
  panels). \textit{Upper Panels}: 
  Frequency-dependent thermalization radius given by
  eq. (\ref{eq:102}). Note that for the highest frequencies $\gtrsim
  20$~eV (left) and $\gtrsim 10$~eV (right), the opacity (due to HeII/HI
  Ly$\alpha$ and their bound-free transition) is so high that $r_{\rm
    th,\nu}$ is beyond our radial grid, so our results are
  unreliable. The trapping radius (eq. \ref{eq:23}) and thermalization
  radius (eq. \ref{eq:rth}) are shown as dotted and dashed
  lines. \textit{Middle Panels}: Spectrum of the escaping photons. We
  mark the three observational windows: UV 
  ($1700$--$2900\rm\, \AA$), Optical ($3000$--$7000\rm\, \AA$), and NIR
  ($0.8$--$2.5\rm\, \mu m$). The shallower behavior $\nu L\nu\propto
  \nu^{1.5}$ (left panel) or $\propto \nu^{1.8}$ (right panel) in the NIR is
  caused by the increasing $r_{\rm th,\nu}$ towards lower frequencies
  (due to free-free opacity, see eq. \ref{eq:rth-nu}), which is a
  robust prediction of our model. For comparison, 
  we show two blackbody (BB) spectra at temperature $T(\rth)$ (black dashed) and 
  $T(\rtr)$ (orange dashed). Data points are the 
  SEDs for two TDEs ASASSN-14li \citep[cyan, left
  panel,][]{2016MNRAS.455.2918H} and PS1-10jh 
  \citep[red, right panel,][]{2012Natur.485..217G} near peak
  luminosity. \textit{Lower Panels}: The 
  (artificially broadened) emissivity at different radii, from $\rtr$
  (uppermost) to $30\rtr$ (lowermost).
}\label{fig:optical-spectrum}
\end{figure*}

Then, the two characteristic radii $\rtr$ and $\rth$ determine the radial profile
of the radiation temperature $T(r)$, which has three power-law
segments: $T\propto r^{-2/3}$ ($\rin<r<\rtr$), $T\propto r^{-3/4}$ ($\rtr < r <
r_{\rm th}$, assuming $\rth < r_{\rm scat}$), and $T=\rm const$
($r>\mr{max}(\rtr, \rth)$). Note that in the case where $\rtr > \rth$,
the middle segment does not exist. The mean photon energy the observer
sees is given by $2.7k_{\rm B}T[\mr{max}(\rtr, \rth)]$. With the radiation
temperature $T(r)$, energy density $U(r)$, and density $\rho(r)$ at
each radius (for a logarithmic radial grid), we use Cloudy to compute
the degree of ionization for each chemical species and their
energy-level population, under Solar abundance.

We make use of the volumetric emissivity $j_\nu(r)$ (for a logarithmic
frequency grid) output from Cloudy. At radius $r>\rtr$, the energy of
photon are still significantly modified by electron scattering. This 
is because the local intensity distribution is anisotropic with an
outwards diffusive flux. This intensity anisotropy means  
that, at a given radius, an electron scatters more outward-going
photons than inward-going ones, and hence photons overall exert a
force on this electron. Since the electron is moving outwards
at velocity $v$, this force due to photon scattering is doing work to 
accelerate the electron (of course, this electron is dynamically
coupled with a proton such that the actual acceleration is small). The
net effect of the photon-electron momentum transfer is that, photons
lose a fraction $\sim$$v/(c\tau)$ of energy over each scattering (see
Appendix C). Since it takes $\tau^2$ scatterings for each photon to
escape, the total amount of energy loss is $\sim$$\tau v/c$. We are interested in
the region at $r>\rtr$ or $\tau<c/v$, so photons lose energy by less
than a factor of 2 (and hence overall adiabatic cooling is not important) but this
energy shift is important for the transport of line photons by
effectively broadening the lines \citep{2000ApJ...530..757P,
  2016ApJ...827....3R}. We take a broadening factor of 
\begin{equation}
  \label{eq:101}
  \sigma = \mr{max}(\tau v/c, v/\sqrt{2}c),
\end{equation}
and perform a Gaussian kernel smoothing over the Cloudy output of
$j_\nu(r)$ at each radius.

Now we have all the ingredients to calculate the specific luminosity
of the escaping photons from the wind. For each frequency $\nu$, the 
thermalization radius $r_{\rm th,\nu}$ is given by the equilibrium
between emission and absorption, i.e.
\begin{equation}
  \label{eq:102}
  j_\nu(r_{\rm th,\nu}) c t_{\rm diff} = B_\nu(T_{\rm BB}(r_{\rm th,\nu})),
\end{equation}
which is equivalent to the effective absorption optical depth
$\tau_{*,\nu}\simeq \sqrt{\tau_{\rm a,\nu} (\tau_{\rm a,\nu} +
  \tau_{\rm s})}\simeq 1$ \citep{1979rpa..book.....R}.
Then, the specific luminosity is roughly given by   
\begin{equation}
  \label{eq:103}
  L_\nu \simeq 4\pi r_{\rm th,\nu}^3 4\pi j_\nu(r_{\rm th,\nu}) \simeq 4\pi r_{\rm th,\nu}^2 {4\pi 
  B_\nu(T_{\rm BB}(r_{\rm th,\nu})) \over \tau_{\rm s}(r_{\rm th,\nu})},
\end{equation}
where $\tau_{\rm s}(r_{\rm th,\nu}) = \kappa_{\rm s} \rho (r_{\rm th,\nu}) 
r_{\rm th,\nu}$. As shown in Fig. \ref{fig:optical-spectrum}, our
model can reproduce the optical and UV spectral-energy distributions
(SEDs) of typical TDE candidates such as ASASSN-14li and
PS1-10jh. One robust prediction of our wind reprocessing model is that
the SED in the NIR band is softer than that in the optical-UV, typically
$L_{\nu}\propto \nu^{\sim 0.5}$. This can be explained as follows.

The absorption opacity in the NIR continuum is dominated by free-free
transitions \citep[ignoring the Gaunt factor,][]{1979rpa..book.....R}
\begin{equation}
  \kappa_{\rm low\, \nu}\simeq (1.1\times10^{17}\mr{\,cm^2\,g^{-1}})\,
  \rho T^{-3/2} (h\nu/\mr{eV})^{-2},
\end{equation}
where the density $\rho$ and temperature $T$ are in units of $\rm
g/cm^3$ and $\rm K$, respectively. In the limit $\kappa_{\rm
  low\,\nu}\ll \kappa_{\rm s}$, the effective opacity is given by
$\kappa_{\rm *,\nu}\simeq \sqrt{\kappa_{\rm low\,\nu}\kappa_{\rm s}}$
\citep{1979rpa..book.....R}, so the frequency-dependent thermalization
radius is given by $\kappa_{\rm *,\nu}\rho r_{\rm th,\nu}\simeq 1$, i.e.,
\begin{equation}
\label{eq:rth-nu}
  r_{\rm th,\nu}\simeq (3.9\times10^{14}\mr{\,cm})\,
  (h\nu/\mr{eV})^{-1/2} K^{3/4}
  T_{4.5}^{-3/8},
\end{equation}
where $T=10^{4.5}T_{4.5}\rm\,K$ is the electron temperature at the
thermalization radius (the final results depends very weakly on
$T$). The above equation agrees reasonably well with
Fig. \ref{fig:optical-spectrum}. 

At frequencies with $r_{\rm th,\nu} < \rtr$, thermalization occurs
below the trapping radius, and the escaping specific luminosity is
given by $L_{\nu} = 4\pi r_{\rm tr}^2 (4\pi)
B_{\nu}(T(\rtr))/\tau_{\rm s}(\rtr)$, which has a blackbody shape
at temperature $T(\rtr)$. However, at frequencies with $r_{\rm 
th,\nu} > \rtr$, thermalization occurs above the trapping radius, and
eq. (\ref{eq:103}) gives
\begin{equation}
\label{eq:low_frequency}
  \begin{split}
      \nu L_\nu \simeq (2.4\times10^{41}\mr{\,erg\,s^{-1}})\,
      (h\nu/\mr{eV})^{3/2} K^{5/4} T_{4.5}^{-1/8},
  \end{split}
\end{equation}
which applies for at low frequencies (such that $r_{\rm
th,\nu} > \rtr$)
\begin{equation}
  h\nu < (5.2\mr{\,eV})\, K^{-1/2} T_{4.5}^{-3/4} v_9^{-2}.
\end{equation}
This behavior $L_{\nu}\propto \nu^{0.5}$ should be observable
in the NIR \citep[see Figs. 4 and 5 of][]{2016ApJ...827....3R}.
This effect is analogous to the radio/infrared free-free
absorption in the wind of Wolf-Rayet stars
\citep{1975MNRAS.170...41W, 2007ARA&A..45..177C}. The weak dependence
on the electron temperature $T^{-1/8}$ means that
eq. (\ref{eq:low_frequency}) can be used to measure the ``wind
parameter'' $K\propto \dot{M}/v$ for 
individual TDEs, similar to measuring the mass-loss rate from Wolf-Rayet
stars\footnote{The CIO is likely clumpy (due to e.g. episodic mass
  ejection), so a further correction for the volume filling factor
  $f_{\rm V}<1$ is needed \citep{1959ApJ...129...26O}.}.

\subsubsection{Other pieces of information --- lines and X-rays}

The observed H$\alpha$ and HeII emission lines have complex and
sometimes double-peaked or boxy structures 
\citep[e.g.][]{2014ApJ...793...38A, 2016MNRAS.455.2918H,
2018arXiv180802890H, 2018arXiv180907446B}. They have been
modeled with the reprocessed emission from an elliptical disk
\citep{2017MNRAS.472L..99L, 2018arXiv180802890H}. However, these
elliptical disks may be highly unstable on timescales $\sim$months
because each annulus undergoes apsidal precession at a different
rate. In our picture, the emission line profiles are mainly controlled 
by the bulk motion of the line formation region of the CIO (at a few
times the trapping radius $\rtr$), which can either be blue- or red-shifted
depending on the observer's line of 
sight. We also note that, if the line formation region has large
scattering optical depth, the line profile may be further modified by
Comptonization \citep{2018ApJ...855...54R}. The N III and O III
emission lines in some TDEs, e.g. AT2018dyb
\citep{2019arXiv190303120L}, are probably due to Bowen fluorescence,
which requires a large flux of (unseen) EUV photons. 

The partial sky coverage of the CIO provides a unification of
the diverse X-ray properties of optically selected TDEs.
When the line of sight to the inner accretion disk is not blocked by
the CIO, the observer should 
see optical emission as well as the EUV or soft X-ray  emission 
from the inner accretion disk or its wind \citep{2009MNRAS.400.2070S, 
  2018ApJ...859L..20D, 2019MNRAS.483..565C}. When the line of
sight is only blocked by the 
region of the CIO with modest optical depth, the observer may see
blueshifted absorption lines from high ionization species
\citep[e.g.][]{2018MNRAS.473.1130B, 2018arXiv180907446B}. When the
line of sight is blocked by the highly optically thick region of CIO,
the observer only sees optical emission initially. Then, as the
CIO's mass outflowing rate drops with time, the trapping radius
shrinks and hence the EUV and soft X-ray photons from the inner
disk suffer less adiabatic loss. As a result, the soft X-ray
flux (on the Wien tail) should gradually rise and the spectrum hardens
on timescales of $\sim1\,$yr \citep{2017ApJ...836...25M,
  2017ApJ...851L..47G, 2018MNRAS.480.5689H}.

\subsubsection{Radio emission from non-jetted TDEs}
In this subsection, we discuss the radio emission from the adiabatic
shock driven by the CIO into the circum-nuclear medium (CNM). As shown
in Fig. \ref{fig:Ecio}, the CIO has kinetic energies from $E_{\rm
  k}\sim$$10^{50}\rm\,erg$ up to 
$\sim$$10^{52}\rm\,erg$ and mean speed between $v_0\sim$$0.01c$ and
$\sim$$0.1c$. In the following, we simplify the complex CIO structure
as a thin shell covering a solid angle $\Omega$ within which the density
and velocity distributions are uniform. We assume that the ambient
medium has a power-law density profile in the radial direction $n = n_{\rm
  pc}r_{\mr{pc}}^{-k}$ ($k<3$), where $r_{\rm pc} = r/\mr{pc}$. We
also ignore sideway expansion of the shocked region since
$\Omega\sim 2\pi$, so the system is one dimensional.

When the CIO reaches a radius $r$, the total number of shocked
electrons from the CNM is given by  
\begin{equation}
  \label{eq:26}
  N(r) = \int^r \Omega r^2 n(r) \d r = {\Omega \over 3-k
    } N_{\rm pc}r_{\rm pc}^{3-k},
\end{equation}
where $N_{\rm pc} \equiv n_{\rm pc}\times (1\mr{\,pc})^3$ is a
reference number of electrons. We ignore the
acceleration of particles by the reverse shock (driven 
into the ejecta) because it is much weaker than the forward shock
(driven into the CNM). The deceleration radius $\rdec$
is given by $E_{\rm k} = (1/2) N(\rdec) \mp v_0^2$ ($\mp$ being proton
mass), which means 
\begin{equation}
  \label{eq:27}
  r_{\rm dec,pc}^{3-k} = {3-k \over \Omega} {2E_{\rm k}\over N_{\rm pc}
  \mp v_0^2}.
\end{equation}
We smoothly connect the free-expansion phase with the Sedov-Taylor
phase by using the following velocity profile
\begin{equation}
  \label{eq:28}
  v(r) = v_0\,\mr{min}\left[1, (r/\rdec)^{(k-3)/2}\right],
\end{equation}
and hence the shock reaches radius $r$ at time
\begin{equation}
  \label{eq:33}
  t(r) = {\rdec \over v_0}\, \mr{min} \left[ {r\over \rdec},
{2\over 5-k} \left(r \over \rdec\right)^{5-k\over 2} + {3-k \over 5-k}\right].
\end{equation}
The electron number density in the shocked region is $4n(r)$ and the
mean energy per proton is $(1/2)\mp v(r)^2$, so the thermal energy density
is $2n(r)\mp v(r)^2$. We assume that a fraction $\epsB\ll 1$ of the
thermal energy is shared by magnetic fields, so the magnetic field
strength is
\begin{equation}
  \label{eq:29}
  B(r) = \left[16\pi\, \epsB n(r) \mp v(r)^2\right]^{1/2}.
\end{equation}

We assume that electrons share a fraction $\epse\ll1$ of the thermal
energy and that they are accelerated to a power-law momentum
distribution with index $p$. We expect particle acceleration from
non-relativistic shocks to give $2< p<3$, both 
theoretically \citep{1978MNRAS.182..147B, 1987PhR...154....1B,
2001RPPh...64..429M, 2015PhRvL.114h5003P, 2014ApJ...783...91C} and
observationally \citep{1998ApJ...499..810C, 2014BASI...42...47G,
  2014ApJ...796...82Z}. For fast shocks where the mean energy per
electron $\epse \mp v^2/2\gg \me c^2$ ($\me$ being electron mass),
the majority of the particle number and kinetic energy are both
concentrated near a relativistic minimum momentum $\gg \me c$. For 
slow shocks where $\epse \mp v^2/2\lesssim \me c^2$, most
particles have non-relativistic momenta but the majority of kinetic
energy is in mildly relativistic particles with Lorentz factor
$\gamma\sim 2$. We are interested in the number density of
ultra-relativistic electrons. These two regimes above can be smoothly
connected by assuming a power-law Lorentz factor distribution
$\d N/\d \gamma \propto \gamma^{-p}$ above the minimum Lorentz factor
\citep{2006ApJ...638..391G, 2013ApJ...778..107S} 
\begin{equation}
  \label{eq:31}
  \gamma_{\rm m} =\mr{max}\left[2, {p-2 \over p-1} {\epse \mp v(r)^2
      \over 2 \me c^2} \right].
\end{equation}
Then the normalization is given by the total energy of these
relativistic electrons being $\epse N(r)\mp v(r)^2/2$, i.e.,
\begin{equation}
  \label{eq:32}
  \d N/\d\gamma = {\epse \mp v(r)^2 \over 2 \me c^2}
{(p-2) N(r) \over \gamma_{\rm m}^2} (\gamma/\gamma_{\rm
    m})^{-p}. 
\end{equation}

An electron of Lorentz factor $\gamma\gg 1$ has characteristic synchrotron
frequency
\begin{equation}
  \label{eq:34}
  \nu(\gamma) = {3\over 4\pi} {\gamma^2 eB\over \me c},
\end{equation}
where $e$ is the electron charge. Since the peak specific power is
$P_{\nu,\rm max}\simeq e^3B/\me c^2$,
the specific luminosity at frequency $\nu$ in the optically thin
regime is given by 
\begin{equation}
  \label{eq:35}
  L_\nu \simeq \gamma {\d N\over \d\gamma} {e^3 B \over \me c^2}.
\end{equation}
The synchrotron self-absorption frequency $\nu_{\rm a}$ and the
corresponding Lorentz factor $\gamma_{\rm a}$ are defined where the 
optical depth $\alpha_{\nu_{\rm a}} \Delta\ell_{\rm r}\sim 1$
($\Delta\ell_{\rm r}$ being the radial thickness of the emitting
region). The volumetric emissivity at $\nu_{\rm a}$ is given by
$j_{\nu_{\rm a}} = \alpha_{\nu_{\rm a}} 2kT \nu_{\rm a}^2/c^2$ (in the
Rayleigh-Jeans limit $h\nu_{\rm a}\ll kT$), where $kT \simeq 
\gamma_{\rm a}\me c^2$ is the temperature of electrons responsible for
absorption. Assuming $\gamma_{\rm a}>\gamma_{\rm m}$ (which will later
be shown to be true for non-relativistic shocks), we can write the
specific luminosity as $4\pi j_{\nu_{\rm a}}\Omega r^2\Delta \ell_{\rm r}$,
and hence
\begin{equation}
  \label{eq:36}
  L_{\nu_{\rm a}} = \left. L_{\nu} \right|_{\nu_{\rm a}}\simeq 4\pi \Omega
  r^2 {2kT \nu_{\rm a}^2 \over c^2}.
\end{equation}

\begin{figure}
  \centering
\includegraphics[width = 0.42\textwidth,
  height=0.45\textheight, trim=0.0cm 0.2cm 0.0cm
  .5cm]{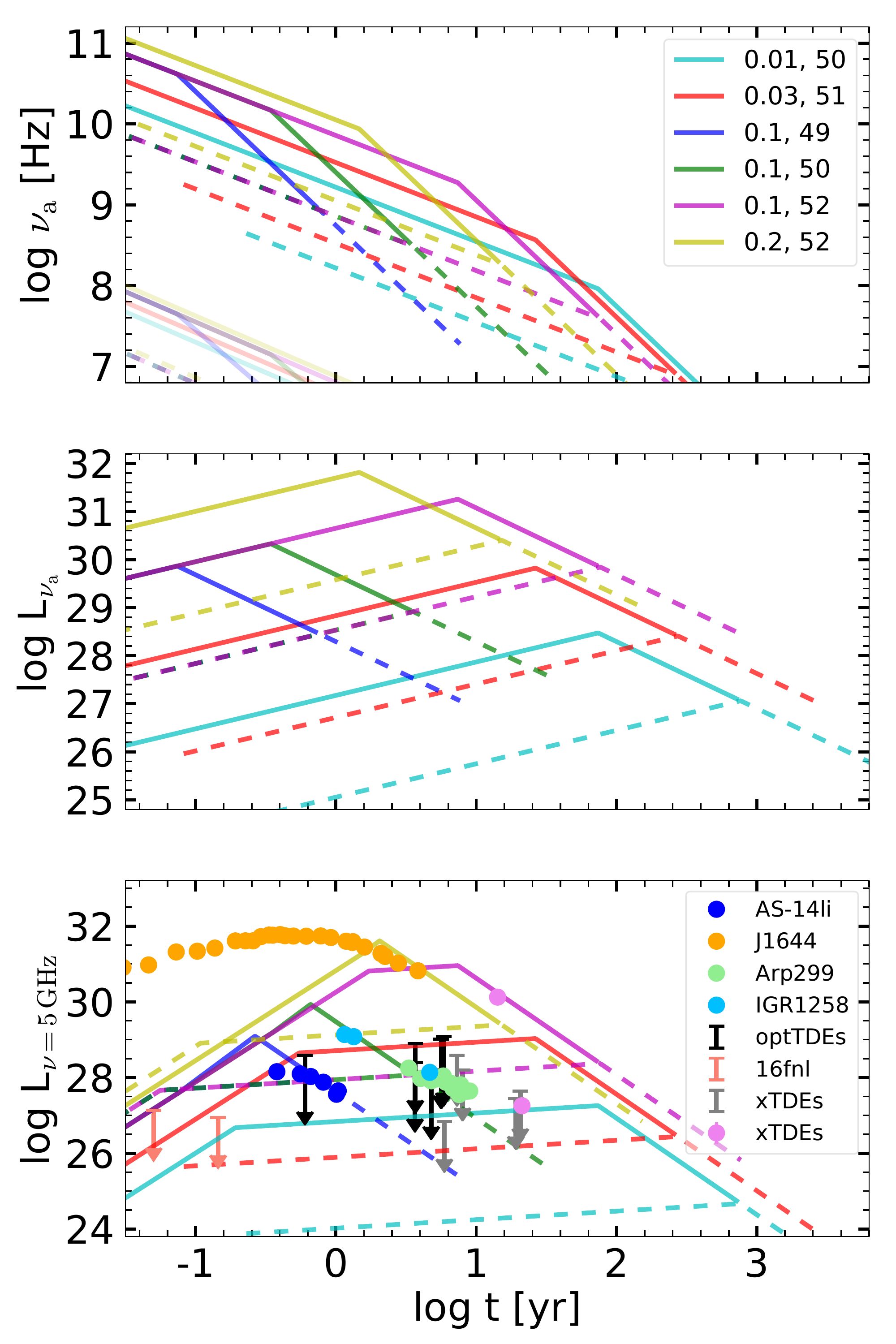}
\caption[caption]{Radio emission from the CIO interacting with the
  CNM for the case of $k$=1.5, $p$=2.4, $\epse$=0.1, $\epsB$=0.01, and
  $\Omega$=2$\pi$. The solid and dotted curves are for $n_{\rm
  pc}$=100 and $3\rm\,cm^{-3}$. \textit{Upper panel}: The
evolution of the synchrotron self-absorption frequency $\nu_{\rm a}$
as a function of time for 
  different cases with initial velocities $v_0/c=0.01,\ 0.03,\ 0.1,\
  0.2$ and kinetic energies $\mr{log}\,E_{\rm k}\mr{/erg} = 49,\
  50,\ 51,\ 52$. The fainter curves show $\nu(\gamma_{\rm m})$ which
  is always much smaller than $\nu_{\rm a}$ (and in some cases always below
  $10^6\,$Hz). \textit{Middle panel}: The evolution of the peak 
  specific luminosity (in $\rm erg\,s^{-1}\, Hz^{-1}$) with time for
  the cases indicated in the upper panel.
\textit{Lower panel}: The
  lightcurve at $5\,$GHz for the cases indicated in the upper
  panel. We also show the measured fluxes or upper limits at $5\,$GHz
  for several TDEs in the lower panel, including ASASSN-14li
  \citep[blue circles,][]{2016ApJ...819L..25A}, Swift J1644+57
  \citep[gold circles,][]{2013ApJ...767..152Z, 2018ApJ...854...86E}, Arp299
  \citep[green circles,][]{2018Sci...361..482M}, IGR1258 \citep[light
  blue circles,][]{2015ApJ...809..172I, 2017ApJ...842..126P}, iPTF16fnl
  \citep[red upper limits,][]{2017ApJ...844...46B},
  other optical selected TDEs \citep[black upper
  limits,][]{2013A&A...552A...5V}, X-ray selected
  TDEs \citep[light blue circles and grey upper
  limits,][]{2013ApJ...763...84B}. 
}\label{fig:radio}
\end{figure}

Combining eqs. (\ref{eq:35}) and 
(\ref{eq:36}), we see that the Lorentz factor $\gamma_{\rm a}$
corresponding to the self-absorption frequency is determined by
\begin{equation}
  \label{eq:37}
\left.  {\d N\over \d \gamma}\right|_{\gamma_{\rm a}} \simeq {9\Omega
  r^2 B\over 2\pi e} \gamma_{\rm a}^4,
\end{equation}
which gives
\begin{equation}
  \label{eq:38}
      \gamma_{\rm a}^{4+p} \simeq {2\pi (p-2) \over 9 (3-k)} {e
        \mc{N}_{\rm pc} \over B} {\epse \mp v(r)^2 \over 2 \me c^2}
      r_{\rm pc}^{1-k} \gamma_{\rm m}^{p-2}, 
\end{equation}
where we have defined a reference column density $\mc{N}_{\rm pc} =
n_{\rm pc} \times 1\,$pc. If $\nu (\gamma_{\rm m})\ll \nu_{\rm a}$ and
synchrotron/inverse-Compton cooling are negligible, the synchrotron
spectrum when the shock is at radius $r$ is given by
\citep{2002ApJ...568..820G}
\begin{equation}
  \label{eq:39}
  L_{\nu} \simeq 
  \begin{cases}
   L_{\nu_{\rm a}} (\nu/\nu_{\rm a})^{5/2},\ &\mr{for}\
   \nu(\gamma_{\rm m}) <\nu < \nu_{\rm
      a},\\
     L_{\nu_{\rm a}} (\nu/\nu_{\rm a})^{(1-p)/2} ,\ &\mr{for}\ \nu > \nu_{\rm
      a}.
  \end{cases}
\end{equation}
In Fig. \ref{fig:radio}, we show the radio emission from CIO colliding
with the CNM for a number of cases. We denote the average velocity
$v_0$ in units of $c$ and the kinetic energy (with unit $\rm erg$) in
log-scale.  The three cases with $(v_0, E_{\rm k}) = (0.01, 50),\
(0.03, 51),\ (0.1, 52)$ are motivated by the mean velocities and
kinetic energies in  Fig. \ref{fig:Ecio}. The case with $(v_0, E_{\rm
  k}) = (0.2, 52)$ is for comparison with that with $(v_0,
E_{\rm  k}) = (0.1, 52)$, and we see that an outflow
with higher velocity generates brighter and earlier-peaked radio
emission. The cases with $(v_0, E_{\rm
  k}) = (0.1, 49)$ and (0.1, 50) are motivated by the fact
that the CIO velocity profile is non-uniform with a fraction of the
mass moving faster than the mean velocity. We find that the faster
portion of the ejecta generates bright radio emission at early
time. For each combination of $(v_0, E_{\rm
  k})$, we take two different CNM density normalizations
$n_{\rm pc} = 100$ and $3\rm\,cm^{-3}$. As expected, we find that, for
higher CNM densities, the radio emission is brighter and peaks earlier.

We also show the data from several TDEs for comparison but do 
not intend to search for the best-fit parameters for individual
cases. The upper limits for iPTF16fnl \citep{2017ApJ...844...46B}
reported at $15\,$GHz have been scaled by a factor of
$(5/15)^{(1-p)/2}$ (assuming $\nu_{\rm a}>5\,$GHz). The upper limits 
for the X-ray selected TDEs reported at $3\,$GHz
by \citet{{2013ApJ...763...84B}} are not scaled. 

Even though we keep the following parameters fixed $k$=1.5,
$p$=2.4, $\epse$=0.1, $\epsB$=0.01, and $\Omega$=2$\pi$, the radio
luminosity and duration are extremely diverse. Generally, we expect
TDEs with CIO to have some radio emission at the level of ASASSN-14li
lasting for years up to centuries. We also note that radio emission
from the jetted TDE Swift J1644+57 \citep{2011Sci...333..203B,
  2011Natur.476..421B, 2011Natur.476..425Z} is much brighter (and
peaks earlier) than that from the CIO, because this source was powered
by a relativistic jet pointing towards the observer. For off-axis jetted
TDEs, the radio emission due to the CIO may be mistaken as a
signature of jets (a possible way of distinguishing between them is to
resolve the motion of 
the radio emitting region by long-baseline interferometry).

Another possible source of wide-angle outflow is the wind expected from
super-Eddington accretion in TDEs with BH masses $M\lesssim 10^7\msun$ 
\citep{2009MNRAS.400.2070S, 2014MNRAS.439..503S,
  2017arXiv170902845J}. In fact, the super-Eddington wind may be more
powerful than the CIO, because the energy efficiency of the CIO is
only $\sim$$\rg/\ri\sim0.001$ to 
0.01. Thus, we expect TDEs with strong super-Eddington wind to
generate radio emission comparable to or even brighter than that in
our $(v_0, E_{\rm k}) = (0.2, 52)$ case. Late-time radio
observations can potentially test whether super-Eddington
accretion flows generate jets or winds. We note that the unbound tidal
debris typicall has very small solid angle
\citep{2014ApJ...783...23G}, so its radio emission (and 
reprocessing of the high-energy photons from the disk) is much weaker
than that of the CIO. It is less likely that the radio emission
from ASASSN-14li is caused by the unbound tidal debris
\citep{2016ApJ...827..127K}, unless the star was in a very deeply
penetrating $\beta\gg 1$ orbit \citep{2019arXiv190302575Y}.

Finally, we note that the CIO may interact with a pre-existing
accretion disk (or the broad line region), if the BH was active before
the TDE. If the accretion disk gas is sufficiently dense, the shocks
become radiative and bright optical emission like in PS16dtm
\citep{2017ApJ...843..106B} may be generated.

\subsection{TDE demographics}
TDE demographics, in terms
of the total TDE rate as a function of BH mass and properties of the
disrupted star, has been considered by \citet{2016MNRAS.455..859S} and
\citet{2016MNRAS.461..371K}.
In this section, we focus on the rate of 
optically bright TDEs only, based on the picture that the CIO reprocesses the disk
emission from the EUV into the optical band. 

We differentiate the TDE 
rate with three parameters, stellar mass $m_* = M_*/\msun$,
impact parameter $\beta$, and BH mass $M$, in the following way
\begin{equation}
  \label{eq:40}
  {\d \dot{n} \over \d m_*\, \d \beta\, \d \mr{log}M} = \mc{R} M_6^{\alpha}
  m_*^{-1/12} r_*^{1/4} {\d n_*\over \d m_*} {\d P \over \d \beta} {\d
  n_{\rm BH}\over \d \mr{log}M},
\end{equation}
where $\mc{R}$ is the normalization rate per BH in units of
$\mr{yr}^{-1}$, the normalized stellar mass function satisfies $\int
(\d n_*/\d m_*) \d m_* = 1$, the probability distribution of
the impact parameter has
also been normalized $\int (\d P / \d \beta)\d \beta = 1$, and the
BH mass function (BHMF) $\d n_{\rm BH}/\d \mr{log}M$ has unit $\rm
[Mpc^{-3}\, dex^{-1}]$. The factor $m_*^{-1/12} r_*^{1/4}$ is
because stars with a larger tidal radius are slightly preferred roughly by a
factor of $\rt^{1/4}$ \citep{2012ApJ...757..134M}.

The power-law dependence on  
the BH mass $M^{\alpha}$ depends, in a non-trivial way, on the stellar
density and velocity 
profiles near the center of individual galaxies. The index $\alpha$
is empirically derived by combining the surface brightness profiles
of a sample of galaxies with BH masses inferred from galaxy scaling
relations \citep[e.g.][]{1999MNRAS.309..447M, 2004ApJ...600..149W, 
2016MNRAS.455..859S}. There is a core-cusp bimodal distribution of
central surface brightness profiles of early-type galaxies used for TDE rate
calculations \citep{2007ApJ...664..226L}. The most recent work by
\citet{2016MNRAS.455..859S} gives $\alpha\simeq -0.25$ for samples of
only\footnote{The TDE rates for cusp galaxies are typically $\sim$10
  times higher than that for core galaxies of the same BH mass.} cusp
or core galaxies. For comparison, we also show the results for
$\alpha\simeq -0.5$ which do not affect our conclusions
qualitatively. We caution that the above 
studies typically assume a spherically symmetric and time-independent
galactic potential, nearly isotropic stellar velocity distribution
(except for the loss cone), and the refilling of the loss cone by
two-body relaxation only. Other factors, such as massive perturbers,
aspherical potential, binary BHs, resonant relaxation, may strongly
affect the estimated TDE rate \citep[e.g.][]{2013ApJ...774...87V,
  2013CQGra..30x4005M}. Therefore, we leave the  
normalization factor $\mc{R}$ as a free parameter, which roughly means
the (per-BH) rate of TDEs for M-dwarf stars disrupted by
$\sim$$10^6\msun$ BHs. 

In loss-cone dynamics, the probability distribution for the impact parameter $\d
P/\d\beta$ has two regimes. In the ``pinhole'' regime (far from
the BH), the change in 
stars' angular momentum per orbit $\Delta \ell$ is much larger than
the size of the loss-cone $\ell_{\rm lc}\simeq
\sqrt{2\rg \rt}$, so $\d P/\d\beta$ simply depends on the ``area''
of the loss cone per unit change in $\beta$,
i.e. $\d P/\d\beta\propto \beta^{-2}$. In the ``diffusive'' regime (near
the BH), $\Delta \ell\ll \ell_{\rm lc}$ and hence stars are always disrupted
near the boundary of the loss-cone with minimum penetration depth,
i.e. $\d P/\d\beta$ is nearly a $\delta$-function. The fraction of
TDEs in the pinhole regime $f_{\rm pin}$ depends on the detailed
stellar density profile near the BH and has large uncertainty at
each BH mass. Following \citet{2016MNRAS.461..371K}, we take 
\begin{equation}
  \label{eq:41}
  f_{\rm pin} \simeq \left(1 + M_7^{1/2}\right)^{-1},
\end{equation}
which is very similar to the fitting result by
\citet{2016MNRAS.455..859S} in the range of BH masses of 
interest. Then the probability distribution of 
$\beta$ is given by
\begin{equation}
  \label{eq:42}
    {\d P\over \d \beta} \simeq
  \begin{cases}
f_{\rm pin} \beta^{-2}\beta_{\rm
      min},\ &\mr{for}\ \beta > \beta_{\rm  min}\\
  (1-f_{\rm pin})\,\delta (\beta - \beta_{\rm min}),\ &\mr{for}\ \beta
  \approx \beta_{\rm  min}.
  \end{cases}
\end{equation}
According to eq. (\ref{eq:15}), the minimum impact parameter is $\beta_{\rm
  min} \simeq 0.6\,\xi_*^{-1}$, which includes relativistic tidal forces for the
Schwarzschild spacetime \citep{2012PhRvD..85b4037K} and $\xi_*$
depends on the star's internal structure. We note that $\xi_*$ is not
well measured in general relativity even for polytropic
stars. Hydrodynamic simulations of disruptions with
polytropic or realistic stellar structures in the
Newtonian limit \citep[$\rg\ll \rp$,][]{2013ApJ...767...25G,
  2017A&A...600A.124M, 2019arXiv190208202G} show that the star 
loses about half of the mass when $\xi_*\simeq 0.5$ (for
polytropic index 4/3) or $\xi_*\simeq 1.0$ (for polytropic index
5/3). The former is appropriate for radiative stars with $m_*>1.2$ and
the latter is good for convective stars with $m_*<0.3$ \citep[see a
similar treatment by][]{Phinney89}. For stars in between
$0.3<m_*<1.2$, we take a linear interpolation in log$\,m_*$
space. Thus,
\begin{equation}
  \beta_{\rm min}\simeq
  \begin{cases}
    0.6,\ & \mr{if}\ m_* < 0.3,\\
    0.6 + \mr{log}\,(m_*/0.3),\ & \mr{if}\ 0.3 < m_* < 1.2,\\
    1.2,\ &  \mr{if}\ m_* > 1.2.
  \end{cases}
\end{equation}
We also note that the maximum impact parameter is taken to be
infinity, because a star's orbit can have arbitrarily low angular
momentum. The effect of stars being swallowed by the event horizon
will be taken into account later when integrating over the BHMF.

\begin{figure}
  \centering
\includegraphics[width = 0.48\textwidth,
  height=0.25\textheight]{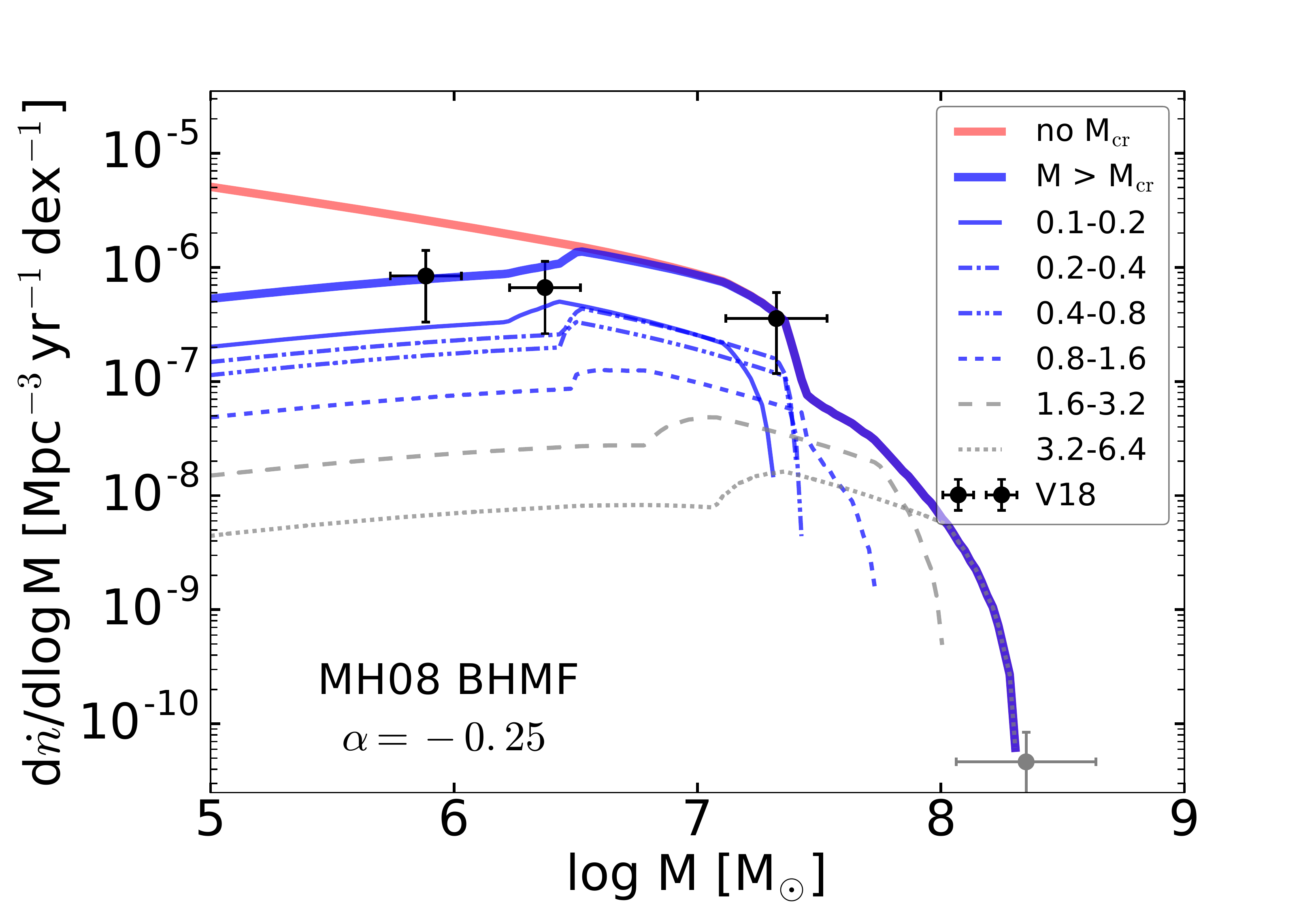}
\includegraphics[width = 0.48\textwidth,
  height=0.25\textheight]{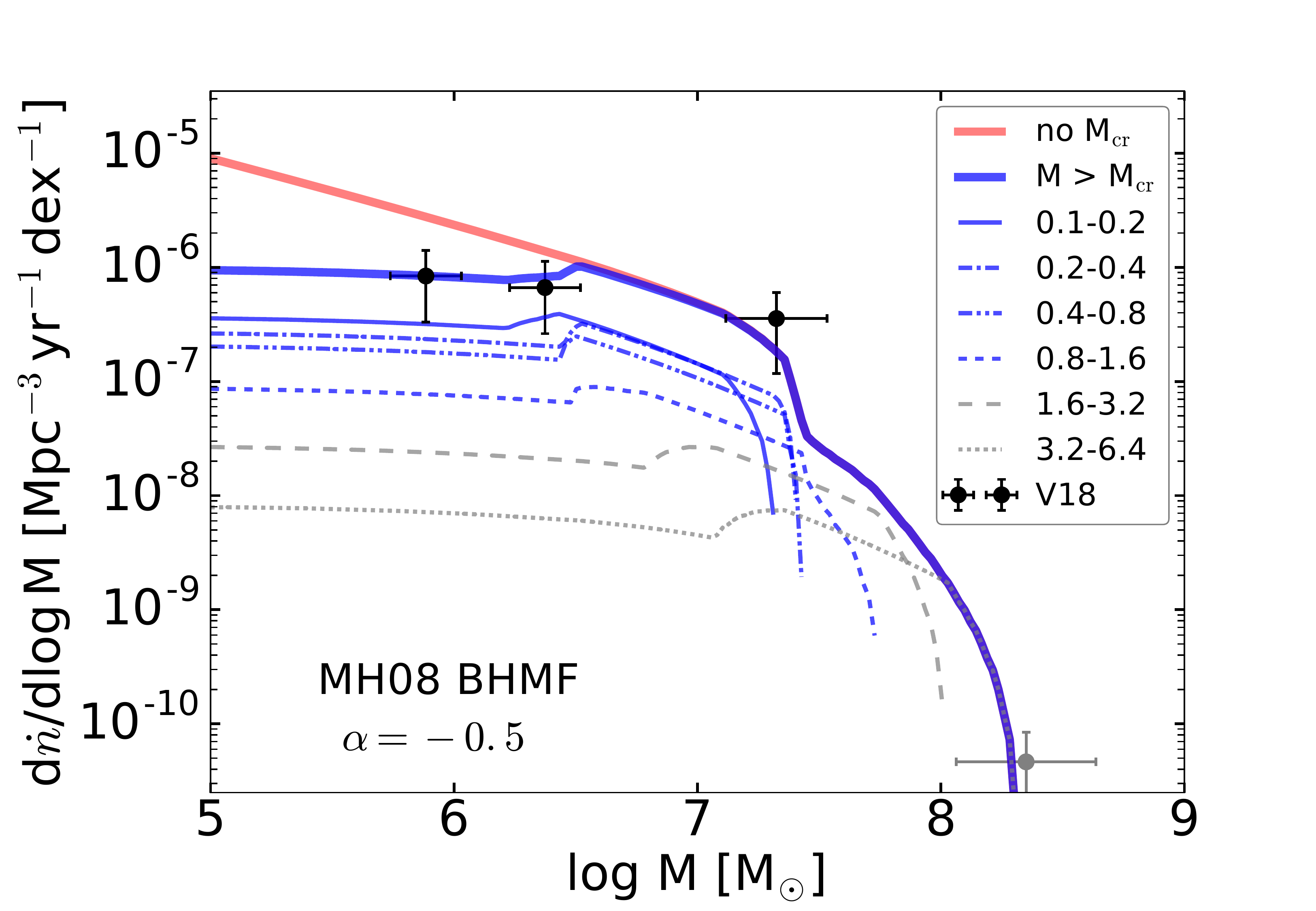}
\caption[caption]{The rate of optically bright TDEs as a function of
  BH mass is shown in thick blue curve, based on the BHMF of
  \citet{2008MNRAS.388.1011M}. The upper and lower panels are for
  $\alpha=-0.25$ and $-0.5$, respectively. The other parameters are
  fixed: $\mc{R} = 6\times10^{-4}\rm\,yr^{-1}$, $\eta = 1$, $m_{\rm
    *,max}=6.4$. The total TDE rate 
  without requiring $M>M_{\rm cr}$ (or the launching of CIO) is shown
  in red curve. We decompose the TDE rates into six (logarithmic)
  stellar mass bins as shown in thinner curves: $m_*\in (0.1, 0.2)$,
  $(0.2, 0.4)$, $(0.4, 0.8)$, $(0.8, 1.6)$, $(1.6, 3.2), (3.2,
  6.4)$. Including up to the $(1.6, 3.2)$ or $(3.2, 6.4)$ bin means we
  are considering a stellar population of relatively young age
  $\sim$$500\,$Myr or $\sim$$100\,$Myr, respectively. If we take away
  these two high-mass bins, then the stellar population has age
  $\sim$$3\,$Gyr. The observationally inferred rates by
  \citet[][V18]{2018ApJ...852...72V} are shown for comparison. The
  grey point near $\mr{log}M\sim 8.3$ only contains the TDE candidate
  ASASSN-15lh.
}\label{fig:demo1}
\end{figure}

We take the Kroupa initial mass function
\citep[IMF,][]{2001MNRAS.322..231K} truncated at $m_{\rm*,
  max}$ (related to the age of the stellar population)
\begin{equation}
  \label{eq:43}
  {\d n_* \over \d m_*} = 
  \begin{cases}
    a_1 m_*^{-1.3},\ & \mr{if}\ m_{\rm *,min} < m_* < 0.5,\\
    a_2 m_*^{-2.3},\ & \mr{if}\ 0.5 < m_* < m_{\rm *,max},\\
    0,\ & \mr{otherwise}.
  \end{cases}
\end{equation}
The two constants $a_1$ and $a_2$ are given by the continuity at
$m_*=0.5$ and normalization $\int (\d n_*/\d m_*) \d m_* = 1$. We
ignore compact stellar remnants since they are fewer in number and
are typically swallowed as a whole for $M\gtrsim 2\times
10^5\msun$. We also ignore red giants, because they have long fallback
time $P_{\rm min}\sim 11\mr{\,yr}\, M_7^{1/2} m_*^{-1}(r_*/10)^{3/2}$
(and even longer circularization time) and do not have an
optically thick layer of gas to reprocess the hard disk emission into
the optical band (see \S 5.2.2). The rate of TDEs contributed by
binary stars is lower than that from single stars by a factor of
$\sim$$f_{\rm bi}f_{\rm pin}(R_*/a)^{3/4}\ll 1$, where $f_{\rm bi}$ is
the binary fraction near the galactic center and $a$ is the semimajor
axis of the binary orbit. Tidal breakup of the binary has a larger Roche radius
$r_{\rm T,b}\simeq (a/R_*)\rt$ and hence occurs at a higher rate than
that for single stars by a factor of $\sim$$(a/R_*)^{1/4}$
\citep{2012ApJ...757..134M}. However, stellar disruption is only
possible at high impact parameter $\beta \gtrsim a/R_*$ in pinhole
regime, which means the disruption rate is a 
factor of $f_{\rm pin}R_*/a$ smaller than the tidal breakup rate.

The Kroupa IMF extends down to $m_*=0.08$
and then becomes shallower $\d n_*/\d m_*\propto m_*^{0.3}$ for lower
mass brown dwarfs. However, TDEs of such low-mass objects likely do
not generate much optical emission, the reason being as follows. The mass of the
reprocessing CIO can be estimated by $M_{\rm CIO}\sim \rho A \Delta t$
\citep{2018ApJ...865..128L}, where $\rho\simeq 
(\kappa r_{\rm ph})^{-1}$ is the gas density, $A \simeq L_{\rm
  opt}/\sigma_{\rm SB}T^4\simeq \Omega r_{\rm ph}^2$ is the surface
area of the optical photosphere, $\kappa$ is the effective absorption
opacity, $r_{\rm ph}$ is the photospheric 
radius, $L_{\rm opt}$ and $T$ are the optical luminosity and blackbody
temperature, and $\Delta t$ is the peak duration. Since half of the
star's mass is in unbound tidal debris and only half of the bound mass
may be ejected as CIO, we obtain a lower limit for the star's mass
$M_*\gtrsim 4M_{\rm CIO}$. Putting in conservative numbers, we obtain
\begin{equation}
  \label{eq:44}
  M_*\gtrsim (0.18\,\msun)\, {L_{\rm opt,43}^{1/2} v_9\over
    T_{4.5}^2 \kappa_{-2}} \sqrt{\Omega\over 2\pi} {\Delta t\over
  10\rm\,d}.
\end{equation}
Fast transients with $\Delta t\lesssim 10\,$d and $L_{\rm opt}\lesssim
10^{43}\rm\,erg\,s^{-1}$ are increasingly likely
to have been missed by current surveys. In the following, we take the
conservative minimum stellar mass of $m_{\rm *,min}=0.1$. Larger $m_{\rm *,min}$
will lead to lower rates of optically bright TDEs.

We plug eqs. (\ref{eq:42}), (\ref{eq:43}) and a given BH mass function into
eq. (\ref{eq:40}) and calculate the integrated volumetric TDE rate
\begin{equation}
\label{eq:45}
\begin{split}
      \dot{n} =&\, \mc{R} \int_{m_{\rm *,min}}^{m_{\rm *,max}} \d m_*\, {r_*^{1/4}
\over m_*^{1/12}} {\d n_*\over \d m_*} \int_{\beta_{\rm min}}^{\infty}
\d \beta\, {\d P\over \d\beta} \\
&\, \int_{M_{\rm cr}}^{M_{\rm max}} \d
\mr{log}M\, M^{\alpha} {\d n_{\rm BH}\over \d \mr{log}M},
  \end{split}
\end{equation}
where the minimum BH mass for CIO launching $M_{\rm cr}$ is given by
eq. (\ref{eq:19}) and the BH mass above which the entire star gets
swallowed is given by eq. (\ref{eq:13}).

\begin{figure}
  \centering
\includegraphics[width = 0.48\textwidth,
  height=0.25\textheight]{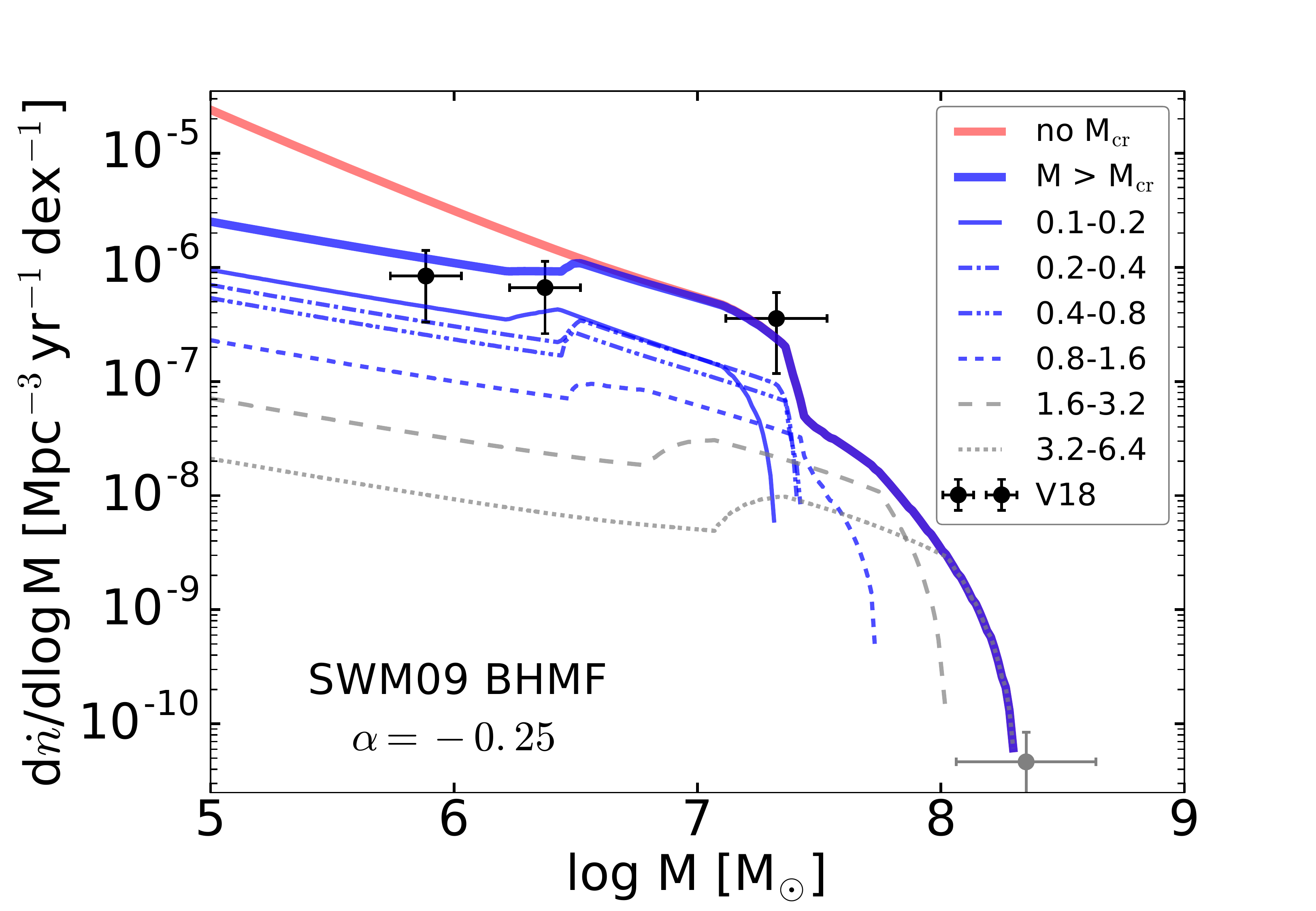}
\includegraphics[width = 0.48\textwidth,
  height=0.25\textheight]{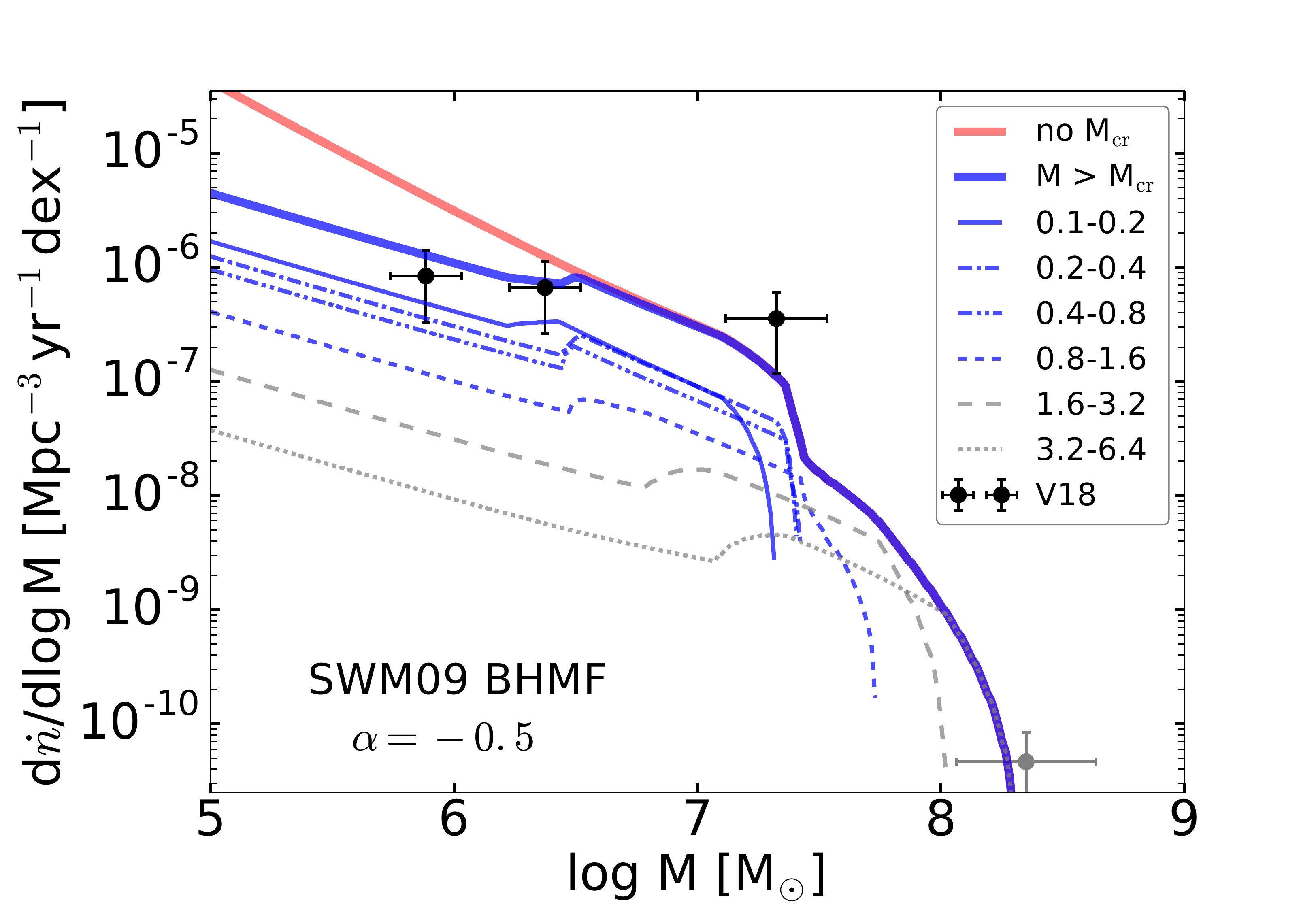}
\caption[caption]{The rate of optically bright TDEs as a function of
  BH mass, for the BHMF by \citet{2009ApJ...690...20S}. The rate
  normalization constant is $\mc{R}=3\times10^{-4}\rm\,yr^{-1}$. All
  other parameters are the same as in Fig. \ref{fig:demo1}.
}\label{fig:demo2}
\end{figure}

The BHMF for $M\lesssim 10^{6.5}\msun$ is highly 
uncertain even in the local Universe. Evolutionary models are
constructed by inferring BH growth by the ``observed'' bolometric
luminosity function of active galactic nuclei (AGN). Various treatments of
bolometric corrections, radiative efficiency of the accretion disks,
and AGN duty cycles may give different results. In this paper, we take
two different BHMFs for the local Universe by
\citet[][MH08]{2008MNRAS.388.1011M} and
\citet[][SWM09]{2009ApJ...690...20S}, as shown in Fig. \ref{fig:BHMF}
in the Appendix. The main difference between the two 
lies in the low-mass end: the MH08 mass function is nearly flat
while the SWM09 mass function rapidly diverges\footnote{We see that
  TDE demographics provide a valuable, direct probe of the BH mass
  function on the low-mass end.} as $\d n_{\rm
  BH}/\d \mr{log}M \propto M^{-0.6}$. Fig. \ref{fig:demo1} and
Fig. \ref{fig:demo2} shows the TDE demographics for these two BHMFs,
respectively. The BHMF can also be directly
calculated by applying correlations between BH mass, bulge luminosity
and stellar velocity distribution for galaxies in the local Universe,
as done\footnote{The two methods of obtaining
  the BHMF are not independent. Typically, the radiative efficiency
  of AGN is calibrated by the total BH mass density
  in the local Universe inferred from galaxy scaling relations
  \citep{1982MNRAS.200..115S, 2004MNRAS.351..169M}.} by
\citet{2004MNRAS.351..169M}. We also tried using 
their BHMF and found that it gives similar results as the MH08 mass
function, as shown in Fig. \ref{fig:demo3} in the Appendix.

On the low BH-mass side, the predicted rate of optically bright TDEs
is nearly flat with respect to the BH mass. This is 
because those TDEs with $M<M_{\rm cr}$ have been filtered out due to
insufficient amount of CIO being launched. Our results
roughly agree with the rate given by \citet{2018ApJ...852...72V}, which
was based on the ``V/V$_{\rm max}$'' method and the BH masses are
inferred from galaxy scaling relations with updated stellar velocity dispersion
by \citet{2017MNRAS.471.1694W}. We also show the total TDE rate without
requiring $M>M_{\rm cr}$ (red curves), which rises more rapidly towards the
low-mass end. This is because TDEs favor
smaller BHs by the factor $M^{\alpha}$ (we have taken
$\alpha=-0.25$ or $-0.5$) and the BHMF
model of \citet{2009ApJ...690...20S} diverges towards the low-mass end
(the MH08 model has a shallower behavior). Unfortunately, the current
small-number statistics are not able to discriminate between the two
scenarios (shown in blue and red curves) at a significant confidence level. 

Thus, our picture predicts that the 
majority of TDEs by BHs with $M\lesssim 10^6\msun$ are not optically bright
and will hence be missed by current optical transient surveys. The
rate of optically bright TDEs is a factor of $\sim$10 or
more\footnote{ In Figs. \ref{fig:demo1} and \ref{fig:demo2}, if we
  take away the $m_*\in (0.1, 0.2)$ bin due to 
  insufficient mass for the reprocessing layer, the rate of optically
  bright TDEs will be lower by a factor of $\sim$2 (but the overall
  shape of the rate as a function of BH mass stays nearly the
  same). We also note that there could be a large population of TDEs
  hidden from optical view by dust extinction
  \citep{2018MNRAS.477.2943W}.} 
lower than the total TDE rate. Some of
these missing TDEs should be observable by wide field-of-view soft X-ray
surveys like eROSITA \citep{2011MSAIS..17..159C} and Einstein Probe
\citep{2015arXiv150607735Y}. Our picture can be tested by comparing
the detection rates of TDEs in the X-ray and optical bands, although
one should keep in mind that low BH-mass TDEs may have a long
circularization timescale due to weak apsidal precession (see \S5.1).


On the high BH-mass end, the optically
bright TDE rate is strongly suppressed due to stars being swallowed
by the event horizon, which has been used as a supportive evidence for
the existence of BH event horizon \citep{2017MNRAS.468..910L} and that
the observed candidates are actually TDEs
\citep{2018ApJ...852...72V}. We note that the grey data point near
$\mr{log}M\sim8.3$ only contains the TDE candidate
ASASSN-15lh, whose 
nature is still being debated \citep{2016Sci...351..257D, 2016NatAs...1E...2L,
  2018A&A...610A..14K}. In our picture, it can be explained by
disruption of a relatively massive star $m_*\sim 5$ by a non-spinning
BH. Disruption of a Sun-like star by a rapidly spinning BH is also
possible, because $\beta_{\rm min}$ can be
smaller than $0.6\,\xi_*^{-1}$ for a prograde orbit
\citep{2012PhRvD..85b4037K, 2016NatAs...1E...2L}.

Finally, we note that rare post-starburst galaxies are over-represented in
the current sample of TDE host galaxies by a factor of $\sim$20 to $\sim$100
\citep{2014ApJ...793...38A, 2016ApJ...818L..21F, 2017ApJ...850...22L,
  2018ApJ...853...39G}, which may be due to higher stellar
density concentration near the galactic centers
\citep[e.g.][]{2017ApJ...850...22L, 2018MNRAS.480.5060S}. Our method
also applies to the group of post-starburst galaxies (with a higher rate
normalization constant $\mc{R}$), as long as their BHMF is
similar to that of the entire galaxy population. An important difference is that
the age of the stellar population near the centers of post-starburst
galaxies may be significantly younger than that for the other normal galaxies,
ranging from 100 Myr to 1 Gyr. This will affect the TDE rate on the 
high BH-mass end. Another potential difference is that the pinhole
fraction $f_{\rm pin}$ may be lower for more cuspy (steeper) stellar
density distribution in post-starburst galaxies
\citep{2018MNRAS.480.5060S}.

\section{Discussion}
In this section, we discuss a number of issues that require further
thoughts in future works.

(1) The stream self-intersection may be delayed due to Lense-Thirring (LT)
precession, if the BH's spin is misaligned with the angular momentum
of the stellar orbit \citep[e.g.][]{1994ApJ...422..508K,
  2013ApJ...775L...9D, 2015ApJ...809..166G, 2016MNRAS.461.3760H}. For
highly eccentric orbits, the angle by which the orbital angular momentum vector
precesses over one period is $(\Delta \omega)_{\rm LT}\approx 4\pi
a(\rg/2\rp)^{3/2}\sin i$ (to leading post-Newtonian
order), where $a$ is the dimensionless spin of the BH and $i$ is the
inclination angle ($i=0$ for spin-orbit alignment). For a given orbit,
we express the maximum ratio of the stream width over the distance to
the BH as $(H/r)_{\rm max} = f_{\rm H}\beta R_*/\rt$, 
where $f_{\rm H}$ describes possible broadening of the
stream due to apsidal/LT precession\footnote{Without LT precession, the ratio
  between the velocity perpendicular to the orbital plane $v_\perp$
  and the velocity within the orbital plane $v_{\para}$ is of order
  $v_{\perp}/v_{\para}\sim (M_*/M)^{1/3}\ll 1$. However, for strong
  LT precession $(\Delta \omega)_{\rm LT}\gtrsim v_\perp/v_\para$, a
  fraction of the $v_{\para}$ component is aligned 
  with the direction of vertical compression, so the stream width after
  the bounce may be broader than in the case without LT
  precession. Strong apsidal precession can also cause the tidal
  compression in the orbital plane to be oblique and hence part of the
orbital velocity may be dissipated near the pericenter.},
hydrogen recombination, and magnetic 
fields. Then, intersection may be avoided for a particular orbit when
$(\Delta \omega)_{\rm LT}\gtrsim (H/r)_{\rm max}$, i.e.
\begin{equation}
\label{eq:47}
  M_6^{4/3} m_*^{1/6} r_*^{-3/2} \gtrsim 1\, f_{\rm H}\, a^{-1} \beta^{-1/2}.
\end{equation}
We can see that TDEs by slowly spinning $a\ll 1$ low-mass $M_6\lesssim
1$ BHs are expected to have prompt intersection between the first and 
second orbits (as shown in Fig. \ref{fig:orbit}). For rapidly spinning
high-mass BHs, intersection may be avoided promptly (if $f_{\rm H}\sim
1$) but will eventually occur with a delay. From the point of view of an
observer who defines $t=0$ as the moment of stream intersection, the
delay itself is not important, since the mass flux of the
stream stays unchanged. On the other hand, as long as the intersection
occurs between two adjacent orbits \citep[the n-th and the
n+1-th, as found by][]{2015ApJ...809..166G}, the intersection
radius ($r_{\rm I, LT}$) under LT precession is roughly the same as
that without LT precession ($r_{\rm I}$). This is because for most TDEs 
the apsidal precession angle ($3\pi\rg/\rp$) is much larger than 
the LT precession angle. Thus, the intersection radius $\ri$, angle
$\t{\theta}_{\rm I}$ and velocity $\t{v}_{\rm I}$ calculated in the
Schwarzschild spacetime are similar to those for spinning BHs.
Therefore, our model for the hydrodynamical collision process, including
redistribution of angular momentum/specific energy and the 
possibility of launching the CIO, should be largely applicable.

(2) TDE demographics on the high BH-mass end depends on the
spin distribution. In the case where the star's initial angular
momentum is parallel to the BH's spin angular momentum, the pericenter
radius of the marginally bound parabolic orbit is $r_{\rm mb} =(1 +
\sqrt{1-a})^2\rg$ \citep{1972ApJ...178..347B},
where $-1<a<1$ is the spin parameter of the BH ($a<0$ for retrograde
orbits). The marginal disruption case
corresponds to $r_{\rm p} \approx r_{\rm mb}\approx 5^{1/3}\xi_*\rt$,
which gives the maximum mass for Kerr BHs hosting TDEs
\begin{equation}
  M_{\rm max,Kerr} = (8.9\times10^{7}\msun) \left(2\over 1 +
    \sqrt{1-a} \right)^3 {\xi_*^{3/2} 
  r_*^{3/2}\over m_*^{1/2}}.
\end{equation}
We can see that the $M_{\rm max,Kerr}$ is strongly affected by
the BH spin only when $a\gtrsim 0.5$. The maximum BH mass is also
affected by the age of the stellar population ($m_{\rm *,max}$,
stellar interior structure $\xi_*$, and the number of evolved
subgiants). It may be difficult to extract the information on the BH
spin distribution from TDE rate on the high BH-mass end. We also note
that the critical mass $M_{\rm cr}$ (above 
which significant amount of CIO is launched) is mainly affected by the
spin-independent de Sitter term of the apsidal precession, so the TDE
demographics on the low BH-mass end should be insensitive to the BH
spin distribution.


(3) We have assumed that the two colliding streams have the
same cross-section and that there is no offset in the transverse
direction. This is reasonable provided that (i) all processes 
occurring when the fallback stream passes near the pericenter $r\sim
\rp$ before the collision are largely reversible\footnote{This means that, if
we denote the two colliding ends as $\mc{C}_1$ and $\mc{C}_2$ (in
chronological order) and reverse the velocity at $\mc{C}_2$, the
stream will evolve back to the conditions at $\mc{C}_1$ (except for
the velocity being in the opposite direction).} and that (ii) the
angular stream widths $H/r$ are larger than than the amount of LT
precession per orbit. However, there could
be many irreversible processes occurring near the pericenter, including:
(i) apsidal and LT precession causing the tidal compression to be
oblique (instead of perpendicular to the orbital velocity); (ii) mass
loss along with the bounce following the tidal compression; (ii)
viscosity causing exchange of angular momentum between adjacent
shear layers. The general relativistic evolution of the fallback
stream over multiple orbits is 
still an open question, mainly because the extremely large aspect
ratio makes it a challenging task for numerical simulations. If these
irreversible processes are indeed important and eq. (\ref{eq:47}) is
satisfied, our current model needs two additional parameters: the
ratio of the cross-sections of 
the two colliding streams and the fractional offset in the transverse
direction. The hydrodynamics of the stream-stream collision and the subsequent
expansion of the shocked gas may be largely modified. This is out of
the scope of the current work and should be studied in the future.

(4) The energy radiated in the UV-optical band is typically
$\lesssim$$10^{51}\,$erg, which is much smaller than the energy budget of
the system, even assuming radiatively inefficient accretion
\citep{2015ApJ...806..164P, 2018ApJ...865..128L}. In our 
picture, this ``missing energy'' puzzle may be explained in two
possible scenarios. The first is that the disk bolometric
emission is capped near the Eddington level for an extended amount of
time\footnote{The late-time (5-10 yrs) UV-optical emission from a
  number of TDEs reported by \citet{2018arXiv180900003V} supports this
  scenario, but the it is also possible that the late-time excess is
  due to dust scattering echo (which has been seen in many supernovae).}
($\gg P_{\rm min}$) but the CIO reprocesses the disk emission in to
the UV-optical band only for a timescale of order $P_{\rm min}$ (then the
reprocessed emission moves into the EUV and soft X-ray as the trapping
radius shrinks). Since the observed peak 
UV-optical luminosity is also near the Eddington limit
\citep[e.g.][]{2017MNRAS.471.1694W}, the 
efficiency for reprocessing, defined as the observed UV-optical
luminosity divided by the intrinsic disk luminosity, is required to be
of order unity in this case. The second scenario is that the disk
bolometric emission significantly exceeds the Eddington limit
\citep[as in the simulations by][]{2017arXiv170902845J}, but the
reprocessing efficiency is much 
less than unity. The reason for a low reprocessing efficiency
could be that, if $\rtr\gg \rin$, photons are trapped in the expanding CIO and hence
their energy is adiabatically lost in the form of $PdV$ work.
However, detailed radiation-hydrodynamic simulations are needed to
distinguish between these two scenarios. 


\section{Summary}
We have described a semi-analytical model for the dynamics of TDEs,
including the properties of the fallback stream before the
self-intersection and the fate of the shocked gas after the
intersection. We circumvent the computational challenge faced by previous
TDE simulation works by assuming that the post-disruption bound stream
follows the geodesics in the Schwarzschild spacetime until the
self-intersection. Then we 
numerically simulate the (non-relativistic) hydrodynamical collision
process in a local box at the intersection point. Since the
cross-sections of the two colliding streams are much smaller than the
size of the orbit and the streams are pressureless (or cold) before
the collision, the collision process and the expanding structure of
the shocked gas are self-similar. This allows us to explore a wide
range of TDE parameter space in terms of the stellar mass, BH mass,
and impact parameter. Our method provides a way for
global simulations of the disk formation process by injecting gas at
the intersection point according to the velocity and density
profiles (eqs. \ref{eq:12} and \ref{eq:14}) shown in this paper.

The most important observational implication is that a large fraction
of the fallback gas can be launched in the form of a collision-induced
outflow (CIO) when the BH mass is above a critical value 
$M_{\rm cr}$ (eq. \ref{eq:19}). We propose that the CIO is
responsible for reprocessing the accretion disk emission from the EUV
or soft X-ray to the optical band. This picture can naturally explain
the large photospheric radius of $\sim$$10^{14}$--$10^{15}\,$cm (or low blackbody
temperature of a few$\times10^4\,$K), and the typical widths of the H
and/or He emission lines. We predict the CIO-reprocessed spectrum
in the infrared to be $L_{\nu}\propto \nu^{\sim 0.5}$, shallower than
a blackbody. A blackbody fit to the optical SED, as commonly done in
the literature, may underestimate the true color temperature. Our
picture is different from that of \citet{2015ApJ...806..164P} in that
the radiation energy ultimately 
comes from the accretion flow rather than the stream collision \citep[which is
shown to be nearly adiabatic,][]{2016ApJ...830..125J}. Our model is 
also different from that of \citet{2016MNRAS.461..948M} in that we
identify the physical origin of the ``reprocessing
layer'' and that this layer is aspherical.

The partial sky coverage of the CIO provides a natural unification of
the diverse X-ray behaviors of the optically selected TDEs. Depending
on the observer's line of sight, an optically bright 
TDE may show strong X-ray emission (when the inner disk is not veiled)
or weak/no X-ray emission (when the inner disk is veiled), which
agrees with the large range of X-ray to optical peak flux ratios:
$\sim$$10^{-4}$ for iPTF16fnl \citep{2017ApJ...844...46B}, $\sim$$10^{-2}$ for
AT2018zr \citep{2018arXiv180902608V}, and $\sim$1 for ASASSN-14li
\citep{2016MNRAS.455.2918H}. As the CIO's mass outflowing rate drops
with time, the X-ray fluxes for veiled TDEs may gradually rise with
time, as observed in ASASSN-15oi and -15lh \citep{2017ApJ...836...25M,
  2017ApJ...851L..47G, 2018MNRAS.480.5689H}. Our picture is different from
those of \citet{2018ApJ...859L..20D} and \citet{2019MNRAS.483..565C}
which describe that the X-ray to optical flux ratio is controlled by
the observer's viewing angle with respect to rotational axis of the
accretion disk (instead of the CIO's outflowing direction).

In cases where the CIO is launched (BH mass $M>M_{\rm cr}$), the rest
of the fallback gas is left in more tightly bound orbits with higher
(sometimes negative) specific angular momentum than the original star,
and hence the circularization process is expected to occur on timescale of
order $\sim$$P_{\rm min}$ after the onset of intersection. If this is
confirmed by future simulations, then it explains the rise/fade
timescale ($\sim$months) of optically bright 
TDEs. We note that the circularization radius of the accreting gas may
be different from $2\rp$ \citep[as generally assumed in the
literature, e.g.][]{1988Natur.333..523R, 2009MNRAS.400.2070S, 
  2014ApJ...784...87S}. Another unexpected prediction is that, in some
cases, the accretion disk rotates in the opposite direction as that of
the initial star.

The total kinetic energy of the CIO spans a wide range from
$\sim$$10^{50}$ up to a few$\times10^{52}\,$erg (in rare cases). The
mass-weighted mean speed varies from $\sim$$0.01c$ to
$\sim$$0.1c$. The shocks driven into the ambient medium by this
outflow can produce radio emission with highly diverse timescales and peak
luminosities, depending on the density profile of the ambient medium,
CIO's velocity and energy, and microphysics of particle
acceleration/magnetic field amplification by the shocks. The radio
emission from ASASSN-14li and a few other TDE candidates may be from
the afterglow of the CIO (instead of the unbound tidal debris, which
typically has a much narrower solid angle).

We also find that the volumetric rate of optically bright TDEs is nearly flat
with respect to the BH mass in the range $M\lesssim 10^7 \msun$. This
is because TDEs with $M<M_{\rm cr}$ have been filtered out due to lack
of significant amount of CIO. Our results roughly agree with the BH
mass function of optically selected TDEs obtained by
\citet{2018ApJ...852...72V}. This 
filtering leads to an optical TDE rate that is a factor of $\sim$10 or
more lower than the total TDE rate (without requiring $M<M_{\rm cr}$).  
For TDEs by BHs with $M<M_{\rm cr}$, the stream self-intersection
becomes less and less efficient at dissipating the orbital energy and
other mechanisms such as MHD turbulence may be responsible for driving
the formation of a circular disk. The circularization timescale of these
TDEs may be much longer than $P_{\rm min}$. Some of them should be
observable by wide field-of-view X-ray surveys like 
eROSITA \citep{2011MSAIS..17..159C} and Einstein Probe
\citep{2015arXiv150607735Y}. Our model can be tested by comparing
the rates of TDEs in the X-ray and optical bands.


\section{acknowledgments}
We are grateful for the discussions with with Eliot Quataert and Tony
Piro on the radiative transfer in a reprocessing wind, with Brian
Metzger and Dan Kasen on the absorption opacity, and with Sterl
Phinney on the effects of stellar interior structure. We also thank
Phil Hopkins, Bing Zhang, Shri Kulkarni, Brad Cenko, Sjoert van
Velzen, and Jim Fuller for useful conversations. We are indebted to
Brian Metzger and Pawan Kumar for reading an earlier version of the
manuscript and providing 
valuable comments. We acknowledge the Texas Advanced Computing Center
(TACC) at The University of Texas at Austin for providing HPC
resources that have contributed to the research results 
reported within this paper. This research benefited from interactions
at the ZTF Theory Network Meeting, funded by the Gordon and Betty
Moore Foundation through Grant GBMF5076 and by the National Science
Foundation under Grant No. NSF PHY-1748958. WL was supported by the
David and Ellen Lee Fellowship at Caltech.

\bibliographystyle{mnras}
\bibliography{refs}

\appendix{}
\section{Supplemental Figures}
In the Appendix, we provide a number of figures to support the
main content. Their descriptions are in the captions.



\begin{figure}
  \centering
\includegraphics[width = 0.48\textwidth,
  height=0.25\textheight]{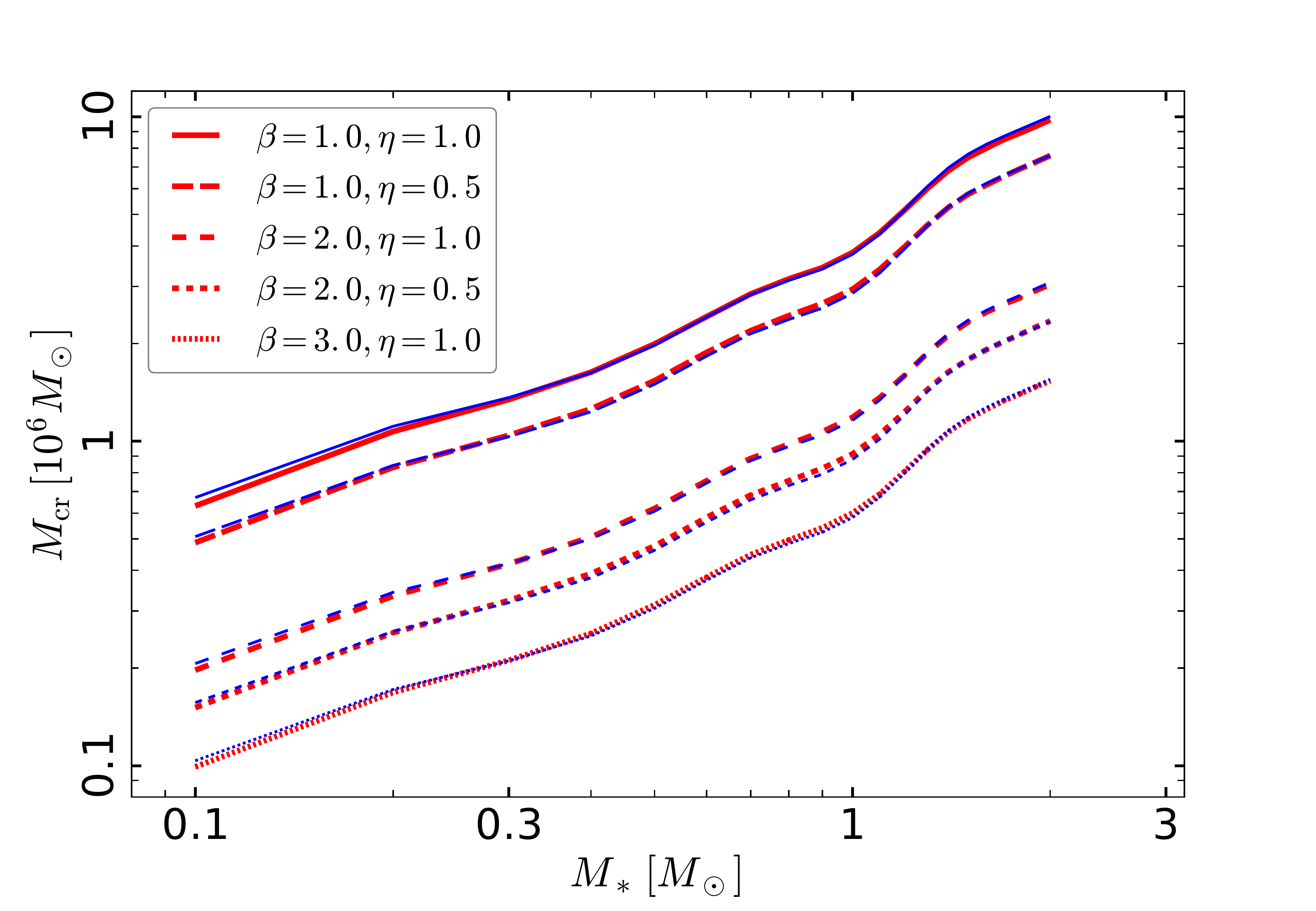}
\caption{The critical mass above which the unbound fraction exceeds
  $20\%$, for a number of cases with parameters indicated in the
  legend. The red curves from numerical calculations almost overlap
  with the blue curves given by eq. (\ref{eq:19}).
}\label{fig:Mcrit}
\end{figure}

\begin{figure}
  \centering
\includegraphics[width = 0.48\textwidth,
  height=0.25\textheight]{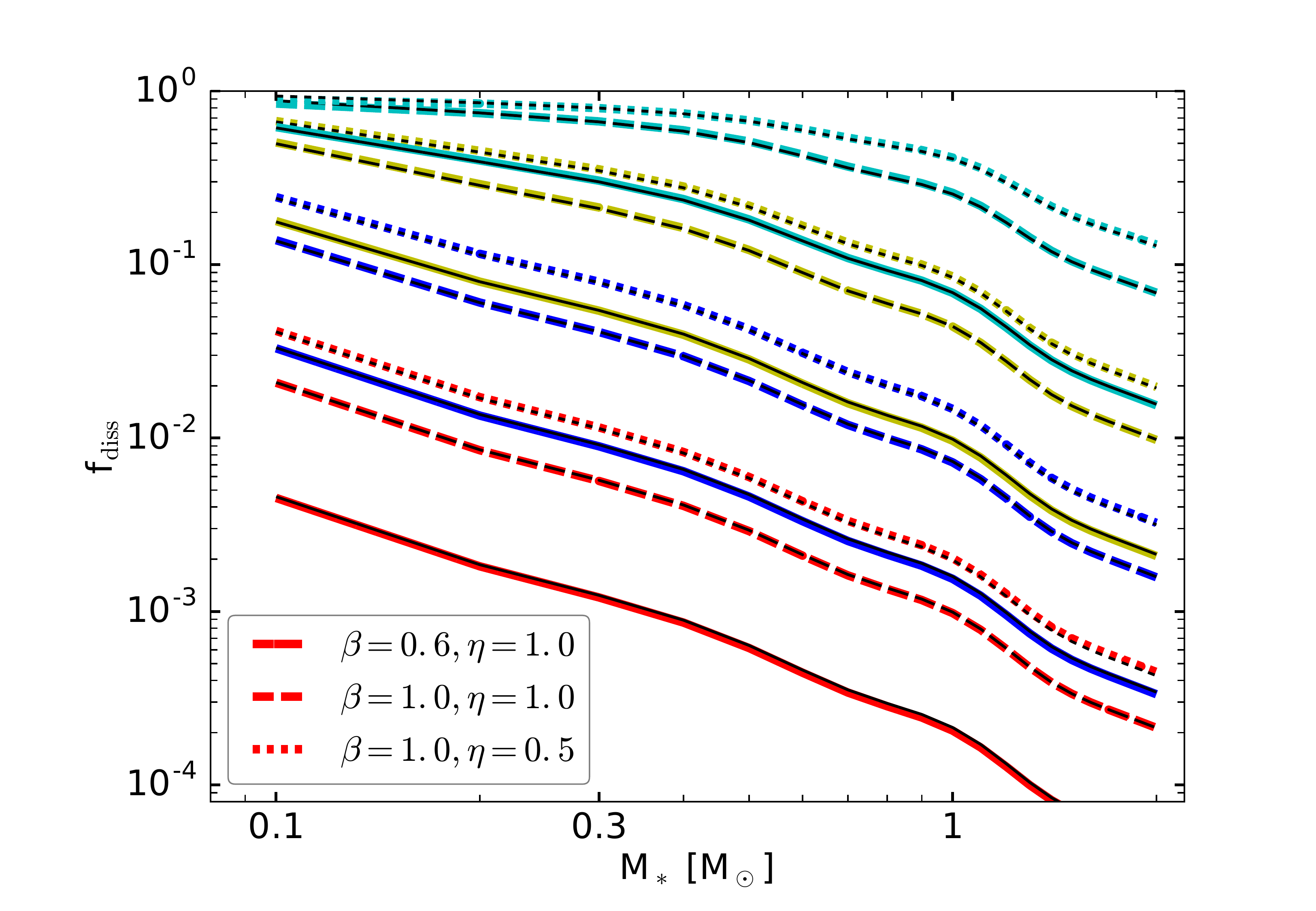}
\caption{The efficiency of orbital energy dissipation $f_{\rm diss}$
  as defined in eq. (\ref{eq:9}), for $M_6 = 0.03$ (red), 0.1 (blue),
  0.3 (yellow) and 1 (cyan curves). For each BH mass, we consider
  three cases with $(\beta,\eta) = (0.6, 1)$ [solid], $(1, 1)$
  [dashed], and $(1, 0.5)$ [dotted curves]. These thick colored curves
  from numerical calculations overlap with the black thin curves,
  which are from the analytical expression in eq. (\ref{eq:20}).
}\label{fig:fdiss}
\end{figure}

\begin{figure}
  \centering
\includegraphics[width = 0.48\textwidth,
  height=0.27\textheight]{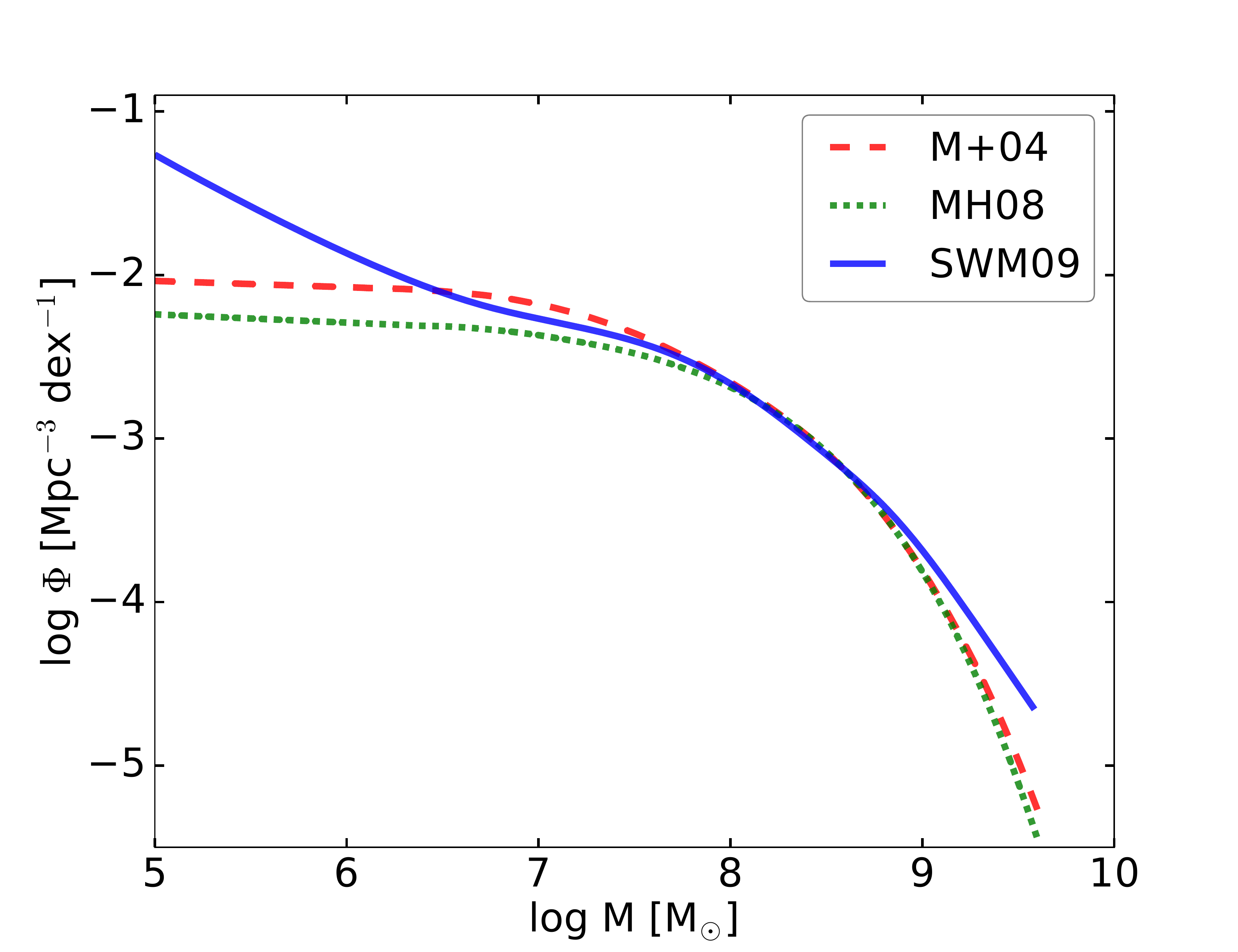}
\caption[caption]{The BHMFs from \citet[][M+04]{2004MNRAS.351..169M}, 
  \citet[][MH08]{2008MNRAS.388.1011M}, and
  \citet[][SWM09]{2009ApJ...690...20S} used in this paper are shown in
  red (dashed), green (dotted), and blue (solid) curves.
}\label{fig:BHMF}
\end{figure}


\begin{figure}
  \centering
\includegraphics[width = 0.48\textwidth,
  height=0.25\textheight]{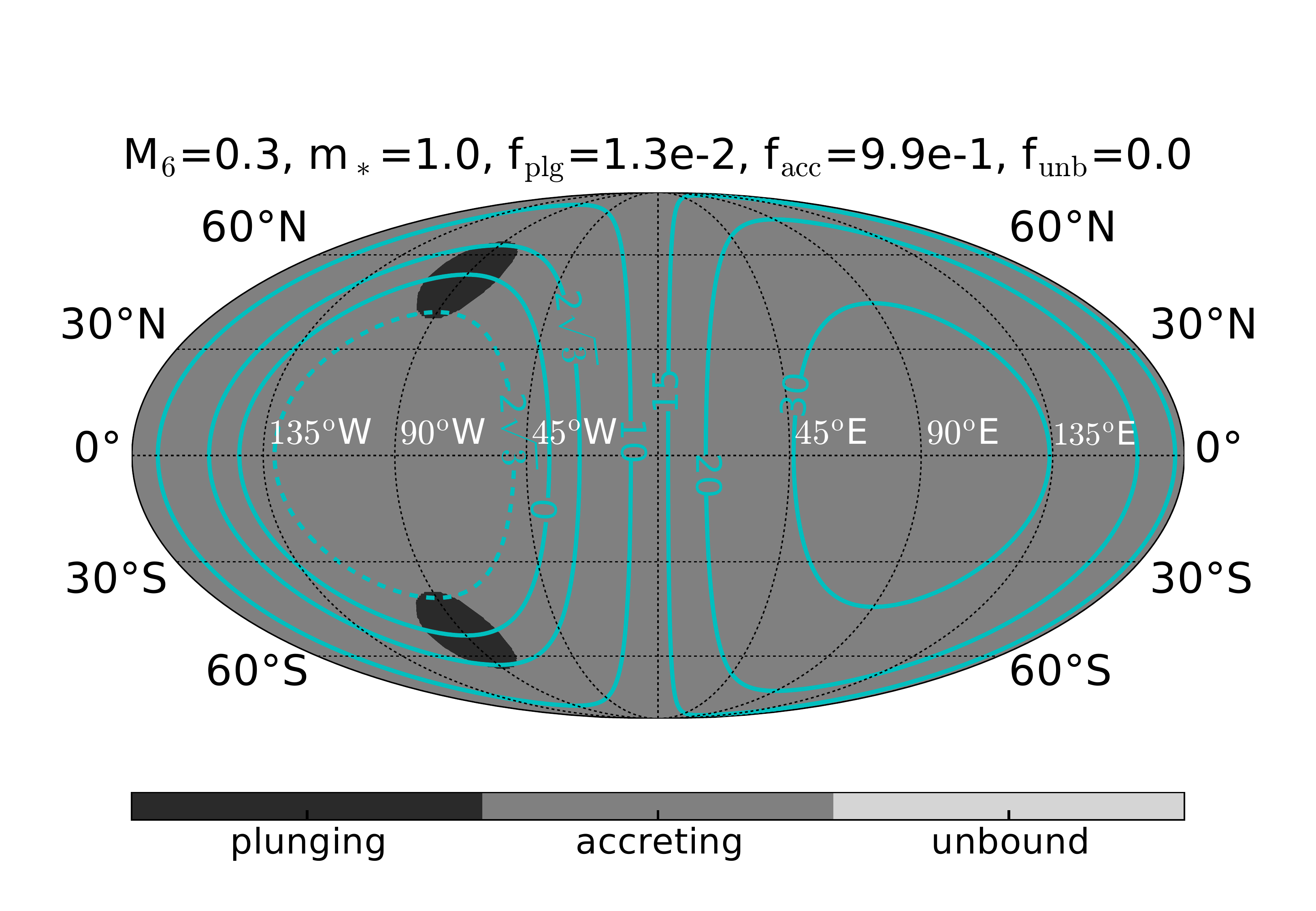}
\includegraphics[width = 0.48\textwidth,
  height=0.25\textheight]{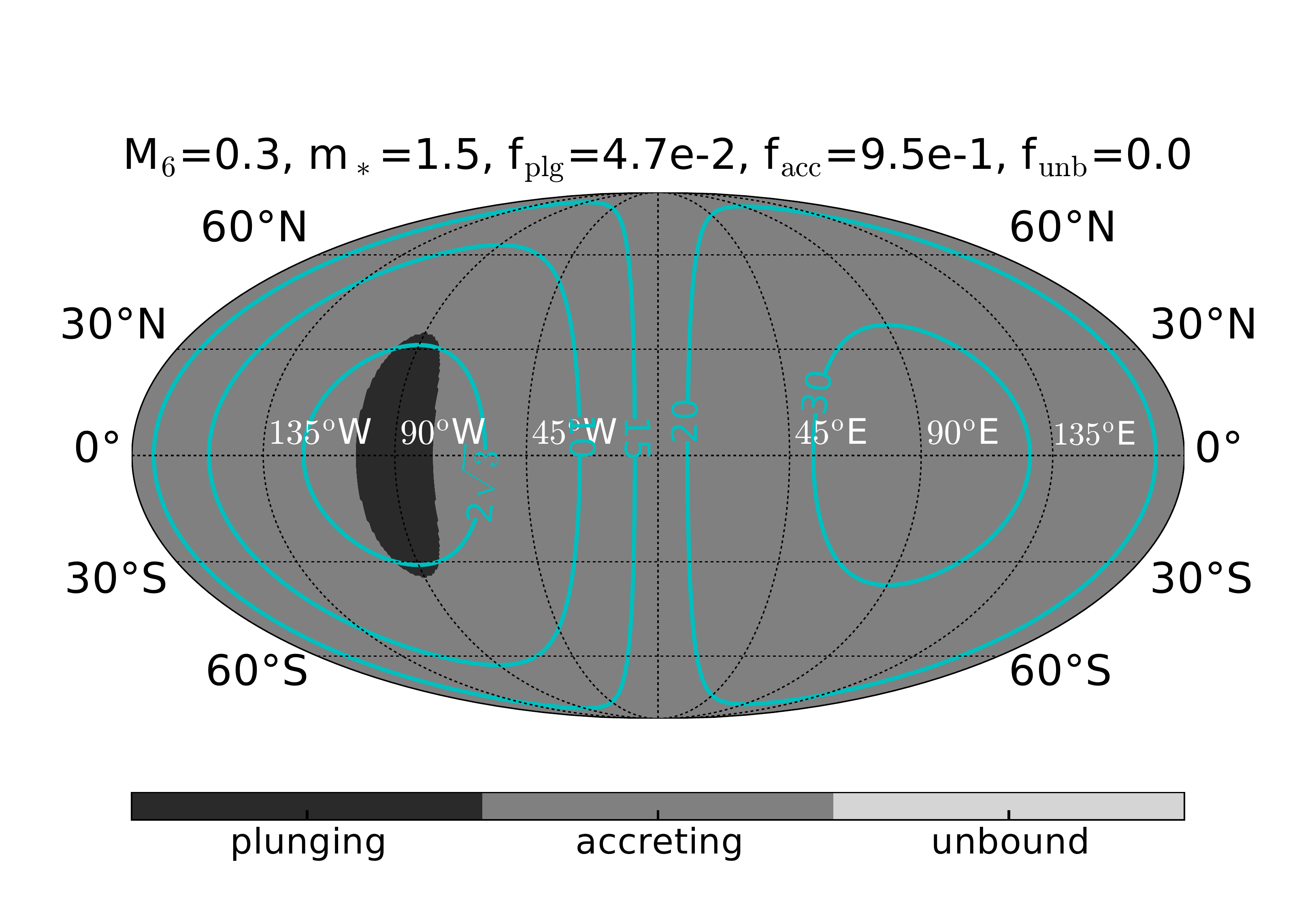}
\caption{The map of fate for the shocked gas expanding in different
  directions for two cases with stellar masses $m_* = 1.0$ (upper
  panel) and $1.5$ (lower panel). The other conditions are the
  same: BH mass $M_6=0.3$, impact parameter $\beta=1.0$, and
  orbital energy parameter $\eta=1.0$. In these two cases, the 
  plunging regions are far from the poles ($\bthe\sim 0$ or
  $\pi$), because the velocity before the collision has comparable
  $\hat{r}$ and $\hat{\phi}$ components: $\t{v}_\phi \sim \t{v}_r$
  (see the third panel of Fig. \ref{fig:intersection}).
  The specific angular momenta of the pre-disruption star are $\ell_0
  \approx 6.5\rg$ (upper panel) and $8.0\rg$ (lower panel).
}\label{fig:fatemap3}
\end{figure}

\begin{figure}
  \centering
\includegraphics[width = 0.48\textwidth,
  height=0.25\textheight]{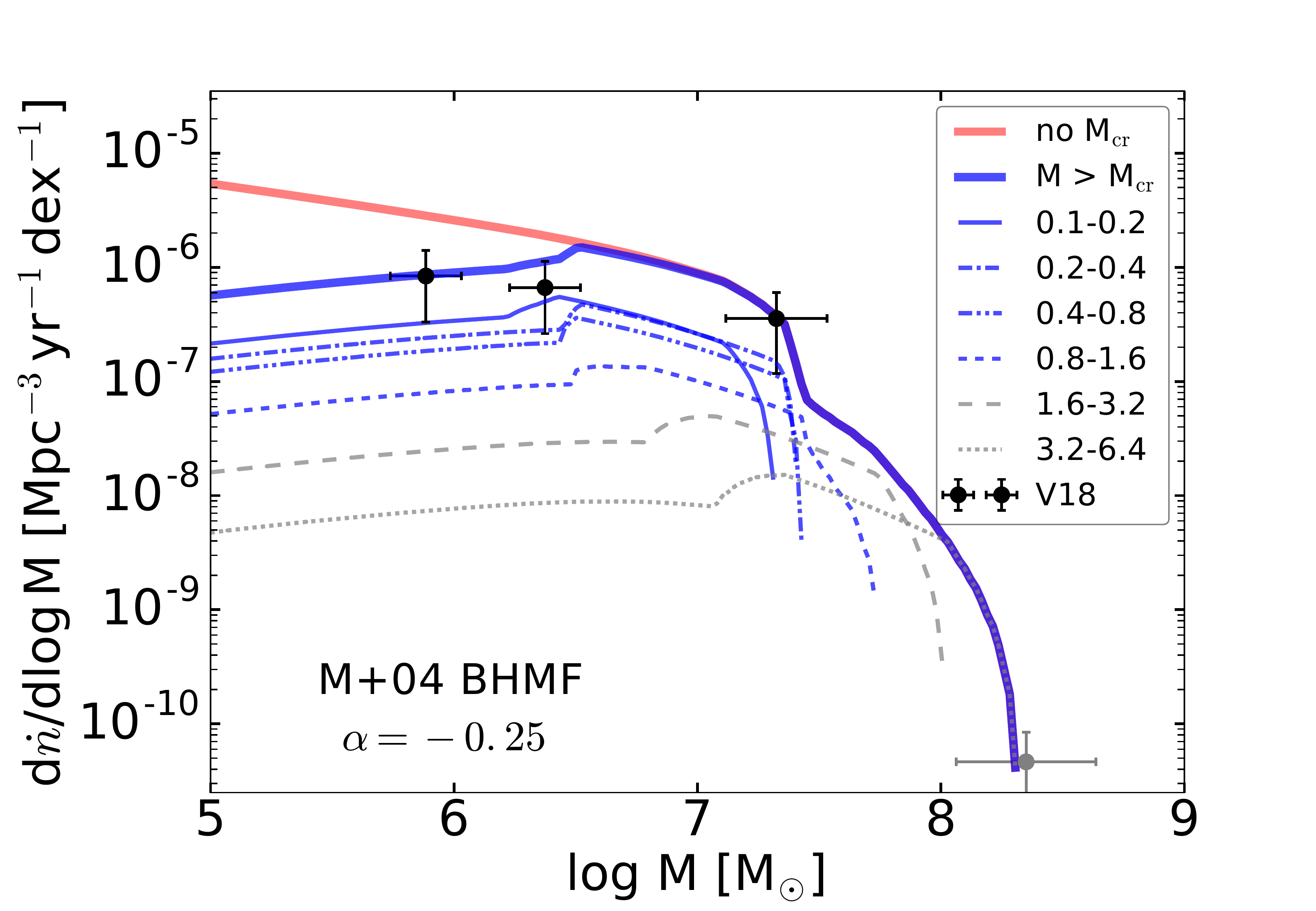}
\includegraphics[width = 0.48\textwidth,
  height=0.25\textheight]{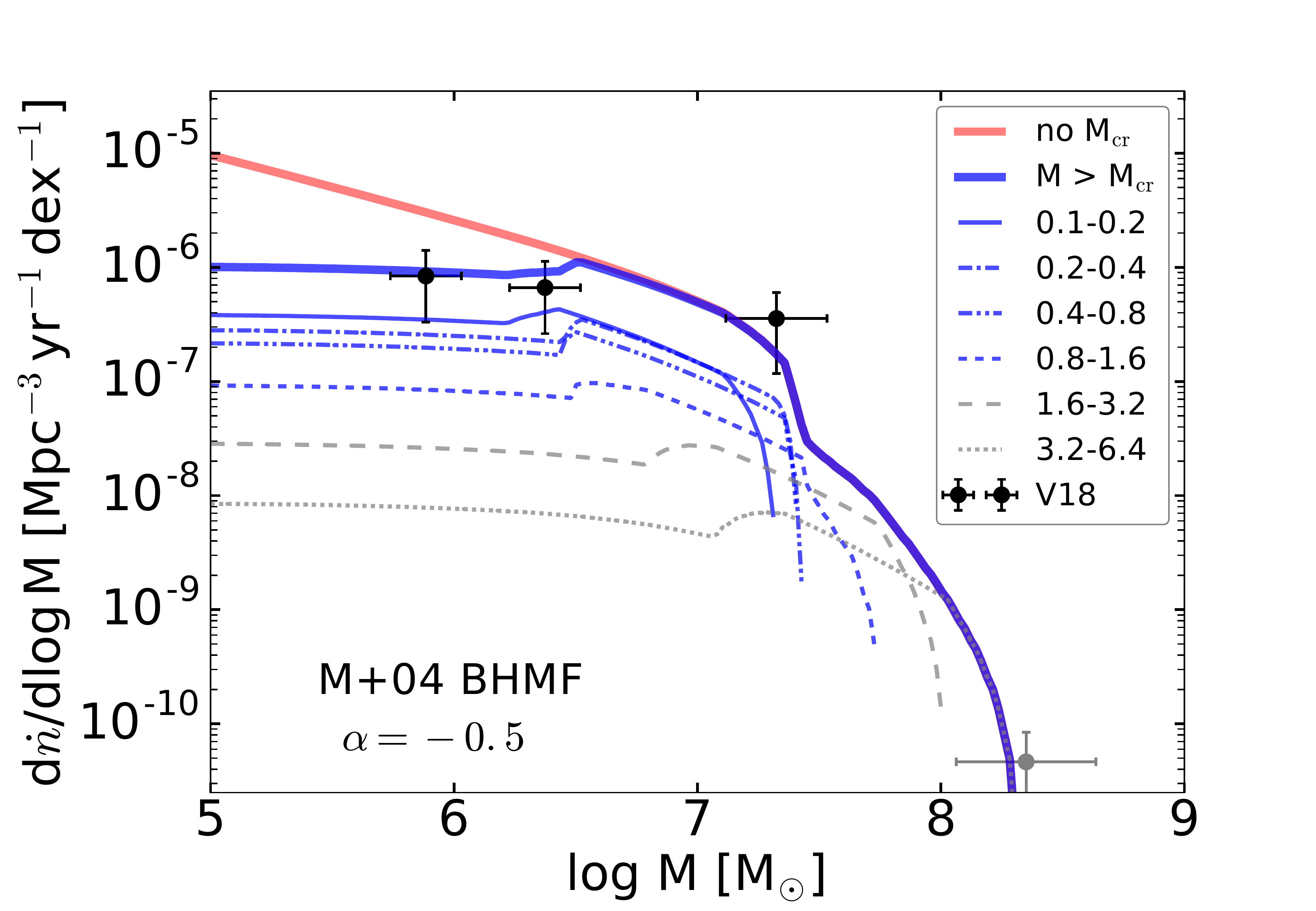}
\caption[caption]{The rate of optically bright TDEs as a function of
  BH mass, for the BHMF by \citet{2004MNRAS.351..169M}. The rate
  normalization constant is $\mc{R}=4\times10^{-4}\rm\,yr^{-1}$. All
  other parameters are the same as in Fig. \ref{fig:demo1}.
}\label{fig:demo3}
\end{figure}

\section{Low Radiative Efficiency of Self-intersection Shocks}

We justify the usage of adiabatic equation of
state for our hydrodynamic simulation in \S 3. At the collision point,
the Thomson scattering optical depth of the stream in the transverse
direction is given by $\tau_{\rm st}\simeq 2\kappa_{\rm s} \rho H$,
where $H$ is the radius of the stream (assuming cylindrical
shape), $\rho$ is the gas density, and $\kappa_{\rm s} =
0.34\rm\,cm^2\,s^{-1}$ is the scattering opacity. The mass flowing
rate of the stream is $\dot{M}_{\rm fb} 
\simeq \pi H^2 \rho v$, and hence
\begin{equation}
  \tau_{\rm st} \simeq {2\kappa_{\rm s} \dot{M} \over \pi H
    v} \sim 6\times10^3 {\dot{M}_{\rm fb}\over \rm\msun\,yr^{-1}} {10
    R_{\odot} \over H} {0.1c\over v},
\end{equation}
where we have used \textit{conservative} values for the flow velocity
$v$, transverse radius $H$, and mass flowing rate (see
eq. \ref{eq:21}). After the collision, the pressure of the shocked gas
is dominated by radiation, and adiabatic expansion 
converts internal energy back into (roughly spherical) bulk motion
over a radius of $r$$\,\sim\,$a few$\times H$.  Beyond this radius, the gas
expands with nearly constant velocity $v$ and the density drops with
radius as $\rho\propto r^{-2}$ (see
Fig. \ref{fig:massflux-radial}). Radiation is advected by 
the expanding shocked gas until the photon trapping radius $r_{\rm
  tr}\simeq \kappa_{\rm s}\dot{M}/(4\pi c)$. The radiative efficiency,
i.e. the ratio between the emergent luminosity and the total
kinetic power, is roughly given by
\begin{equation}
  \left(H\over r_{\rm tr}\right)^{2/3}\sim 5\times10^{-2}
  \left(\dot{M}_{\rm fb}\over
    \rm\msun\,yr^{-1}\right)^{-{2/3}} \left(H \over 10 
    R_{\odot} \right)^{2/3}.
\end{equation}
We see that the stream self-intersection is radiatively inefficient.

\section{Photon Down-scattering in Diffusion Region}
Consider photons diffusing through a scattering slab in the $\hat{\b{x}}$
direction. For a general form of angular dependence for the intensity
\begin{equation}
  I(x,\mu) = \sum_{n=0}^{\infty} I_n(x) P_n(\mu),\ \ \mu \equiv
  \cos\theta\in [-1, 1],
\end{equation}
where $I = \int \d \nu I_\nu$, $P_n(\mu)$ are the Legendre
polynomials (only the first three terms are important here)
$P_0 = 1$, $P_1=\mu$, $P_2 = (3\mu^2-1)/2$, and the 
orthonormality gives $\int P_m P_n \d \mu = 2\delta_{mn}/(2n + 1)$
($\delta_{mn}$ being the Kronecker delta). The energy density, flux,
and pressure of the radiation field are given by the different moments
of intensity $U = (2\pi/c)\int I(\mu) \d \mu = 4\pi I_0/c$, $F = 2\pi
\int I(\mu)\mu \d\mu = 4\pi I_1/3$, and $P = (2\pi/c)\int I(\mu)\mu^2
\d \mu = (4\pi/3c)(2I_2/5 + I_0)$. Now take an electron moving at
velocity $\beta c$ and Lorentz factor $\gamma$ in the $\hat{\b{x}}$
direction. We Lorentz transform the radiation field from the lab frame
to the comoving frame of the electron, where
quantities are denoted with a prime ($'$). In the comoving frame, the
electron gains momentum $\Delta p'$ over time $\Delta t'$ due to
scattering
\begin{equation}
  \begin{split}
   {\Delta p'\over \Delta t'} &= {2\pi \sigma_{\rm T} \over c} \int \d
  \mu' \mu' \int \d \nu' I'_{\nu'}(\mu')\\
  & = {2\pi \gamma^2 \sigma_{\rm T} \over c} \int \d \mu I(\mu)
  (\mu-\beta) (1-\beta\mu),
  \end{split}
\end{equation}
where $\sigma_{\rm T}$ is the Thomson cross-section. Making use of the
orthonormality relations, the integral above can be expressed in terms
of $I_0$, $I_1$, and $I_2$ and hence $U$, $F$, and $P$. Going back to the lab
frame, the electron gains energy at a rate given by
\begin{equation}
  {\Delta E\over \Delta t} = \beta c {\Delta p'\over \Delta t'} =
  \gamma^2 \beta \sigma_{\rm T}c \left[(1+\beta^2){F\over c} - \beta
    (U+P) \right],
\end{equation}
which is at the expense of radiation energy. Therefore, the fractional
energy loss of a photon under each scattering (over a timescale
$\lambda_{\rm mfp}/c$) is given by
\begin{equation}
  {\delta \nu\over \nu} = -{\Delta E\over \Delta t} {\lambda_{\rm
      mfp}\over c} {n_{\rm e}\over U} = -\gamma^2\beta
  \left[(1+\beta^2){F\over Uc} - \beta \left(1 + {P\over U}\right)
\right],
\end{equation}
where the mean free path is $\lambda_{\rm mfp} = (n_{\rm e}
\sigma_{\rm T})^{-1}$. In the limit of isotropic radiation field $F=0$
and $P/U=1/3$, the result $\delta \nu/\nu = 4\gamma^2\beta^2/3$ agrees
with \citet[][eq. 7.16a]{1979rpa..book.....R}. In the diffusion region
of the (non-relativistic) CIO, $\rtr <r <r_{\rm 
  scat}$, the diffusive flux is given by $F \simeq Uc/\tau_{\rm 
  s}\gg \beta$ ($\tau_{\rm s}$ being the scattering optical depth), so
the fractional energy shift per scattering is
\begin{equation}
  \delta \nu/\nu \simeq \beta F/(Uc) = \beta/\tau.
\end{equation}
Since a typical photon undergoes $\sim$$\tau^2$ scatterings before
escaping, the cumulative fractional energy shift is $\tau^2\delta
\nu/\nu \sim \beta \tau$. This justifies eq. (\ref{eq:101}).

\label{lastpage}
\end{document}